\documentclass{jfm}
\usepackage{amsmath,bm}
\usepackage{amsfonts} 
\usepackage{graphicx}
\usepackage{newtxtext}
\usepackage{newtxmath}
\usepackage{natbib}
\usepackage{accents}
\usepackage{caption}
\usepackage{subcaption}

\usepackage{hyperref}
\usepackage{placeins}

\usepackage{xr}
\makeatletter

\newcommand*{\addFileDependency}[1]{
\typeout{(#1)}
\@addtofilelist{#1}

\IfFileExists{#1}{}{\typeout{No file #1.}}
}\makeatother

\newcommand*{\myexternaldocument}[1]{%
\externaldocument{#1}%
\addFileDependency{#1.tex}%
\addFileDependency{#1.aux}%
}

\myexternaldocument{main_suppl}

\hypersetup{
    colorlinks = true,
    urlcolor   = blue,
    citecolor  = black,
}

\newcommand{\be}{\begin{equation}}
\newcommand{\ee}{\end{equation}}
\newcommand{\bea}{\begin{eqnarray}}
\newcommand{\eea}{\end{eqnarray}}
\newcommand{\lb}{\label}

\newcommand{\bn}{{\bf n}}
\newcommand{\br}{{\bf r}}
\newcommand{\bu}{{\bf u}}
\newcommand{\bw}{{\bf w}}
\newcommand{\bx}{{\bf x}}

\newcommand{\bJ}{{\bf J}}
\newcommand{\bS}{{\bf S}}
\newcommand{\bT}{{\bf T}}
\newcommand{\bU}{{\bf U}}
\newcommand{\bW}{{\bf W}}

\newcommand{\bomega}{{\mbox{\boldmath $\omega$}}}
\newcommand{\bsigma}{{\mbox{\boldmath $\sigma$}}}
\newcommand{\bSigma}{{\mbox{\boldmath $\Sigma$}}}
\newcommand{\btau}{{\mbox{\boldmath $\tau$}}}
\newcommand{\boxi}{{\mbox{\boldmath $\xi$}}}
\newcommand{\barphi}{{\mbox{\boldmath $\varphi$}}}

\newcommand{\grad}{{\mbox{\boldmath $\nabla$}}}
\newcommand{\btimes}{{\mbox{\boldmath $\times$}}}
\newcommand{\bdot}{{\mbox{\boldmath $\cdot$}}}
\newcommand{\bdots}{{\mbox{\boldmath $:$}}}
\newcommand{\bzed}{{\mbox{\boldmath $0$}}}

\newcommand{\vertiii}[1]{{\left\vert\kern-0.25ex\left\vert\kern-0.25ex\left\vert #1 
    \right\vert\kern-0.25ex\right\vert\kern-0.25ex\right\vert}}

\newcommand{\RomanNumeralCaps}[1]
\linenumbers

\makeatletter
\newcommand\useleqno{\renewcommand\@eqnnum{\hb@xt@.01\p@{}%
                      \rlap{\normalfont\normalcolor
                        \hskip -\displaywidth(\theequation)}}}

\title{Onsager's ``Ideal Turbulence'' Theory}

\author{Gregory Eyink
  \corresp{\email{eyink@jhu.edu}}}
  
\affiliation{Department of Applied Mathematics, The Johns Hopkins University, Baltimore, Maryland, USA}

\begin{document}
\maketitle

\begin{abstract}
Lars Onsager in 1945-1949 made an exact analysis of the high Reynolds-number limit for individual turbulent flow realizations modeled by incompressible Navier-Stokes equations, motivated by experimental observations that dissipation of kinetic energy does not vanish. I review here developments spurred by his key idea, that such flows are well-described by distributional or ``weak'' solutions of ideal Euler equations. 1/3 H\"older singularities of the velocity field were predicted by Onsager and since observed.  His theory describes turbulent energy cascade without probabilistic assumptions and yields a local, deterministic version of the Kolmogorov $4/5$th law. The approach is closely related to renormalization group methods in physics and envisages ``conservation-law anomalies'', as discovered later in quantum field theory. There are also deep connections with Large-Eddy Simulation modeling. More recently, dissipative Euler solutions of the type conjectured by Onsager have been constructed and his $1/3$ H\"older singularity proved to be the sharp threshold for anomalous dissipation. This progress has been achieved by an unexpected connection with work of John Nash on isometric embeddings of low regularity or ``convex integration'' techniques. The dissipative Euler solutions yielded by this method are wildly non-unique for fixed initial data, suggesting ``spontaneously stochastic'' behavior of high-Reynolds number solutions. I focus in particular on applications to wall-bounded turbulence, leading to novel concepts of spatial cascades of momentum, energy and vorticity to or from the wall as deterministic, space-time local phenomena. This theory thus makes testable predictions and offers new perspectives on Large-Eddy Simulation in presence of solid walls. 
\end{abstract}


\section{Introduction}\label{intro}

This entire essay shall be concerned with a theory of ``ideal turbulence'' which was proposed 
by Lars Onsager in a 1949 paper entitled ``Statistical Hydrodynamics''.  The proposal was set forth 
by Onsager in his characteristic laconic style in the final paragraph of that paper, which I  quote here 
in full:
\begin{quotation}
\noindent 
``It is of some interest to note that in principle, turbulent dissipation as described could take 
place just as readily without the final assistance by viscosity. In the absence of viscosity, 
the standard proof of the conservation of energy does not apply, because the velocity field 
does not remain differentiable! In fact it is possible to show that the velocity field in such `ideal'
turbulence cannot obey any LIPSCHITZ condition of the form
\begin{equation*}
(26)\hspace{100pt} 
|\overrightarrow{v}( \overrightarrow{r^{{\!\,}_\prime}}+\overrightarrow{r})-
\overrightarrow{v}( \overrightarrow{r^{{\!\,}_\prime}})|
< ({\rm const.}) r^n, 
\hspace{100pt} 
\end{equation*} 
for any order $n$ greater than $1/3$; otherwise the energy is conserved. Of course, under the circumstances, 
the ordinary formulation of the laws of motion in terms of differential equations becomes inadequate and 
must be replaced by a more general description; for example, the formulation (15) in terms of FOURIER 
series will do. The detailed conservation of energy (17) does not imply conservation of the total energy if the 
number of steps in the cascade is infinite, as expected, and the double sum of 
$Q(\overrightarrow{k}, \overrightarrow{k'})$ converges only conditionally.'' --- L. \cite{onsager1949statistical} 
\end{quotation} 
In fact, the germ of these remarks were contained in a short abstract on fluid turbulence that Onsager published 
four years earlier,  where he noted that ``...various experiments indicate that the viscosity has a negligible effect 
on the primary process; hence one may inquire about the laws of turbulent dissipation in an ideal fluid.''  
\citep{onsager1945distribution}. Although he never published an argument justifying his 
``it is possible to show'' assertion, it is now known that Onsager had indeed derived a mathematical identity 
which implies his conclusion and which he communicated by letter to Theodore von K\'arm\'an and 
Chia-Chiao Lin in 1945 \citep{eyink2006onsager}. Although Onsager's innovative ideas on this subject were long 
overlooked and conflated with the related theory of \cite{kolmogorov1941local,kolmogorov1941dissipation},
it is now understood  that Onsager's work essentially refined and extended the concepts of Kolmogorov,
anticipating ideas in Large-Eddy Simulation (LES) modelling, modern field-theoretic notions of conservation-law 
anomalies and renormalization-group invariance, and the concept of weak Euler solutions in the mathematical 
theory of partial differential equations. 

{ \begin{figure}
\centering
\includegraphics[width=.4\textwidth]{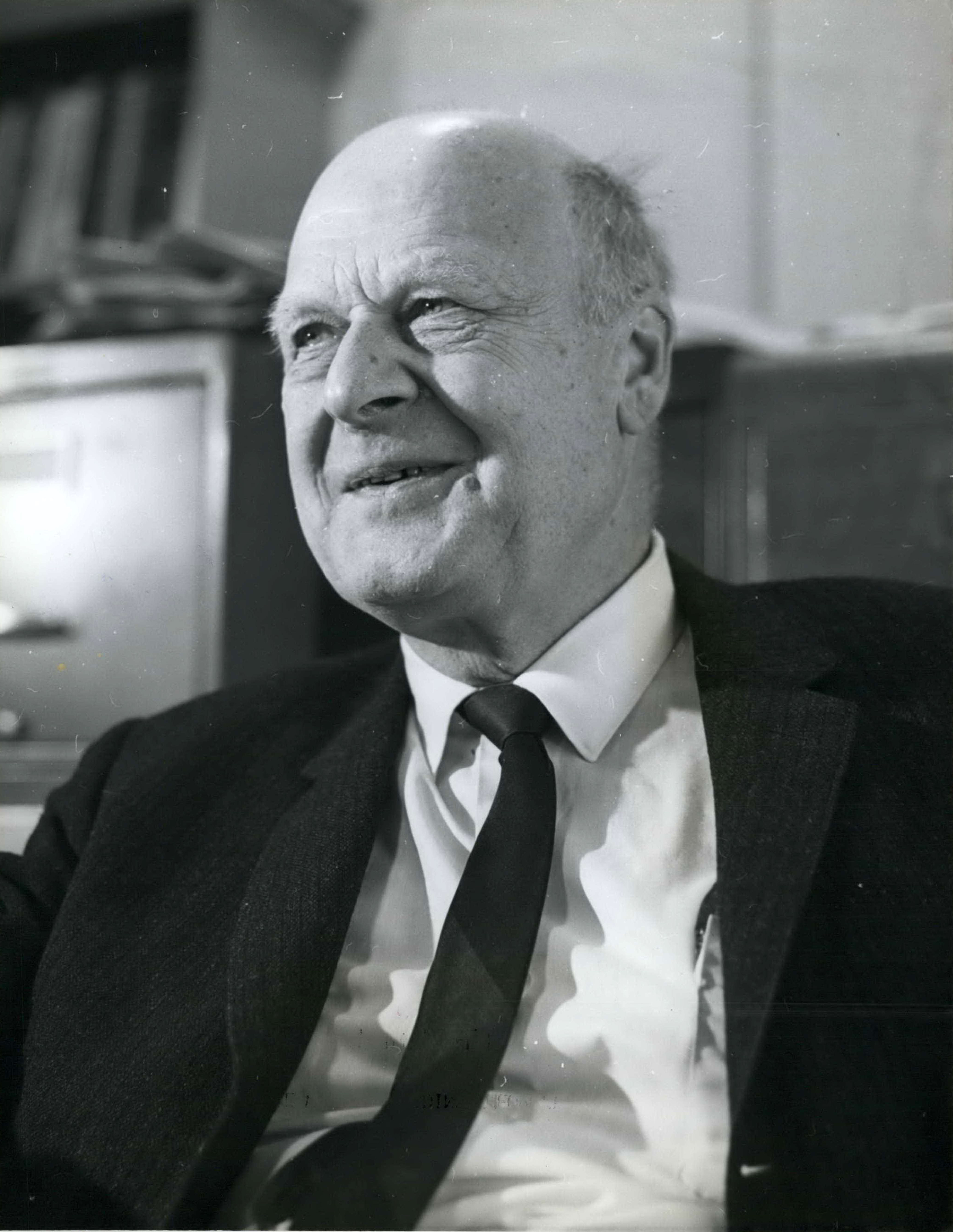}
\caption{Lars Onsager (1903-1976). Photograph published originally in {\it Svenska Dagbladet}
on December 6, 1968 shortly after the announcement of Onsager's award of the Nobel Prize in Chemistry, 
reproduced on license from ZUMA Press.}
\label{correlate}
\end{figure}} 

I first became aware of Onsager's proposal in 1990 when I was a postdoc at Rutgers University 
and Uriel Frisch, visiting for the spring semester, delivered a series of seminars on hydrodynamic turbulence. 
Frisch had rediscovered some of Onsager's key ideas on his own \citep{sulem1975bounds} and he only learned 
of Onsager's prior work himself from Robert Kraichnan in 1972 (Frisch, private communication). Onsager's 
$1/3$ H\"older claim was discussed briefly at the end of Frisch's Rutgers lectures, which followed closely 
an expository article he published that same year \citep{frisch1990turbulence}. (Interestingly, the reference to Onsager's 
1949 paper and the $1/3$ H\"older claim were cut from the published article, presumably because of length 
restrictions.) I looked up Onsager's paper and was immediately impressed by his remarks because, unlike 
Kolmogorov's arguments for 1/3 power-law scaling based upon statistical assumptions and dimensional  
analysis, Onsager claimed that the 1/3 exponent could be derived on a purely dynamical basis  for individual 
solutions of the fluid equations. In fact, we now know that Onsager's 1/3 result is not equivalent to that 
of \cite{kolmogorov1941local,kolmogorov1941dissipation}, e.g. being completely compatible with 
inertial-range intermittency, which Onsager had already anticipated in 1945  \citep{eyink2006onsager}. After consulting 
with two Rutgers experts on mathematics of Navier-Stokes equations, Giovanni Gallavotti and Vladimir Scheffer, 
I was surprised to learn that Onsager's claims were unknown to the PDE community in 1990 and that nothing 
was established about their validity. The situation is now very different and Onsager's ideas have become the focus 
of major developments in the mathematical theory of PDE's, connecting surprisingly with ideas of John Nash  
on a completely different problem of isometric embeddings of Riemannian manifolds \citep{delellis2019turbulence}.
These developments by many people have in turn attracted renewed attention to Onsager's ideas in the fluid mechanics 
community, being indeed the subject of an earlier Perspectives essay by \cite{dubrulle2019beyond}. 

The purpose of the present work is to explain Onsager's ideas in a pedagogical and straightforward manner.
Although I  shall follow somewhat the chronological development of the theory, the emphasis in this work 
is on science and not history. Most of the research on Onsager's theory so far has been by mathematicians 
and this may have led to the impression among many that the subject is an esoteric branch of pure mathematics.
Although the rigorous development leads indeed to some non-trivial mathematical issues, the theory is 
directly motivated by experimental observations and by intuitive physical ideas. Some of these points have been 
explained in my earlier unpublished short note \citep{eyink2018review} and in my online turbulence course 
notes \citep{eyink2007turbulenceI}. In fact, I believe that Onsager's point of view is one of the easiest to 
teach to beginning students of turbulence. The theory does not cover all turbulent phenomena, as it concerns 
the regime of very high Reynolds numbers, but it is not restricted to the idealized limit of infinite Reynolds 
number only (as has been sometimes misunderstood).  It sheds little light, therefore, on the important problem of transition 
to turbulence. However, one of the great virtues of Onsager's approach is the naturalness with which it extends 
concepts of high Reynolds-number turbulence from the traditional problem of incompressible fluid turbulence to more 
general cases of compressible fluids, relativistic fluids, plasma kinetics, and quantum superfluids. See section 
\ref{sec:open}{\it (d)}.
 
There are, by now, a very large number of researchers who have contributed to this subject, many at a deeper 
mathematical level than I have. There are already several illuminating and insightful reviews written by mathematicians,
such as \cite{delellis2013continuous,delellis2019turbulence,buckmaster2020convex}, but the present essay 
is not designed as such a comprehensive review. Instead, my purpose is to explain 
the subject from the personal perspective of a mathematical physicist working for some years in this area. As with any 
currently active field, all researchers may not agree with my proffered interpretations and points of view. 
I endeavor, however, to give technically correct explanations of results, even those to which I have not contributed myself. 
I do not want to present the subject as a {\it fait accompli}, which it is not, but instead as a living scientific theory 
with many critical questions still open. The main reason that I was excited to write this essay is  that I  believe 
that some of the most fundamental problems in this area remain unsolved and call for the combined efforts, not only 
of mathematicians, but also of fluid mechanicians, computational scientists, turbulence modelers and physicists, both 
theorists and experimentalists. The earlier essay of \cite{dubrulle2019beyond} lucidly explained how analytical tools 
developed in the Onsager theory could be applied to direct analysis of empirical data. I  will be  concerned here 
instead with the complementary issue of how the ``ideal turbulence'' theory connects with the numerical modeling method 
of Large-Eddy Simulation (LES), a topic only briefly treated in that earlier work (\cite{dubrulle2019beyond}, section 8.6). 
Although the two subjects have developed historically with little interaction, they are in fact quite intimately connected.
For example, both LES and ``ideal turbulence'' aspire to describe individual realizations of a turbulent flow. I  shall
focus in this essay in particular on the issue of turbulence-wall interactions, which currently concerns deeply both the LES and 
the mathematical PDE communities and which poses some of the greatest current challenges in turbulence research.
As we shall see, Onsager explored this problem himself and his attempts anticipate some recent developments. 

Since this essay is rather long, it may be useful to summarize briefly its contents and to explain the organization, 
so as to provide the reader some guide for the journey. The bare synopsis of the contents is as follows: 

\vspace{10pt} 

\begin{center} 
\begin{tabular}{l}
{\bf \ref{empiric}. Background} \\
{\bf \ref{away}. Turbulence Away From Walls} \\
{\begin{tabular}{l}
$\quad$
{\it \ref{result}. Onsager's Result} \\
$\qquad$
{\begin{tabular}{l}
\ref{sec:RG-LES}.  Ideal Turbulence, Renormalization Group and Large-Eddy Simulation\\
\ref{sec:DissAnom}. Local Deterministic 4/5th-Law and Dissipative Anomaly \\
\ref{sec:open}. Implications and Open Questions\\
\end{tabular}} \\
$\quad$
{\it \ref{sec:ConjIntro}. Onsager's Conjecture} \\
$\qquad$
{\begin{tabular}{l}
\ref{sec:exist}. Existence of Dissipative Euler Solutions\\
\ref{sec:nonuniq}. Non-Uniqueness for the Initial-Value Problem \\
\ref{sec:InfRe}. The Infinite Reynolds Number Limit\\
\end{tabular}} \\
\end{tabular}}\\
{\bf \ref{walls}. Turbulence Interactions With Solid Walls} \\
$\quad$
{\begin{tabular}{l}
{\it \ref{sec:overture}. Overture on Turbulence and Solid Surfaces} \\
{\it \ref{wall:RG}. Onsager Renormalization Group Analysis} \\
$\ \ \,$
{\begin{tabular}{l}
\ref{wall:UV}. Regularization of Ultraviolet Divergences\\
\ref{wall:coarse}. Coarse-Grained Equations \\
\ref{sec:moment}. Momentum Cascade in Space\\
\ref{sec:energy}. Energy Cascade in Space and in Scale\\
\ref{sec:vorticity}. Vorticity Cascade in Space\\
\ref{sec:wkstrng}. Weak-Strong Uniqueness and Extreme Near-Wall Events\\ 
\end{tabular}} \\
{\it \ref{wall:exist}. Dissipative Euler Solutions and Zero-Viscosity Limit} \\
\end{tabular}}\\
{\bf \ref{sec:conclude}. Prospects} \\
$\quad$
{\begin{tabular}{l}
{\it \ref{sec:true}. How Do We Check If It's True?} \\
{\it \ref{sec:matter}. Why Does It Matter?} \\
{\it \ref{sec:last}. Last Words} \\
\end{tabular}} \\
\end{tabular} 
\end{center} 

\vspace{10pt} 
\noindent 
Our tour through this subject begins in Section \ref{empiric} with a summary of some of the experimental phenomena and the basic 
theoretical assumptions generally made to explain them. The first half of the essay, Section \ref{away}, then treats turbulence away 
from solid walls, such as wakes and jets, which was the main subject of Onsager's original work. As I discuss, Onsager 
himself made an exact and essentially rigorous analysis of this problem, deriving various results on energy cascade,
singularities, scale-locality, etc. The section \ref{result} on this subject is titled ``Onsager's Result'', although he never published 
his own derivations and his results had to be recovered and extended by work of others. We shall examine Onsager's unpublished 
material on turbulence, which are now available, along with his unpublished results on many other subjects, in the online Onsager 
Archive hosted by the Norwegian University of Science and Technology: \url{https://www.ntnu.edu/onsager/lars-onsager-archive}. 
In addition, however, Onsager made more ambitious proposals on existence of dissipative Euler solutions and their emergence 
in the infinite-Reynolds number limit, which were motivated by observations but for which he almost certainly had no analytical 
arguments. Only much later was dramatic progress made on these questions and, as I review in section \ref{sec:ConjIntro}
entitled ``Onsager's Conjecture'', this has involved rather sophisticated mathematical tools.  The second half of our essay,
section \ref{walls}, deals with the subject of wall-bounded turbulence, which is of crucial importance for most terrestrial turbulent 
flows and a keen interest of Onsager's, but which has only recently been seriously tackled by his methods. This section closely 
parallels the previous ones, with section \ref{sec:overture} reviewing particular features of the high-Reynolds limit for 
turbulent-wall interactions and early ideas of Taylor and Onsager on the subject. Then section \ref{wall:RG} describes 
in detail how the analysis pioneered by Onsager applies to wall-bounded turbulence and, in recent work of many researchers, 
leads to a picture of both spatial and scale cascades of momentum, energy and vorticity, but understood as deterministic and 
space-time local processes. The concluding section \ref{sec:conclude} of this essay offers some final remarks about the 
empirical status and future importance of Onsager's theory. 

\section{Background}\label{empiric}

Before I  can discuss any theory, I  must briefly review the ``various experiments'' on turbulent energy dissipation 
mentioned by \cite{onsager1945distribution} that motivated his analysis. The specific work cited by \cite{onsager1949statistical} 
was the experimental study of Hugh L. \cite{dryden1943review} on wake turbulence behind a grid using hotwire anemometry,  
which reported that the decay rate of kinetic energy $Q=-\frac{d}{dt}\frac{3}{2}u^{\prime 2}$ satisfies the scaling law 
\be Q\sim A \frac{u^{\prime 3}}{L} \lb{eq:Taylor} \ee 
with $A\doteq 0.2056$ at sufficiently high Reynolds number $Re=u'L/\nu,$ where $u'$ is the r.m.s. streamwise velocity fluctuation,
$L$ is the velocity integral length, and $\nu$ is the kinematic viscosity of the fluid.  The scaling law \eqref{eq:Taylor} 
seems to have been first hypothesized by G.~I. \cite{taylor1935statistical}, who already argued in \cite{taylor1917observations} 
that kinetic energy can be ``dissipated in fluid of infinitesimal viscosity, when the turbulent motion takes place in three 
dimensions.'' Onsager inferred from the results of \cite{dryden1943review} that the coefficient $A(Re)$ becomes 
constant for $Re\gg 1,$ but the first systematic evidence was obtained by \cite{sreenivasan1984scaling} based on a 
compilation of data from several experiments with different types of grids. See Fig.~\ref{fig_empirical_a}. Note that 
the same hypothesis of the asymptotic $Re$-independence of non-dimensionalized energy dissipation rate was made 
also by \cite{kolmogorov1941local,kolmogorov1941dissipation}.

{\begin{figure}
 \centering
\begin{subfigure}[b]{0.46\textwidth}
\centering
\includegraphics[width=\textwidth]{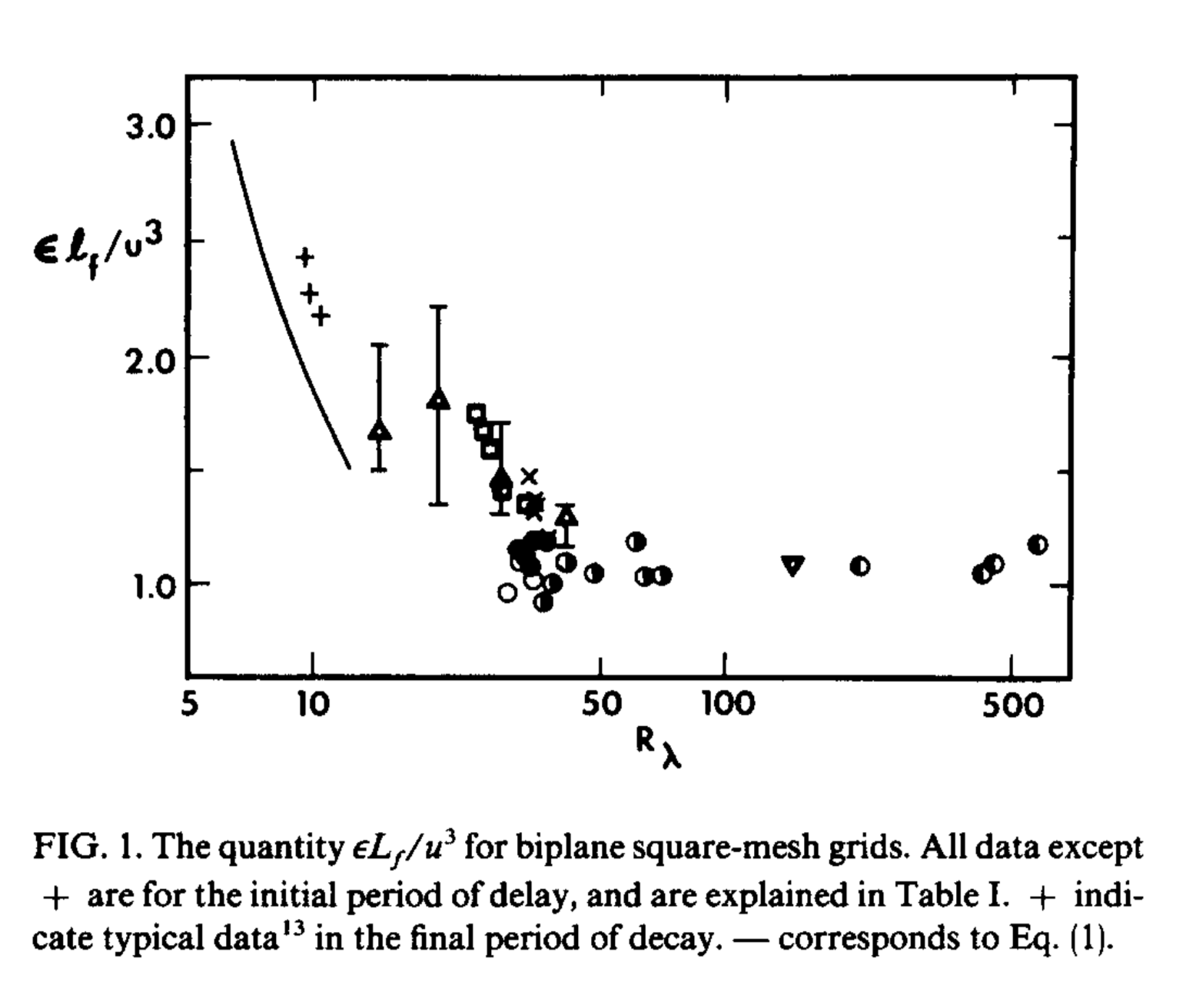}
\caption{Reproduced from K.~R. Sreenivasan, ``On the scaling of the turbulence energy dissipation rate,''
Phys. Fluids {\bf 27} 1048--1051 (1984), 
with permission of AIP Publishing.}
\label{fig_empirical_a}
\end{subfigure}
\hfill
\begin{subfigure}[b]{0.52\textwidth}
\centering
\includegraphics[width=1.05\textwidth]{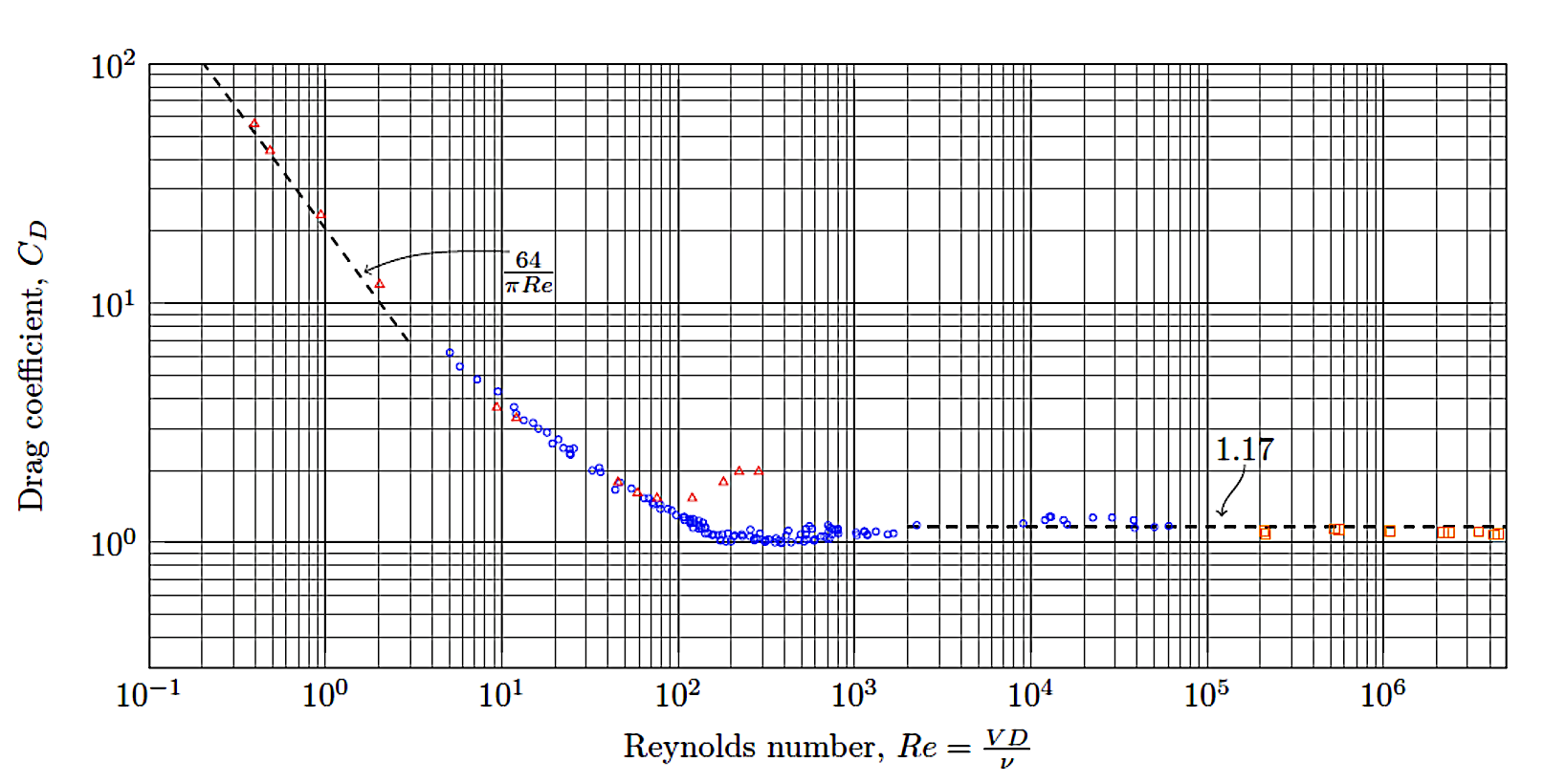}
\vspace{5pt} 
\caption{Drag coefficient for flow past a circular disk oriented normal to the flow. Open red triangles from \cite{hoerner1965fluid},
open circles from \cite{roos1971some}, orange squares from \cite{shoemaker1926resistance}. Reproduced with permission 
from \url{https://kdusling.github.io/teaching/Applied-Fluids}.}
\label{fig_empirical_b}
\end{subfigure}
\caption{Some key pieces of empirical evidence that turbulent energy dissipation is anomalous (non-vanishing) 
at high Reynolds numbers}\label{fig_empirical}
\end{figure}}

Although this observation is basic to several theories of high-Reynolds turbulence and is thus sometimes referred to as 
the ``zeroth law of turbulence'', the experimental situation is in fact more complex and interesting. It is indeed true that 
the result is observed to hold well in wake flows, such as those past bluff or streamlined bodies. One evidence for this is the 
common observation that the drag coefficient of the body 
\be C_D= \frac{F_d}{\frac{1}{2}\rho U_\infty^2 A} \lb{CD} \ee 
tends to a constant value for $Re\gg 1$ where $F_d$ is the drag force, $\rho$ is fluid mass density, $U_\infty$ is external flow velocity
 and $A$ is the frontal area of the body. See Fig.~\ref{fig_empirical_b} for the example of a circular disk and \cite{frisch1995turbulence}, 
 section 5.2, for the connection with dissipation in the wake. 
On the other hand, there is a striking dichotomy for internal flows such as flows through pipes or channels, likewise Taylor-Couette, 
Rayleigh-B\'enard, and von K\'arm\'an flows, and as well  for flat-plate boundary layers, in all of which the strict independence from 
Reynolds number depends upon whether the solid boundary is hydraulically smooth or hydraulically rough. For mathematical readers 
I must clarify that this distinction has nothing to do with the mathematical smoothness of the boundary and ``hydraulic roughness'' 
means simply that the solid boundary has small ripples, ridges, etc. with some characteristic roughness height $k$ that impede the flow. 
Thus, a hydraulically rough surface may in some cases be modeled by a $C^\infty$ manifold. The general observation in these flows 
is that the dimensionless dissipation rate becomes asymptotically independent of Reynolds number only when the solid boundary 
is hydraulically rough. This fact was noted as early as the 18th century by the French engineer  Antoine de Ch\'ezy who attributed 
the seasonal variation of drag in the Parisian water canals to the growth and decline of algae and moss on the side walls. 
A more quantitative observation was made in straight circular pipe flows by \cite{nikuradse1933laws} who studied the friction factor 
\be \lambda = \frac{-\frac{\partial P}{\partial x}D}{\frac{1}{2}\rho \bar{U}^2}, \lb{lambda} \ee 
where $-\frac{\partial P}{\partial x}$ is the applied pressure gradient, $D$ is the pipe diameter and $\bar{U}$ is the mean flow 
velocity. He found that $\lambda(Re)$ slowly decayed to zero as $Re\to\infty$ for a smooth-wall pipe but, with sand-grains glued 
to the wall, instead tended to a positive constant that depended on the grain height $k.$ These same qualitative observations 
have been confirmed in a great variety of internal flows, e.g. see \cite{cadot1997energy}. 
Onsager, by the way, was certainly aware of such observations as indeed he cited in his 1949 paper the work of 
\cite{montgomery1943generalization}, who considered drag laws in pipe flows with both smooth and rough walls and who discussed 
the classic works of \cite{nikuradse1933laws} and others. 

The starting point of the theoretical analysis of \cite{onsager1949statistical} was the incompressible Navier-Stokes equation 
 \be \partial_t\bu +(\bu\bdot\grad)\bu=-\grad p +\nu \Delta\bu, \quad \grad\bdot\bu=0, \lb{NS} \ee
where $p=P/\rho$ is kinematic pressure, in line with the common assumption that all turbulence phenomena 
at low Mach numbers can be described within this continuum approximation. Most of our essay will present results that have 
been obtained from this standard point of view. It is not, however, entirely obvious that this equation should be adequate to describe 
turbulent energy dissipation fully, since the fundamental {\it fluctuation-dissipation relation} of statistical physics implies that molecular 
dissipation phenomena and thermal fluctuations are intrinsically intertwined and must always occur together. Onsager was, of course, 
deeply familiar with the statistical theory of thermal fluctuations. The so-called ``Onsager principle'', which he proposed in his 1931 work 
on reciprocal relations \citep{onsager1931reciprocalII,onsager1931reciprocalI} 
and which was worked out by \cite{onsager1953fluctuations} for the linear regime, is probably the most 
elegant form of the fluctuation-dissipation relation, expressing the probability of a time-history to arise by thermal 
fluctuations directly in terms of the time-integrated dissipation. The prediction that thermal fluctuations should fundamentally modify
the turbulent dissipation range starting at the Kolmogorov length-scale $\eta=\nu^{3/4}/\varepsilon^{1/4}$ (with $\varepsilon$
the mean dissipation rate of kinetic energy per unit mass) was apparently published first in this journal by \cite{betchov1957fine}. 
However, it was not until the work of \cite{landau1959fluid}, and independently \cite{betchov1961thermal}, that a 
stochastic version of the Navier-Stokes equation was formulated that incorporates the fluctuation-dissipation relation. 
For a low-Mach incompressible fluid this equation for fluctuating velocity field $\tilde{\bu}$ takes the form  
\be \partial_t\tilde{\bu} +(\tilde{\bu}\bdot\grad)\tilde{\bu}=-\grad \tilde{p} +\nu \Delta\tilde{\bu}+\grad\bdot\tilde{\btau}
, \quad \grad\bdot\tilde{\bu}=0, \lb{LLNS} \ee
where $\tilde{\btau}$ is a Gaussian random stress field with mean zero and covariance
\be  \langle \tilde{\tau}_{ij}(\bx,t)\tilde{\tau}_{kl}(\bx',t')\rangle =\frac{2\nu k_BT}{\rho}\left(\delta_{ik}\delta_{jl}+ 
\delta_{il}\delta_{jk}-\frac{2}{3}\delta_{ij}\delta_{kl}\right)\delta^3(\bx-\bx')\delta(t-t') \lb{FDR} \ee
which represents the thermal momentum transport. Recently, evidence has emerged from numerical simulation of the stochastic 
equation \eqref{LLNS} and related models which confirm Betchov's prediction \citep{bandak2022dissipation,bell2022thermal}. 
I shall therefore critically examine also in this essay the question whether the deterministic equations \eqref{NS} are sufficient 
to explain all of the experimental observations on turbulent energy dissipation. 

Since it has been traditionally assumed that the Navier-Stokes equations are an adequate model of incompressible fluid turbulence, 
direct numerical simulation (DNS) of those equations has also been used as a tool to study turbulent energy dissipation. 
I am not aware of any systematic DNS study of the Reynolds-dependence of dissipation in 3D wall-bounded flows
(but see \cite{nguyenvanyen2018energy} for a 2D flow). Instead 
most of the attention has been focused on turbulence in a periodic box stirred by a large-scale body force, which has 
been considered a close analogue of the decaying turbulence behind a grid. A great advantage of numerical simulations 
is that the entire flow field is accessible and thus local viscous energy dissipation per mass 
\be \varepsilon(\bx,t)=2\nu S_{ij}(\bx,t)S_{ij}(\bx,t),   \lb{epsilon} \ee 
with strain rate tensor $S_{ij}=\frac{1}{2}\left(\frac{\partial u_i}{\partial x_j}+ \frac{\partial u_j}{\partial x_{i}}\right),$
can be calculated exactly for Navier-Stokes solutions, limited only by numerical resolution and machine precision. Corresponding 
to the prefactor $A$ studied in grid turbulence is the dimensionless dissipation rate 
\be D=\frac{\varepsilon}{u'^3/L} = \frac{2}{Re}\hat{S}_{ij}\hat{S}_{ij} \lb{eq:D} \ee  
where the hat denotes the strain tensor calculated from the dimensionless variables $\hat{\bu}=\bu/u',$ $\hat{\bx}=\bx/L,$ $\hat{t}=t/(L/u').$
Numerical results in \cite{sreenivasan1998update} and \cite{kaneda2003energy} are consistent with the hypothesis that 
the volume- and time-averaged dimensionless dissipation rate $\bar{D}(Re)$ are asymptotically independent of $Re.$ Of course, 
no empirical data could ever strictly verify this hypothesis, as it would be impossible to rule out an extremely slow decay.  

\section{Turbulence Away From Walls}\label{away}

Except for brief mention of pipe flow, \cite{onsager1949statistical} restricted his attention to the ``simplest type of turbulence''
which is the ``nearly homogeneous and isotropic turbulence [produced] by means of a grid in a streaming gas''. I  shall therefore 
consider this topic first in this essay, since most subsequent work has been done on this problem. I  emphasize at the outset, however, 
that no statistical assumptions of homogeneity or isotropy are required for the results mentioned in the final paragraph of Onsager's paper.
The conclusions thus apply more generally to turbulence away from walls, such as in ubiquitous wake flows that are neither homogeneous 
nor isotropic. It is a bit ironic that Onsager suggested a completely deterministic approach to the analysis of turbulent flow in a paper entitled 
``Statistical Hydrodynamics"! 

\subsection{Onsager's Result}\lb{result} 

In the quote that began this essay, \cite{onsager1949statistical} wrote that ``it is possible to show that'', which was his characteristic 
catch-phrase to assert that HE had shown something and to invite others to prove it as well. He never published his 
own calculations (which was also customary), but, as I  shall discuss below, he had worked out the essential steps of a proof that 
energy will be conserved when the velocity field satisfies a H\"older-Lipschitz condition ``for any order $n$ greater than 1/3.'' The first 
published result was given by \cite{eyink1994energy}, who attempted to transform the brief argument using Fourier series sketched by 
\cite{onsager1949statistical} into a rigorous proof. The idea was to show that the triply infinite series of Fourier coefficients which 
arises from the time-derivative of kinetic energy is absolutely summable on the assumption of H\"older regularity with exponent $>1/3.$
This ultimately requires a bound on the spectral energy flux $\Pi(k),$ similarly as in the work of \cite{sulem1975bounds}. It was quickly 
obvious that the $1/3$ claim would hold if the spectral flux were dominated by ``local'' wavevector triads with all three wavectors 
of magnitudes roughly $k.$ This approach is plagued by difficulties, however, not least because there is no necessary and sufficient 
condition for H\"older continuity in terms of absolute Fourier coefficients. I finally managed in 1992 to work out a proof of 
Onsager's $1/3$ claim, but invoking a condition on absolute Fourier coefficients stronger than H\"older continuity. 

At the same time, it appeared from the analysis that most contributions to the energy flux could be bounded even with a weaker assumption 
on the velocity of ``Besov regularity'', which corresponds to H\"older regularity in space-mean sense, closely related to standard structure 
functions \citep{eyink1995besov}. After developing my first proof, I discussed the problem with Weinan E at the Institute for Advanced Study 
in 1992 and, the following year E and his collaborators, Peter Constantin and Edriss Titi, found an extremely simple proof not only of Onsager's 
original claim but also of the natural Besov result \citep{constantin1994onsager}. After being informed privately of this development by E, 
I realized that their method of proof was closely related to LES modelling and renormalization group, and that it could be simply 
explained in those terms \citep{eyink1995local}. It is this argument that I  present first in this section, before considering the alternative 
technical derivation given by Onsager himself. Note that I  refer to the energy conservation statement, however, as "Onsager's result" rather than 
by the term "Onsager's conjecture" used by \cite{constantin1994onsager},  which was at a time when Onsager's own mathematical argument 
was unknown. I  prefer to use the term "Onsager's conjecture" for the deeper statement that inviscid energy dissipation is possible when the 
Euler velocity field has H\"older regularity $\leq 1/3,$ which was only proved much later.   

\subsubsection{Ideal Turbulence, Renormalization Group and Large-Eddy Simulation}\lb{sec:RG-LES} 

The most basic conclusion that can be inferred from the empirical observations of $Re$-independence in \eqref{eq:Taylor} or \eqref{eq:D} 
is that the non-dimensionalized velocity gradients, if interpreted as ordinary classical derivatives, must diverge as $|\hat{\grad}\hat{\bu}|\to\infty$  
when $Re\to\infty.$ This was the starting point of Onsager's own line of argument, who began the short abstract in \cite{onsager1945distribution} 
with the observation: ``The dissipation of energy by turbulence is regarded as primarily a `violet catastrophe'." In the language of modern 
field theory, turbulence exhibits ultraviolet (UV) divergences of velocity gradients in the limit $Re\to\infty.$ In consequence, the equations 
of motion \eqref{NS} (or alternatively \eqref{LLNS}) cannot remain valid in a naive sense in the infinite-$Re$ limit. Just as in field theory, to obtain 
a dynamical description of turbulence as $Re\to\infty$, one must somehow regularize these UV divergences. The simple but 
effective approach of \cite{constantin1994onsager} was to perform what is called ``low-pass filtering'' in engineering, ``spatial coarse-graining'' 
in physics, and ``mollifying'' in mathematics. This method involves the use of a smooth kernel $G$ for space dimension $d$ that satisfies 
\begin{eqnarray*}
(i) \quad G(\br) \geq 0, \qquad 
(ii) \quad G(\br) \rightarrow 0 \,\,\mbox{rapidly for} \,\,\,\,\,|\br| \rightarrow \infty, \qquad 
(iii) \quad \int d^d r \,\,G(\br) =1
\end{eqnarray*}
It is understood that $G$ is centered at $\br = {\bf 0}$, 
$\int d^d r \,\,\br \,\,G(\br) = {\bf 0},$ 
and that $\int d^d r |\br|^2 G(\br) \approx 1$. 
We can then set  
\be G_\ell (\br) \equiv \ell^{-d} G(\br/\ell) \ee
so that all of the above properties hold, except that now $\int d^d r |\br|^2 G_\ell(\br) \approx \ell^2$, with 
$\ell>0$ the regularization length scale. Finally, one defines a coarse-grained velocity at length-scale $\ell$ 
by the formula: 
\be \overline{\bu}_\ell(\bx,t)=\int d^dr\  G_\ell(\br)\, \bu(\bx+\br,t).   \lb{II5} \ee 
spatially averaging over the eddies of size $<\ell.$

This coarse-grained field is roughly analogous to a ``block spin'' in critical phenomena and field theory. 
An identical operation is also employed in the  ``filtering approach'' to turbulence advocated by 
\cite{leonard1974energy} and \cite{germano1992turbulence}, but with a different motivation 
than regularization of divergences.  For Onsager's theory, the important point is that the coarse-graining 
operation \eqref{II5} regularizes gradients, so that $\grad\overline{\bu}_\ell$ remains finite as $\nu\to 0$ 
for any fixed length $\ell>0.$ This may be shown using the simple integration-by-parts identity
\be \grad\overline{\bu}_\ell(\bx,t)=-\frac{1}{\ell}\int d^dr\  (\grad G)_\ell(\br)\, \bu(\bx+\br,t),  
\lb{II7} \ee 
which by Cauchy-Schwartz inequality yields the bound $|\grad\overline{\bu}_\ell(\bx,t)|\leq 
(1/\ell)\sqrt{C_\ell\int d^dr \, |\bu(\br,t)|^2}$ with constant $C_\ell=\int d^dr\ |(\grad G)_\ell(\br)|^2.$
Thus, the coarse-grained gradient is bounded as long as the total kinetic energy remains finite as $\nu\to 0$
(which is necessarily true for freely-decaying turbulence with no stirring).  The price 
of this regularization, as for quantum field-theory divergences, is that a new, arbitrary regularization
scale $\ell$ has been introduced.  

Because divergences have been eliminated, one may now seek a dynamical description in terms 
of the coarse-grained field defined in \eqref{II5}. The equation which is satisfied is easily found to be
\be \partial_t\overline{\bu}_\ell+\grad\bdot\overline{(\bu\bu)_\ell}=
-\grad\overline{p}_\ell+\nu\Delta\overline{\bu}_\ell, \quad \grad\bdot\overline{\bu}_\ell=0,
\lb{II9} \ee 
because the coarse-graining operation commutes with all space- and time-derivatives, and one may 
inquire about its limit as $Re\to\infty.$ For this purpose, one should non-dimensionalize 
the above equation and introduce hats  everywhere, but doing so simply replaces 
the physical viscosity with the dimensionless viscosity $\hat{\nu}=1/Re.$ Thus, as customary in the 
mathematical literature, one may omit the hats on non-dimensionalized variables and consider the 
limit $Re\to\infty$ as a ``zero-viscosity limit''  limit $\nu\to 0.$ I  shall do so hereafter, when this causes 
no confusion.  Because the quantity $\Delta\overline{\bu}_\ell$ in \eqref{II9} remains bounded by 
a similar estimate as \eqref{II7}, one can easily show rigorously that $\nu\Delta\overline{\bu}_\ell\to 0$
as $\nu\to 0$ for fixed $\ell.$ If one indexes the solutions $\bu^\nu$ of the Navier-Stokes equation \eqref{NS} 
by viscosity $\nu,$ then the above results suggest that a limiting field $\bu=\lim_{\nu\to 0}\bu^\nu$ 
will satisfy the equation
\be \partial_t\overline{\bu}_\ell+\grad\bdot\overline{(\bu\bu)}_\ell=
-\grad\overline{p}_\ell, \quad \grad\bdot\overline{\bu}_\ell=0,
\lb{wEuler} \ee 
Since one is dealing with velocity fields in function spaces, a careful statement of this hypothesis 
requires a suitable notion of convergence $\bu^\nu\to \bu.$ It is not hard to show that strong $L^2$ 
convergence suffices, which is the condition that $\lim_{\nu\to 0}\|\bu^\nu-\bu\|_2=0$ where
\be \|\bu\|_p:=\left[\frac{1}{T}\int_0^T dt\, \frac{1}{|\Omega|}\int_\Omega d^dx\, |\bu(\bx,t)|^p\right]^{1/p},\quad p\geq 1 \ee  
As I  shall discuss later (see section \ref{sec:InfRe}), such strong $L^2$-convergence is not guaranteed {\it a priori},
although it is in fact implied by some relatively mild assumptions on the energy spectrum which can be 
tested empirically. For the moment, I  shall simply assume such convergence so that \eqref{wEuler} 
directly follows, but I return to this issue later. 

An important observation is that the validity of the coarse-grained equations \eqref{wEuler} for all lengths $\ell>0$
is equivalent to the condition in mathematical PDE theory that the velocity $\bu$ is a {\it weak solution} 
of the incompressible Euler equations. Readers may be more familiar with the standard formulation of weak
solutions by smearing the dynamical equations with smooth spacetime test functions, so that the equations 
hold in the sense of distributions or generalized functions. A simple proof that ``coarse-grained solutions'', 
satisfying \eqref{wEuler} for all $\ell>0,$ are equivalent to standard weak solutions is given in section 2
of \cite{drivas2018onsager}. Furthermore, these notions of weak solution are equivalent to what 
\cite{onsager1949statistical} meant when he wrote that ``the ordinary formulation of the laws of motion in 
terms of differential equations becomes inadequate and must be replaced by a more general description;
for example, the formulation (15) in terms of FOURIER series will do.''  Indeed, the incompressible Euler 
equations written as an infinite-dimensional set of ODE's for the Fourier coefficients of the velocity field 
yield also standard weak solutions, as discussed by \cite{eyink1994energy} and \cite{delellis2013continuous}. 
All of these equivalent formulations of weak solutions make sense even when spatial derivatives of velocity 
no longer exist in the classical sense, but only in the sense of distributions. 

It may appear odd to some readers that \eqref{wEuler} can be interpreted as ``Euler equations'',  since 
the filtered equations \eqref{wEuler} are generally regarded as unclosed whereas the Euler equations 
are closed PDE's. It is indeed true that the equations \eqref{wEuler} are not closed equations for the 
coarse-grained velocity $\overline{\bu}_\ell$ itself. This fact is usually underlined by introducing the 
{\it turbulent/subscale stress} 
\be \btau_\ell(\bu,\bu) := \overline{(\bu\bu)}_\ell - \overline{\bu}_\ell\overline{\bu}_\ell \lb{tau} \ee
so that the equations \eqref{wEuler} can be rewritten as 
\be \partial_t\overline{\bu}_\ell+\grad\bdot\left[ \overline{\bu}_\ell \overline{\bu}_\ell + \btau_\ell(\bu,\bu)\right] =
-\grad\overline{p}_\ell, \quad \grad\bdot\overline{\bu}_\ell=0 \lb{FEuler} \ee 
with the non-closed term $\btau_\ell$ distinguished. However, the coarse-grained velocity $\overline{\bu}_\ell
=G_\ell*\bu$ and the subscale stress $\btau_\ell(\bu,\bu)$ are both explicit functions of the fine-grained 
velocity field $\bu,$ as emphasized by my notations, and thus the equations \eqref{FEuler} are explicit 
conditions on that field. I  may note that it is quite standard to consider that the large-scales $>\ell$
in a turbulent flow must satisfy the Euler fluid equations. For example, \cite{landau1959fluid}, \S 31, wrote that: 
``We therefore conclude that, for the large eddies which are the basis of any turbulent flow, the viscosity is 
unimportant and may be equated to zero, so that the motion of these eddies obeys Euler's equation.'' 
What needs to be stressed is that the proper interpretation of such commonplace remarks is {\it not} that the 
coarse-grained velocity $\overline{\bu}_\ell$ satisfies the Euler equations in the usual na\"{\i}ve sense,
but instead that $\bu$ is described by a weak Euler solution at those scales, in the sense that \eqref{FEuler}
holds. Although this point is elementary, it is frequently misunderstood and the source of common errors. 

There is another very important conceptual point which must be emphasized about the coarse-grained
equations \eqref{FEuler}. Some critics have argued that these equations are unphysical and not appropriate
as the basis for a fundamental theory of turbulence because both the length-scale $\ell$ and the filter 
kernel $G$ are arbitrary.  For example, \cite{tsinober2009informal} wrote 
``The filter decomposition is formally more general than the Reynolds decomposition. However, the 
former is one among many decompositions, so to say, of a technical nature'' (p.379) and also 
``After all Nature may and likely does not know about {\it our} decompositions.'' (p.114). These are very 
shrewd remarks. In fact, I  agree completely with these concerns of \cite{tsinober2009informal} 
that Nature should not care about such arbitrary choices. Interestingly, this same problem has 
arisen in another area of physics, relativistic quantum field-theory, which is also plagued with similar 
UV divergences. In that case also arbitrary regularizations are required to eliminate the divergences
and these introduce a new arbitrary length scale, equivalent to $\ell$ or, more commonly,  
a related energy scale $\mu=c\hbar/\ell,$ with $c$ the speed of light and $\hbar$ Planck's constant. 
In the renormalized field theory, fundamental parameters of the theory such as coupling constants
$\lambda(\mu)$ become dependent on this arbitrary scale, in the same manner that $\btau_\ell$ becomes 
dependent upon $\ell$ . Elementary particle physicists in the 1950's were also worried that 
predictions of the theory should not depend upon such arbitrary choices and the concept of 
{\it renormalization group (RG) invariance} arose as the commonsense demand that any observable consequence 
of the theory should be independent of $\mu.$ For a very clear discussion, see \cite{gross1976applications}, 
section 4.1. What makes RG invariance interesting and important is that non-trivial consequences 
can be deduced precisely by varying $\mu$ and demanding that physical results be $\mu$-independent. 
What we shall see is that Onsager anticipated such arguments in the 1940's and that his 1/3 H\"older claim
for turbulent velocities is an exact non-perturbative consequence of such RG invariance.

\begin{figure}
 \centering
\begin{subfigure}[b]{0.32\textwidth}
\centering
\includegraphics[width=\textwidth]{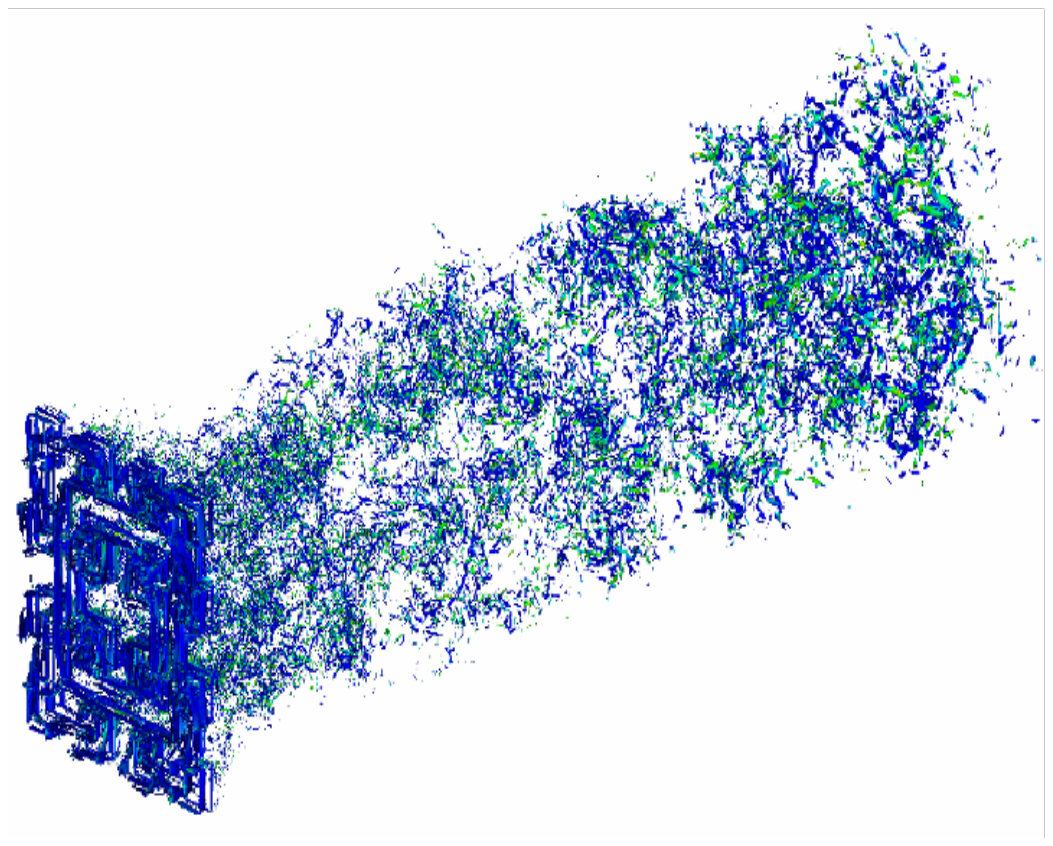}
\caption{Fine-grained}
\label{fig_fine}
\end{subfigure}
\hfill
\begin{subfigure}[b]{0.32\textwidth}
\centering
\includegraphics[width=\textwidth]{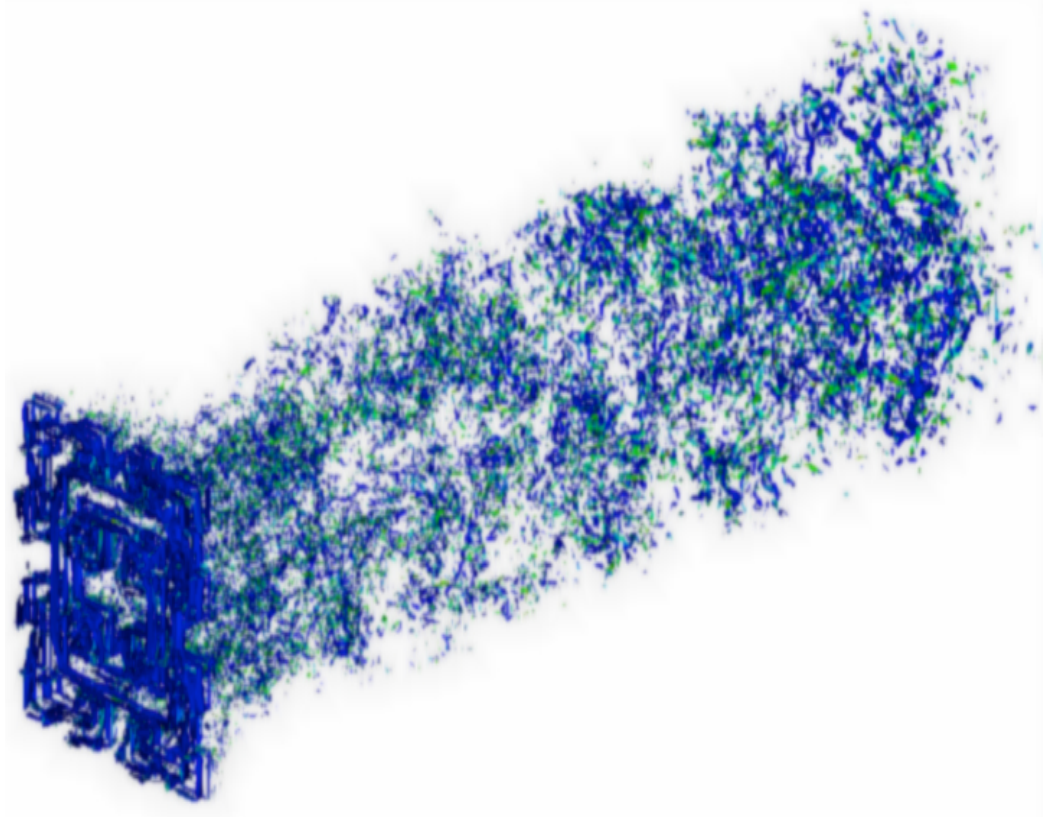}
\caption{Coarse-grained}
\label{fig_coarse}
\end{subfigure}
\hfill
\begin{subfigure}[b]{0.32\textwidth}
\centering
\includegraphics[width=\textwidth]{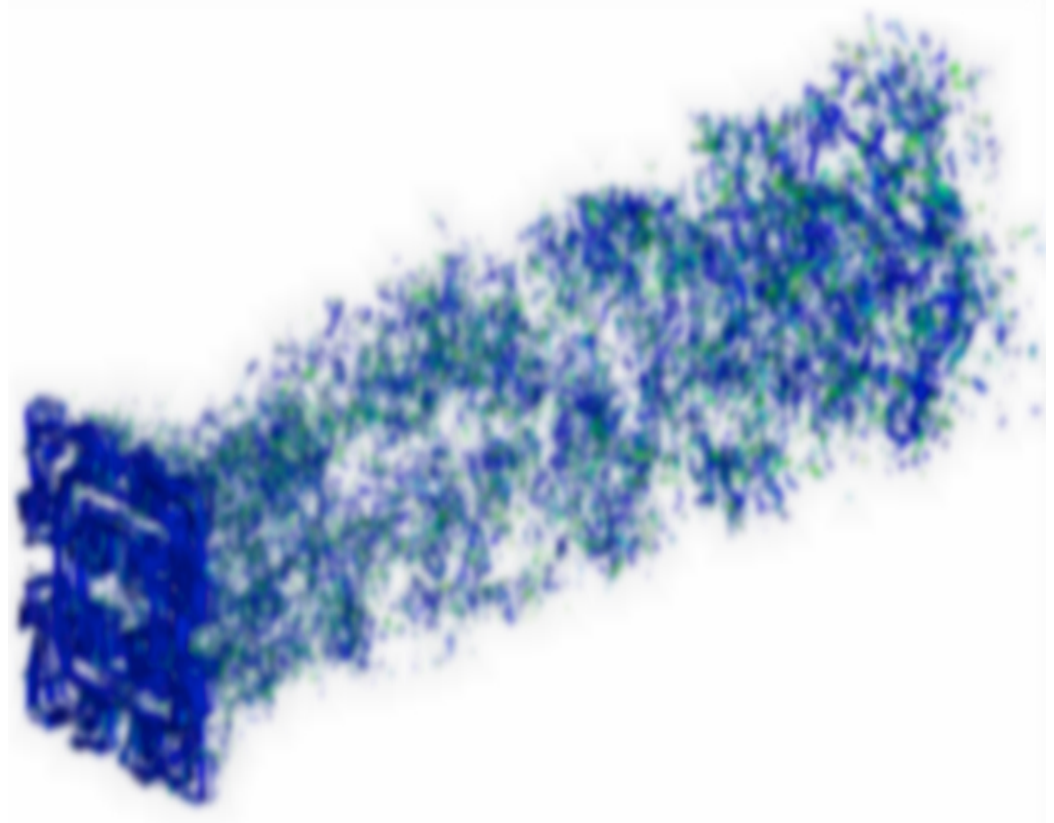}
\caption{Further coarse-grained}
\label{fig_coarser}
\end{subfigure}
\caption{Schematic illustration of spatial coarse-graining for turbulent flow past a grid: (a) Fine-grained flow resolved to 
the dissipation scale $\eta$; (b) Flow coarse-grained at a length-scale $\ell>\eta$; (c) Flow coarse-grained further
at scale $\ell'>\ell.$ At each stage, eddies smaller than the coarsening scale are unresolved and ignored.
Panel (a) is reproduced from S. Laizet et al., ``Low Mach number prediction of the acoustic signature of fractal-generated 
turbulence,'' Internatl. J. Heat \& Fluid Flow {\bf 35}, 25--32 (2012), 
and panels (b),(c) modified by image filtering, with permission from Elsevier. 
}\label{fig_coarsening}
\end{figure}

To underscore the nature of the argument, I  emphasize that coarse-graining is a purely passive operation
---``removing one's spectacles''---which changes no physical process. The effect of the coarse-graining 
operation in \eqref{II5}  is illustrated by Fig.~3
, which shows the velocity field observed at successively 
coarser spatial resolutions $\ell$. Although the dynamics of the velocity field resolved at the Kolmogorov 
scale $\ell=\eta$ are described rather well by the Navier-Stokes equation \eqref{NS}, the coarse-grained 
velocity field $\overline{\bu}_\ell$ at scales $\ell\gg \eta$ is described by 
the highly non-Newtonian equation \eqref{FEuler}. While the description of the dynamics changes with resolution $\ell,$ nevertheless 
objective facts cannot depend upon the ``eyesight'' of the observer. Thus, an energy dissipation rate which is 
non-vanishing in the limit as $\nu\to 0$ implies that kinetic energy will decrease over a fixed interval of time $[0,t],$  
as observed by experiment, and this fact cannot depend upon the arbitrary scale $\ell.$
An observation at resolution $\ell$ can only miss some kinetic energy of smaller eddies, since by convexity 
\be \frac{1}{2}|\overline{\bu}_\ell(\bx,t)|^2\leq \frac{1}{2}\overline{\left(\left|\bu(\bx,t)\right|^2\right)}_\ell, 
\lb{II8} \ee 
and it then follows that 
$E_\ell(t):=(1/2)\int d^dx\, |\overline{\bu}_\ell(\bx,t)|^2\leq (1/2)\int d^dx\, |\bu(\bx,t)|^2=E(t).$ If kinetic 
energy continues to decay even in the limit as $\nu\to 0,$ then such persistent energy decay 
must be seen also by the ``myopic'' observer who observes fluid features only at space-resolution 
$\ell.$  As I  now show, however, the persistent energy decay observed at the fixed length-scale 
$\ell$ with $\nu\to 0$ is not due to molecular viscosity acting directly at those scales. 

The local kinetic energy balance at length-scales $\ell$ in the inertial-range obtained for the limit $\nu\to 0$ 
is calculated straightforwardly from \eqref{FEuler} to be 
\bea 
&& \partial_t\left(\frac{1}{2}|\overline{\bu}_\ell|^2\right)
+\grad\bdot\left[\left( \frac{1}{2}|\overline{\bu}_\ell|^2+\overline{p}_\ell\right)\overline{\bu}_\ell
+\btau_\ell\bdot\overline{\bu}_\ell\right] 
=-\Pi_\ell\cr
&& \lb{II15} \eea
where the quantity on the right side of the equation is given (with ${\bf A}\bdots{\bf B}=\sum_{ij} A_{ij}B_{ij}$) by 
\be \Pi_\ell(\bx,t) = -\grad\overline{\bu}_\ell(\bx,t)\,\bdots\,\btau_\ell(\bx,t),  \lb{II16} \ee 
and represents the ``deformation work''  of the large-scale strain acting 
against small-scale stress, or the ``energy flux'' from resolved scales $>\ell$ to 
unresolved scales $<\ell.$ The mechanism of loss of energy by  the inertial-range eddies is thus ``energy cascade'', a 
term first used in this connection by \cite{onsager1945distribution}.  A key observation is that the stress-tensor 
$\btau_\ell(\bu,\bu)$ may be rewritten in terms of {\it velocity-increments} $\delta \bu(\br;\bx,t)=\bu(\bx+\br,t)-\bu(\bx,t),$ 
as 
\be \btau_\ell(\bu,\bu)=\langle \delta \bu\,\delta \bu\rangle_\ell
-\langle \delta \bu\rangle_\ell\,\langle \delta \bu\rangle_\ell, \lb{II19} \ee 
where $\langle f\rangle_\ell(\bx,t):=\int d^dr\, G_\ell(r) f(\br;\bx,t).$ This formula was 
originally obtained in \cite{constantin1994onsager} in a slightly different form, 
and as above in \cite{eyink1995local} as a physical re-interpretation of their result. Equation \eqref{II19} 
is easy to verify by direct calculation, but it can be simply understood as due to the 
invariance of the 2nd-order cumulant $\btau_\ell(\bu,\bu)$ to shifts of $\bu$ by 
vectors that are ``non-random'' with respect to the average $\langle\cdot\rangle_\ell$
over displacements $\br,$ i.e. that are independent of $\br.$ This allows $\bu(\bx+\br,t)$ 
in the definition \eqref{tau} of $\btau_\ell(\bu,\bu)$ to be replaced with $\delta\bu(\br;\bx,t),$
yielding the formula \eqref{II19}. Similarly, one may rewrite eq.\eqref{II7} for coarse-grained 
velocity-gradients in terms of increments as 
\be \grad\overline{\bu}_\ell(\bx,t)=-\frac{1}{\ell}\int d^dr\  (\grad G)_\ell(\br)\, \delta\bu(\br;\bx,t),  
\lb{II20} \ee 
using the fact that $\int d^dr\, (\grad G)_\ell(\br)=\bzed.$ 


As an immediate application of these formulas, one can rederive the prediction of \cite{onsager1949statistical} 
that H\"older singularities $h\leq 1/3$ are required in the velocity field in order for energy dissipation 
to persist in the limit $\nu\to 0.$ Indeed, assuming for some constant $C>0$ that 
\be |\delta \bu(\br;\bx,t)|\leq C|\br|^h, \lb{II21} \ee 
then it is straightforward to show using \eqref{II16},\eqref{II19} and \eqref{II20} that 
\be \Pi_\ell(\bx,t)=O\left(\ell^{3h-1}\right). \lb{II22} \ee
As is clear from \eqref{II15}, persistent energy decay at resolution 
length $\ell$ can only occur if $\int d^dx\, \Pi_\ell(\bx,t)>0$ as $\nu\to 0.$ On the other hand, 
the resolution scale $\ell$ is completely arbitrary. For any fixed $\ell$ one can take $\nu$ sufficiently 
small so that the ``ideal equations'' \eqref{wEuler} or \eqref{FEuler} hold to any desired accuracy at that scale, 
and then sequentially further decreasing $\ell$ one can correspondingly decrease $\nu$. If the H\"older regularity \eqref{II21} 
held for all $(\bx,t)$ with $h>1/3,$ then clearly by \eqref{II22} it would follow that $\int d^dx\, \Pi_\ell(\bx,t)\to 0$ 
as $\ell\to 0.$ This is a contradiction, since the rate of decay of energy must be independent of the 
arbitrary length-scale of resolution $\ell$ as $\ell\to 0.$ Just as in \cite{onsager1949statistical}, 
we thus infer that there must appear H\"older singularities $h\leq 1/3$ 
in the limit as $\nu\to 0,$ or $Re\to \infty$.  As I  shall discuss in the next section \ref{sec:DissAnom}, 
Onsager had given a very similar argument to that above in his unpublished work.   

As an aside, I  remark that the RG character of this argument can be made even more explicit
by a somewhat different technical approach of \cite{johnson2020energy,johnson2022physics},
where it is observed for a Gaussian kernel $G(\br)=\exp(-r^2/2)/(2\pi)^{3/2}$ that 
$\frac{\partial}{\partial \ell^2} \overline{\bu}_\ell=\frac{1}{2}\Delta\overline{\bu}_\ell$ and thus 
 \be \frac{\partial}{\partial \ell^2} \btau_\ell=\frac{1}{2}\Delta\btau_\ell 
 +(\grad\overline{\bu}_\ell)^\top \grad\overline{\bu}_\ell. \lb{RGflow} \ee
In this approach, the subscale stress that ``renormalizes'' the bare Navier-Stokes dynamics can in fact be obtained 
by solving an equation that evolves $\btau_\ell$ in the scale parameter $\ell$, analogous to the RG flow
equations in high-energy physics and critical phenomena \citep{wilson1975renormalization,gross1976applications}.  
Solving \eqref{RGflow} with the initial data $\left. \btau_\ell\right|_{\ell=0}=\bzed,$ \cite{johnson2020energy} 
obtained 
\be \btau_\ell = \int_0^{\ell^2} d\lambda^2  \
\overline{((\grad\overline{\bu}_\lambda)^\top \grad\overline{\bu}_\lambda )}_{\sqrt{\ell^2-\lambda^2}}. 
\lb{johnson} \ee 
Assuming the H\"older condition \eqref{II21}  then gives $\grad\overline{\bu}_\lambda=O(\lambda^{h-1})$ as before and,
from the identity \eqref{johnson}, $\btau_\ell =O(\ell^{2h}).$ Thus, the same bound \eqref{II22} is deduced once again 
and this implies the 1/3 H\"older claim of Onsager.  See also \cite{isett2016heat} for a similar approach. 

The paper of \cite{constantin1994onsager} derived in fact a stronger result, by replacing the H\"older regularity originally 
assumed by \cite{onsager1949statistical} with a weaker assumption of Besov regularity. As discussed by 
\cite{eyink1995local}, Besov regularity can be understood in terms of deterministic $p$th-order ``velocity-structure 
functions'' defined for absolute velocity increments and for space-averages over the flow domain $\Omega:$
\be S_p(\br)= \frac{1}{|\Omega|}\int_\Omega d^dx\, |\delta \bu(\br;\bx)|^p. 
\lb{II23} \ee 
Although power-laws in the separation distance $|\br|$ are generally observed empirically for turbulent fluid flows, 
Besov regularity requires only an upper bound 
\be S_p(\br)\leq C_p|\br|^{\zeta_p}, \quad |\br|\leq 1. \lb{Besov} \ee
If this inequality holds along with the modest assumption of finite $p$th-order moments 
\be \frac{1}{|\Omega|}\int_\Omega d^dx\, |\bu(\bx)|^p <\infty, \lb{Lp} \ee
then the velocity field $\bu$ is said to belong to the Besov space $B^{\sigma,\infty}_p(\Omega)$ with 
$\sigma=\sigma_p:=\zeta_p/p.$ Note that the optimal constant 
$C_p = \sup_{|\br|\leq 1} \frac{S_p(\br)}{|\br|^{\zeta_p}} $
in the inequality \eqref{Besov} is related to the so-called ``Besov semi-norm'' in the mathematical 
literature by the formula $\vertiii{\bu}_{B^{\sigma,q}_p}=C_p^{1/p},$ so that the abstract semi-norm 
is given by the prefactor in expected power scaling laws \citep{drivas2019onsager}. Note further that 
the H\"older condition \eqref{II21} is equivalent to the Besov condition \eqref{Besov} for $p=\infty,$
with $\sigma_\infty=h.$

The result of \cite{constantin1994onsager} was that energy dissipation non-vanishing in the limit as $\nu\to 0$
requires $\zeta_p\leq p/3$ for all $p\geq 3$ or, equivalently, $\sigma_p\leq 1/3$ for $p\geq 3.$ 
The argument generalizes that given previously for $p=\infty$ and uses the simple bound from the H\"older inequality
valid for any $p\geq 3$
\be \left| \frac{1}{|\Omega|}\int_\Omega d^dx\ \Pi_\ell(\bx,t)\right| \leq \|\Pi_\ell\|_{p/3} 
\leq \|\grad\overline{\bu}_\ell\|_{p} \|\btau_\ell\|_{p/2}=O\left(\ell^{3\sigma_p-1}\right)\ee 
since $\|\grad\overline{\bu}_\ell\|_{p}=O\left(\ell^{\sigma_p-1}\right)$ and $\|\btau_\ell\|_{p/2}=O\left(\ell^{2\sigma_p}\right).$ 
For mathematical details, see \cite{constantin1994onsager}, \cite{eyink1995local} or \cite{eyink2007turbulenceI}. 
 Just as before, if the Besov regularity \eqref{Besov} held with $\sigma_p>1/3,$
then it would follow that $\int d^dx\, \Pi_\ell(\bx,t)\to 0$ as $\ell\to 0.$ This would yield the same contradiction as previously, since the 
rate of decay of energy must be independent of the arbitrary length-scale of resolution $\ell$ as $\ell\to 0.$ 
Generalizing the statement of \cite{onsager1949statistical}, one can infer that in fact $\sigma_p\leq 1/3$ 
for all $p\geq 3$ in the limit as $\nu\to 0.$ 

Onsager's ideas anticipate the ``multifractal model''  proposed by \cite{frisch1985singularity} for the turbulent velocity field;
see also \cite{frisch1995turbulence} for a comprehensive exposition. In fact, it is worth recalling the first sentence of \cite{frisch1985singularity}:
``A simple way of explaining power law structure function is to invoke singularities of the Euler equations considered 
as limit of the Navier-Stokes equations as the viscosity tends to zero.''  \cite{frisch1985singularity} proposed 
that the turbulent velocity field possesses a spectrum of H\"older exponents $h\in [h_{\min},h_{\max}]$ with each exponent $h$
occurring on a set $S(h)\subset \Omega$ with Hausdorff dimension $D(h).$ The velocity scaling exponents are then 
obtained by the formula  
 \be \zeta_p=\inf_h \{ hp+ (3-D(h))\}, \lb{MF} \ee 
 thus accounting for the observed deviations from the prediction $\zeta_p=p/3$ of 
 \cite{kolmogorov1941local,kolmogorov1941dissipation}. The velocity scaling exponent relevant in this context can be 
 understood as the ``maximal Besov exponent'' $\sigma_p=\zeta_p/p$ for any $p\geq 0$ with
 \be \zeta_p:=\liminf_{|\br|\to 0} \frac{\log S_p(\br)}{\log|\br|}, \lb{zetap} \ee
 a concept which is meaningful even without any power-law scaling \citep{eyink1995besov}.  
 As we shall see later, Onsager had arrived at similar conclusions already 
 by 1945 (but without the modern concept of fractals). Note that $h_p=d\zeta_p/dp$ is the H\"older 
 exponent $h$ that yields the infimum in \eqref{MF} for each $p\geq 0.$ It then follows from \eqref{MF} or even
directly from the concavity of $\zeta_p$ in $p$ \citep{frisch1995turbulence,eyink2007turbulenceI} that 
$h_p\leq \sigma_p$ for all $p\geq 0.$ Thus, the theorem of \cite{constantin1994onsager} implies also 
that $h_p\leq 1/3$ for $p\geq 3$ and the original result of \cite{onsager1949statistical} is equivalent 
to the statement that $h_\infty=h_{\min}\leq 1/3.$ It is important to emphasize that these are predictive
statements which have received subsequent support from empirical determination of the multifractal dimension 
spectrum $D(h)$ both from numerical simulations \citep{kestener2004generalizing} 
and from hot-wire experiments \citep{lashermes2008comprehensive}.  As I shall discuss later, 
the ``ideal turbulence'' theory makes connection also with the alternative multifractal theory 
for the energy dissipation rate 
\citep{kolmogorov1962refinement,mandelbrot1974intermittent,mandelbrot1989multifractal,meneveau1991multifractal}. 

Another successful prediction made by \cite{onsager1945distribution,onsager1949statistical} concerned
the {\it locality} of the energy cascade, which is the statement that the nonlinear flux $\Pi_\ell$ 
defined in \eqref{II16} as a  cubic function of the velocity field $\bu$ is determined predominantly 
by velocity modes or ``eddies'' of scale near $\ell.$ In the words of  \cite{onsager1945distribution}: 
 ``The modulation of a given Fourier component of the motion is mostly due to those others which
belong to wavenumbers of comparable magnitude.''  In his 1945 letter to von K\'arm\'an and Lin
(reproduced in \cite{eyink2006onsager}, Appendix B), Onsager stated more precisely that ``With a 
hypothesis slightly stronger than (14) the motion which belongs to wave-numbers of the same order 
of magnitude as $\underline{k}$ itself will furnish the greater part of the effective rate of shear,''
where the mentioned condition (14) asserts that $\bu$ is square-integrable but that $\grad\bu$ 
is not. A condition of exactly this type is the H\"older condition \eqref{II21}, which may be shown
for any $0<h<1$ to imply locality. In fact, scale locality of energy cascade holds assuming only the 
$p$th-order Besov condition \eqref{Besov} for any $0<\sigma_p<1$ \citep{eyink2005locality,cheskidov2008energy}. 
These predictions have been confirmed in a number of numerical studies, for example, \cite{domaradzki2007analysis,
aluie2009localness,cardesa2015temporal}.

It is crucial that these locality properties hold instantaneously and deterministically, for individual flow 
realizations \citep{kraichnan1974kolmogorov}. The cited mathematical analyses and numerical simulations 
verify that averaging over velocity realizations improves the degree of locality but also that 
averaging is not necessary for local-in-scale interactions to dominate in energy transfer. 
Locality for individual realizations is necessary for another important feature of turbulence, 
the {\it universality} of small-scale statistics. The usual argument for such universal 
statistics has to do with the scale invariance of the Euler equations and the scale-locality of the energy transfer
by which small scales are excited. It is by the latter property that, in the words of \cite{onsager1949statistical},  
``we are led to expect a cascade such that the wave-numbers increase typically in a geometric series, by a factor 
of the order 2 per step.''  If each step in the cascade is chaotic and also of comparable nature to the previous 
steps, because of scale-invariance, then it is reasonable to expect that the details of large-scale flow features
will be lost and superseded by motions intrinsic to the local Euler dynamics. It is further quite obvious that 
scale-locality for energy transfer only in the mean sense would be inadequate for such universality, because 
then long-range communication between large and small scales could exist instantaneously which is 
canceled only by averaging over time or initial data. 

This universality is one of the many features that connects Onsager's ``ideal turbulence'' with the 
Large-Eddy Simulation/LES method of modelling turbulence.  In LES the large scale motions (large eddies) 
of turbulent flow are computed directly and only small scale (sub-grid scale/SGS) motions are modelled, 
resulting in a significant reduction in computational cost compared to direct numerical simulation/DNS
\citep{piomelli1999large,meneveau2000scale}. The justification for the LES scheme is that the turbulent 
small-scales are expected to be statistically similar for every flow and thus may be universally modelled,
while the non-universal large scales are explicitly computed at far lower cost. LES originated in the pioneering 
work of \cite{smagorinsky1963general} and \cite{lilly1967representation}, and it has developed historically with 
little connection to Onsager's theory of ``ideal turbulence'', although the two are quite intimately related. 
At the most superficial level, the analysis used by \cite{constantin1994onsager} to prove Onsager's 1/3 
result is the same as the filtering approach advocated by \cite{leonard1974energy,germano1992turbulence} and widely 
used in LES. Thus, the filtered equations \eqref{FEuler} used to derive Onsager's theorem are the same as those 
employed in LES. However, filtering/coarse-graining is just one specific form of regularization and other schemes are 
possible (e.g. Onsager used also Fourier series as a regularizer). The specific regularization by spatial filtering 
is not intrinsic to either method. More important is that both approaches aim to describe individual flow 
realizations, unlike Reynolds-Averaged Navier-Stokes/RANS modelling which resigns itself to calculation of 
time- or ensemble-averaged velocities only. Furthermore, Onsager's theory justifies the truism that an LES 
model should run even with molecular viscosity set to zero and thus should describe turbulent dissipation 
at infinite Reynolds number.  

{\begin{figure}
\centering
\includegraphics[width=.5\textwidth]{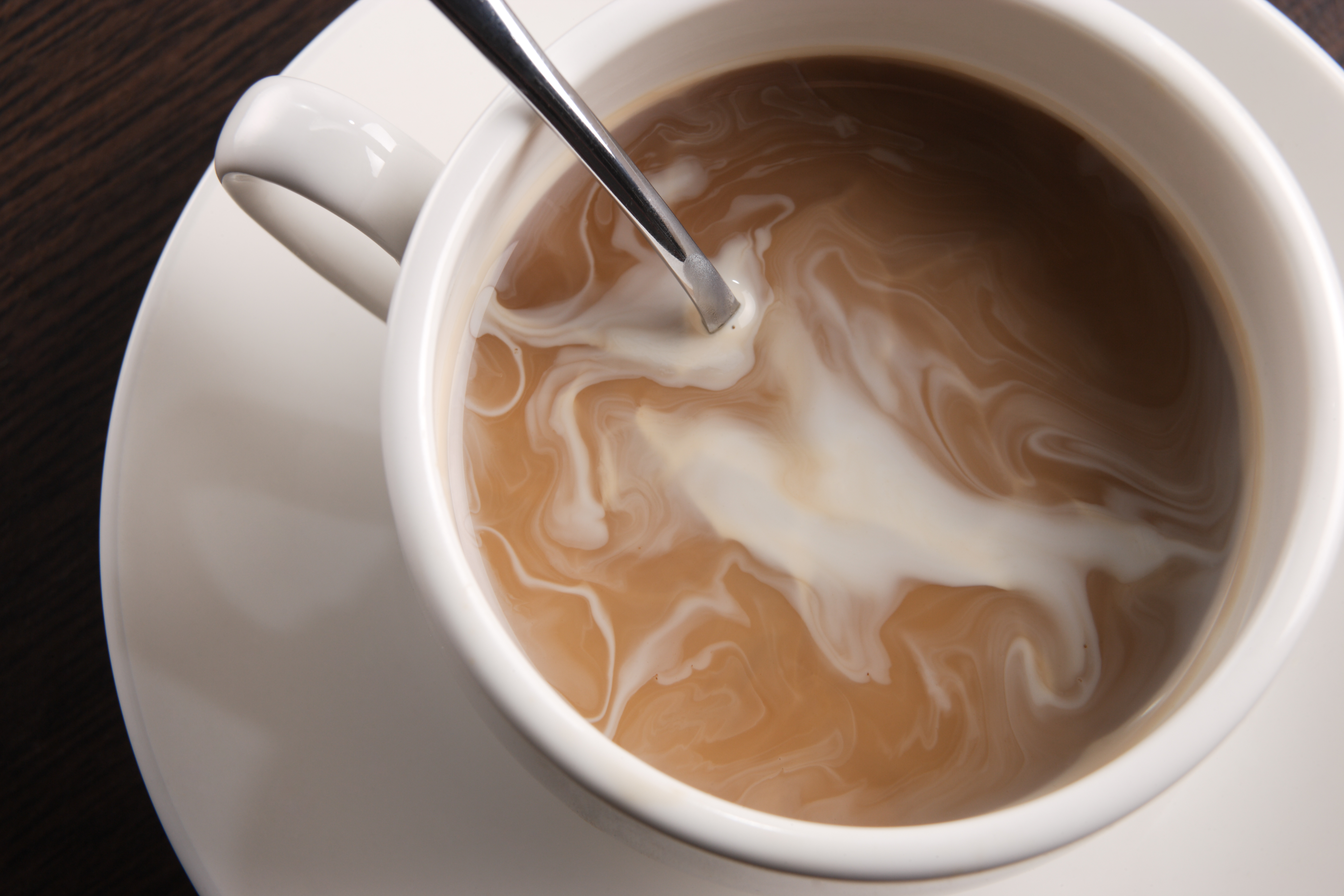}
\caption{Mixing of cream in a mug of coffee, moderately turbulent from mild stirring.
Reproduced on license from Shutterstock.}
\label{fig_coffee}
\end{figure}}

In concluding this section I  want to emphasize the conceptual and practical importance of a deterministic 
description of turbulent flow for individual flow realizations. In much of the scientific literature on fluid turbulence,   
it is often reflexively assumed that any mathematical description must be statistical in nature. However, this 
instinctive reaction does not permit one to discuss even such a commonplace flow as your morning cup of coffee.
See Fig.~\ref{fig_coffee}. This flow has a decent Reynolds number that may be estimated of order 
$Re=UD/\nu=3\times 10^4$ (assuming stirring velocity $U=20$ cm/s, cup diameter $D=6$ cm, and 
kinematic viscosity $\nu=0.4\ {\rm mm}^2/{\rm s}$) and is thus modestly turbulent. However, there is no ensemble in sight!
The coffee is stirred only a few times to mix the cream and thus the flow is decaying, so that one cannot 
average over time. Also, the size of the largest eddies is of the order of the cup diameter and thus one 
cannot average over many integral lengths of the flow. Nevertheless the fluid flow in the cup is turbulent 
each time you stir it and one must be able to describe the specific, individual flow. This need is common 
in many areas of science, such as geophysics and astrophysics, where some specific hurricane
or some specific supernova must be understood. These remarks are not intended to deny the intrinsic  
stochasticity of turbulent flow. It is also true that, when you stir your morning cup of coffee, each time you 
will see a different pattern of the cream, no matter how carefully you try to repeat your actions each day. However, 
in many discussions of turbulence the applications of probabilistic methods are entirely gratuitous. It is only 
by eliminating the superficial and unnecessary resorts to a statistical description that one can  uncover 
where probability theory is truly essential. 

\subsubsection{Local Deterministic $\frac{4}{5}$th-Law and Dissipative Anomaly}\lb{sec:DissAnom} 

One of the landmark results of the statistical approach to turbulence is the celebrated {\it 4/5th-law} derived by 
\cite{kolmogorov1941dissipation}, who invoked statistical hypotheses of local homogeneity and isotropy 
of the flow. In a remarkable work developing Onsager's ideas, \cite{duchon2000inertial} have shown that
the 4/5th law holds not only deterministically for individual flow realizations, not requiring any statistical 
hypotheses, but also that it holds in a much stronger spacetime local form. As we shall see, these developments were also 
foreshadowed by Onsager's own unpublished work.  Here I  shall sketch the essential ideas in the work of 
\cite{duchon2000inertial} and discuss their various ramifications. For a much more detailed explanation, 
see the online course notes of \cite{eyink2007turbulenceI}, Section III.C. 

The analysis of \cite{duchon2000inertial} again attempts to understand the zeroth-law of turbulence or the 
``inertial-dissipation'' of kinetic energy. The basic problem remains that UV divergences must appear in the inviscid
limit, so that the fluid equations can no longer be understood as PDE's in the naive sense. The starting point 
of \cite{duchon2000inertial} is a balance equation for a point-split kinetic energy 
density $\frac{1}{2}\bu(\bx,t)\bdot \bu(\bx+ \br,t)$: 
\begin{eqnarray}
\partial_t (\frac{1}{2} \bu \cdot \bu') &+& \grad \cdot \left[(\frac{1}{2}\bu \cdot \bu')\bu + \frac{1}{2}(p \bu' + p' \bu) + \frac{1}{4}|\bu'|^2 \delta \bu - \nu \grad(\frac{1}{2}\bu \cdot \bu')\right] \cr
&=& \frac{1}{4} \grad_\br \cdot [\delta \bu \,|\delta \bu|^2] - \nu \grad \bu : \grad \bu' + \frac{1}{2}({\bf f} \cdot \bu' + {\bf f'} \cdot \bu), 
\lb{split} \end{eqnarray}
\noindent where I  have introduced the abbreviated notations
\begin{eqnarray*}
\bu &=& \bu(\bx,t),\,\,\, p = p(\bx,t),\,\,\, {\bf f} = {\bf f}(\bx,t)  \cr
\bu' &=& \bu(\bx+ \br,t), \,\,\,p' = p(\bx+ \br,t),\,\,\, {\bf f}' = {\bf f}(\bx+ \br,t)  \cr
\delta \bu &=& \bu' - \bu = \bu(\bx+ \br,t) - \bu(\bx,t) 
\end{eqnarray*}
and I  have permitted a body force ${\bf f}$ in the Navier-Stokes equation, for greater generality.  
Similar exact equations were introduced somewhat later by \cite{hill2001equations,hill2002exact} as 
deterministic versions of the K\'arm\'an-Howarth-Monin relations. However, the balance equation 
\eqref{split} does not yet eliminate UV divergences by point-splitting alone, which requires further multiplying through by a filter kernel $G_\ell(\br)$ and integrating 
over $\br.$ This yields a corresponding balance equation for the regularized energy density $\frac{1}{2} \bu \cdot \bar{\bu}_\ell$:
\begin{eqnarray}
&& \partial_t (\frac{1}{2} \bu \cdot \bar{\bu}_\ell) + \grad \cdot \left[(\frac{1}{2}\bu \cdot \bar{\bu}_\ell)\bu + \frac{1}{2}(p \bar{\bu}_\ell + \bar{p}_\ell \bu) + \frac{1}{4} \overline{(|\bu|^2 \bu)_\ell} - \frac{1}{4} \overline{(|\bu|^2)_\ell} \bu - \nu \grad(\frac{1}{2}\bu \cdot \bar{\bu}_\ell)\right] \cr
&& \hspace{30pt} =\ -\frac{1}{4 \ell} \int d^d \br \,(\grad G)_\ell(\br) \cdot \delta \bu(\br) \,|\delta \bu(\br)|^2 - \nu \grad \bu : \grad \bar{\bu}_\ell + \frac{1}{2}({\bf f} \cdot \bar{\bu}_\ell + \bar{{\bf f}}_\ell \cdot \bu).
\lb{Fsplit} 
\end{eqnarray} 
It is now easy to check that all terms remain bounded in the limit as $\nu\to 0.$ 

The next step of \cite{duchon2000inertial} was to study the convergence as $\nu\to 0$ of the various terms in the regularized energy
balance \eqref{Fsplit}, under the assumption of strong $L^3$-convergence of Navier-Stokes solutions $\bu^\nu$ to some $\bu,$
i.e. $\lim_{\nu\to 0}\|\bu^\nu-\bu\|_3=0.$ Note that $\bu$ is then necessarily a weak Euler solution. As with the prior assumption of strong 
$L^2$-convergence, this even stronger convergence assumption is not guaranteed {\it a priori}, but we shall see that it is implied 
by other empirically observed properties.  Taking strong $L^3$-convergence for granted, it is easy to check that all of the terms proportional 
to $\nu$ in \eqref{Fsplit} in fact disappear in the limit. Furthermore, all of the remaining terms are found to converge in the sense of distributions 
to the same expression but with Navier-Stokes velocity $\bu^\nu$ replaced with the limiting field $\bu$, yielding a modified energy balance 
for the inviscid Euler solution:  
\begin{eqnarray}
&& \partial_t (\frac{1}{2} \bu \cdot \bar{\bu}_\ell) + \grad \cdot \left[(\frac{1}{2}\bu \cdot \bar{\bu}_\ell)\bu + \frac{1}{2}(p \bar{\bu}_\ell + \bar{p}_\ell \bu) + \frac{1}{4} \overline{(|\bu|^2 \bu)_\ell} - \frac{1}{4} \overline{(|\bu|^2)_\ell} \bu
\right] \cr
&& \hspace{30pt} =\ -\frac{1}{4 \ell} \int d^d \br \,(\grad G)_\ell(\br) \cdot \delta \bu(\br) \,|\delta \bu(\br)|^2 + \frac{1}{2}({\bf f} \cdot \bar{\bu}_\ell + \bar{{\bf f}}_\ell \cdot \bu).
\lb{FsplitE} 
\end{eqnarray} 
Note, in particular, that validity in the sense of distributions means that the above balance equation is implicitly smeared with 
a smooth test function $\varphi(\bx,t)$ in spacetime and that all derivatives acting on $\bu$ or related solution fields can be moved over 
to the test function. From these remarks it is then easy to check further that the limit as $\ell\to 0$ of all terms exist directly, 
except for the term containing $(\grad G)_\ell(\br)$. However, because all other terms in the equation converge, this term must 
also converge.  \cite{duchon2000inertial} thus concluded that a local kinetic energy balance holds in the sense 
of distributions for the limiting Euler solution
\be \partial_t (\frac{1}{2}|\bu|^2) + \grad \cdot \left[(\frac{1}{2}|\bu|^2 + p)\bu\right] \equiv - D(\bu) +{\bf f}\bdot\bu \lb{Ebal} \ee
where
\be  D(\bu) = \lim_{\ell \rightarrow 0} \frac{1}{4 \ell} \int d^d \br \,\,(\grad G)_\ell (\br) \cdot \delta \bu(\br) |\delta \bu(\br)|^2
:= \lim_{\ell \rightarrow 0} D_\ell(\bu). \lb{DR} \ee 
The na\"{\i}ve kinetic energy balance of the Euler equations is broken by the term $D(\bu),$ which can be easily shown to vanish  
unless the velocity has Besov regularity exponent $\sigma_p\leq 1/3$ for $p\geq 3.$ Note in fact that the same final equation 
\eqref{Ebal} can be obtained also by considering the coarse-grained energy balance \eqref{II15} with both factors of $\bu$ filtered 
and then taking the limit $\ell\to 0$ in the same manner yields 
\be  D(\bu) = \lim_{\ell \rightarrow 0} \Pi_\ell \lb{DPi} \ee 
so that the ``inertial dissipation'' in fact represents nonlinear energy cascade to infinitesimally small length scales. 

Physically, this loss of energy must arise from viscous dissipation. Notice, however, the equations \eqref{FsplitE}-\eqref{DPi} can 
be derived for {\it any} weak Euler solution, under the sole requirement that $\bu\in L^3.$ Because the notion of weak
Euler solution is time-reversible but $D(\bu)$ is cubic in the velocity field, it follows that any Euler solution with $D(\bu)>0$
yields by time-reversal another Euler solution with $D(\bu)<0.$ It is only Euler solutions obtained in the inviscid limit, or so-called
``viscosity solutions'' of Euler, which can be expected to satisfy this physical expectation. Because coarse-graining is purely optional, 
one should be able to obtain the limiting energy balance \eqref{Ebal} directly in the limit $\nu\to 0.$ This is what \cite{duchon2000inertial} 
did, starting with the familiar kinetic energy balance for Navier-Stokes solutions $\bu^\nu$: 
\be \partial_t(\frac{1}{2} |\bu^\nu|^2) + \grad \cdot \left[(\frac{1}{2}|\bu^\nu|^2 + p^\nu) \bu^\nu - \nu \grad(\frac{1}{2} |\bu^\nu|^2)\right] 
= - \nu |\grad \bu^\nu|^2 +{\bf f}\bdot\bu^\nu.  \lb{NSbal} \ee
(Here I  note in passing that \cite{duchon2000inertial} considered a more general situation where the Navier-Stokes solutions 
are themselves singular and must be interpreted distributionally \`a la \cite{leray1934mouvement}, in which case the ``='' 
in \eqref{NSbal} is replaced by ``$\leq$''. For reasons discussed later, I am not interested physically in this mathematical 
generality.) Now assuming once more the strong $L^3$ convergence $\bu^\nu\to\bu,$ it is easy to see that all of the terms 
on the lefthand side converge distributionally as $\nu\to 0$ to the expected limit on the lefthand side of \eqref{Ebal} and
likewise ${\bf f}\bdot \bu^\nu\to{\bf f}\bdot\bu.$ The only term which cannot be shown directly to converge is the viscous 
dissipation $\nu |\grad \bu^\nu|^2,$ but, since all other terms in the Navier-Stokes balance \eqref{NSbal} converge, so 
also does this term.  One thus derives the same local energy balance \eqref{Ebal} as before for the limiting Euler solution 
but one obtains furthermore the new expression 
\be D(\bu) = \lim_{\nu \rightarrow 0} \nu |\grad \bu^\nu|^2 \geq 0. \lb{match} \ee 
This result can be understood as a matching relation for the ``viscosity solutions'' of Euler equations, which equates 
the energy carried by nonlinear cascade to infinitesimally small scales and the usual energy dissipation by 
molecular viscosity. 

A final important result obtained by \cite{duchon2000inertial} exploited additional freedom in the regularization 
of UV divergences. In fact, not only is the length scale $\ell$ arbitrary, but so also is the precise choice of the 
kernel $G.$ This freedom can be exploited by choosing a rotationally symmetric kernel for which 
$$ \grad G(\br) = \hat{\br} G'(r). $$ 
In that case, the $d$-dimensional integral over $\br$ in the formula \eqref{DR} for $D(\bu)$ can be transformed 
into hyperspherical coordinates so that 
$$D_\ell(\bu) = \frac{1}{4 \ell} \int^\infty_0 r^{d-1} dr \, \int_{S^{d-1}} d \omega_d (\hat{\br}) \hat{\br} \cdot \delta \bu(\br) |\delta \bu(\br)|^2 (G')_\ell (r) $$
where $S^{d-1}$ is the unit hypersphere and $d \omega_d$ is the measure on  in $d$-dimensional solid angles. Using 
the integration by parts identity $\int_0^\infty r^d G'(r)\, dr=d,$ one can infer a new identity 
\be \lim_{r \rightarrow 0} \frac{\langle \delta u_L(\br) |\delta \bu(\br)|^2\rangle_{ang}}{r} = - \frac{4}{d} D(\bu), \lb{43rd} \ee
where $\langle \cdot\rangle_{ang}$ denotes for any function of $\br$ the average with respect to $\omega_d$ 
over the direction $\hat{\br}$ and $\delta u_L(\br)=\hat{\br}\bdot\delta\bu(\br)$ is the longitudinal velocity increment. 
Combined with the previous expression \eqref{match} for $D(\bu)$ in terms of the viscous dissipation, one can see 
that \eqref{43rd} for $d=3$ has the form of the Kolmogorov-Yaglom 4/3rd-law \citep{antonia1997analogy}.  It is well-known 
in the standard derivation by ensemble-averages that the Kolmogorov 4/5th- and 4/15th-laws are equivalent to the 
4/3rd-law under the assumption of statistical isotropy. Since the angle average in the present derivations provides 
isotropy without any statistical hypotheses, some further manipulation \citep{eyink2003local,novack2023scaling} yields 
\begin{eqnarray} 
\lim_{r \rightarrow 0} \frac{\langle\delta u^3_L(\br)\rangle_{ang}}{r} &=& -\frac{12}{d(d+2)} D(\bu) \lb{45th} 
\end{eqnarray} 
\begin{eqnarray}  
\lim_{r \rightarrow 0} \frac{\langle\delta u_L(\br) \delta u^2_T(\br)\rangle_{ang}}{r} &=& -\frac{4}{d(d+2)} D(\bu) \lb{415th} 
\end{eqnarray} 
where $\delta \bu_T(\br)=\delta \bu(\br)-\hat{\br}\delta u_L(\br)$ is the full transverse velocity increment and 
$\delta u_T(\br)$ is the magnitude of any particular (fixed) component.  One can see for $d=3$ that \eqref{45th} 
has the form of the usual 4/5th-law and \eqref{415th} has the form of the 4/15th-law.

Note, however, that the relations \eqref{43rd}-\eqref{415th} are deterministic, holding for individual flow realizations, and are 
furthermore spacetime local in the sense of distributions. The latter statement simply means that  they hold with both sides 
smeared by an arbitrary space-time test function $\varphi(\bx,t),$ taking first $\nu\to 0$ and then $r\to 0.$ Thus, these 
results represent a consider strengthening of the original ensemble-average relations of \cite{kolmogorov1941dissipation}.
An effort was made by \cite{taylor2003recovering} to test these deterministic local relations in pseudospectral numerical 
simulations, but the $512^3$ resolution available at the time was insufficient. It was possible to show that angle-averaging 
greatly improved the validity of the observed $4/5$th-law, attaining results comparable to those at $1024^3$ resolution 
even without time-averaging. Thus, the local 4/5th law appeared in fact to hold instantaneously, without smearing in time, 
beyond what could be proved mathematically. Note that it is only with simulations of  $16384^3$ resolution that 
the 4/5th-law has recently been demonstrated convincingly, exploiting both angle-averaging and time-averaging
\citep{iyer2020scaling}. This brings us close to being able to check the local $4/5$th law in simulations of $32768^3$ resolution 
by dividing the computational cube into eight $16384^3$ subcubes and  verifying that the relation holds with the structure functions 
and mean dissipation calculated by space-averages individually over each subcube. Of course, the result should hold with 
any spacetime test function $\varphi$ held fixed in the limit first $\nu\to 0$ and then $r\to 0,$ but current computational 
limitations seem to make characteristic functions of octants the only choice consistent with the stringent condition
that the local Reynolds number must be high. It is worth remarking that the analogue of 
the deterministic, local 4/5th-law can be rigorously derived for the ``viscosity solutions'' of the inviscid Burgers equation, 
$\partial_t u +\partial_x (\frac{1}{2}u^2)=0.$ In that case the local energy balance holds
$$\partial_t (\frac{1}{2}u^2) + \partial_x (\frac{1}{3} u^3) = - D(u)  $$
with the ``inertial dissipation'' given by  
$$\lim_{r \rightarrow 0} \frac{\langle\delta u^3_L(r)\rangle_{ang}}{|r|} = -12 D(u) $$ 
where $\delta u_L(r):={\rm sign}(r) \delta u(r)$ and $\langle\delta u^3_L(r)\rangle_{ang} = \frac{1}{2}[\delta u^3(+ |r|) - \delta u^3(-|r|)]$. 
See \cite{eyink2007turbulenceI}, Section III, and  \cite{novack2023scaling}. These relations correspond to the ensemble-averaged
``$12$th-law'' for Burgulence, but valid in spacetime local form with ``inertial dissipation'' arising entirely from shock singularities.   

It should be emphasized again that the deterministic versions of the 4/5th-law \eqref{45th} and the 4/15th-law \eqref{415th}, 
and all of the earlier results in this section, require no statistical assumptions whatsoever. Neither 
local homogeneity, local stationarity, isotropy nor any other hypothesis about the flow statistics was invoked 
in their derivation. In particular, these relations will apply in the ``nonequilibrium decay regime'' highlighted 
in the review of \cite{vassilicos2015dissipation} on turbulent energy dissipation. \cite{vassilicos2015dissipation} 
reported observations in a near wake region of various grids and bluff bodies that the time-averaged quantity $C_\epsilon:=
\langle D\rangle=\langle \varepsilon\rangle/(u^{\prime 3}/L)$
scales as $~Re_I^m/Re_L^n$ where in his notations $Re_I=UL_b/\nu$ is the ``global'' or ``inlet Reynolds number'' based
on the inlet flow speed $U$ and a length $L_b$ giving the overall size of the grid or body, whereas $Re_L=u'L/\nu$ 
is the ``local Reynolds number''. In the wake flows considered by  \cite{vassilicos2015dissipation} both the 
rms velocity fluctuation $u'$ and integral length $L$ are statistical quantities that depend upon the longitudinal 
distance $x$ of the observation point from the grid/body. Since wake flows involve interactions with solid walls, 
I shall treat the energy cascade in such flows in section \ref{sec:energy}, but here I note that all of our previous 
results carry over, with possible modifications only directly at the solid surface. As I shall discuss in section \ref{sec:overture},
for flows with walls or solid bodies, I always consider the global Reynolds number $Re_I,$ which appears in the 
non-dimensionalization of the Navier-Stokes equation. As noted already by \cite{vassilicos2015dissipation} (p.108), 
the quantity $C_\epsilon$ at each $x$-location is ``more or less independent of the Reynolds number'' $Re_I$, even in the 
nonequilibrium decay region because ``$m\approx 1\approx n$, and $Re_L$ increases linearly with $Re_I$''.
In fact, $m=n$ corresponds exactly to the ``dissipative anomaly'' conjectured by \cite{onsager1949statistical}  
and all of our previous conclusions follow in the limit $Re_I\to\infty.$ In particular, the local 4/5th-law \eqref{45th} and 
the 4/15th-law \eqref{415th} will hold in the nonequilibrium decay region, not only for time/ensemble-averages
but even for individual flow realizations. \cite{vassilicos2015dissipation} has observed that ``the Richardson-Kolmogorov 
cascade does not seem to be the ... interscale energy transfer mechanism at work'', but the dissipative anomaly for $Re_I\gg 1$ 
must occur by Onsager's local cascade mechanism, which generalizes Kolmogorov's equilibrium picture. I agree
with the observation of \cite{vassilicos2015dissipation} that the 4/5th-law of \cite{kolmogorov1941dissipation} need not be 
valid in general for $r\simeq L.$ In fact, our local, deterministic version \eqref{45th} will generally only hold for $r\ll L_\varphi,$ where 
$L_\varphi$ measures the spatial diameter of the support of the test function $\varphi$ required to make  \eqref{45th} meaningful. 

Amazingly enough, most of the previous developments were foreseen by Onsager in his unpublished work in the 1940's.
When I was on sabbatical at Yale in 2000, I was able to examine Onsager's private research notes available 
there on microfiche and I was astonished to discover that Onsager's own proof of his 1/3 H\"older claim was essentially 
identical to that given by \cite{duchon2000inertial}. These notes are now all available online through the Onsager 
Archive hosted by the NTNU in Trondheim and the reader can peruse them at 
\url{https://ntnu.tind.io/record/121183}. The essential calculations are on pp.14-19 of Onsager's notes in this folder,
where he derives the identity \eqref{FsplitE} but in a space integrated form and with the filtering function $G_\ell(\br)$
denoted $F(\br).$ It is interesting that on p.15, Onsager begins the calculation by writing an overbar, his notation 
for ensemble-average, but then scratches out the overbar in the second line, apparently realizing that volume-averaging 
was sufficient to derive the result. This seems to have been the moment that the theme of my essay was born. Onsager 
communicated his identity to von K\'arm\'an and Lin in his letter of June 1945 (reproduced as Appendix B in 
\cite{eyink2006onsager}) where he gave an argument for $1/3$-scaling by taking $F(\br)$ to be a spherical
tophat filter of radius $a$ and then letting $a\to 0.$ It is further interesting that Onsager in this same letter also foresaw that 
inertial-range intermittency could lead to an energy spectrum having a steeper slope than $-5/3,$ writing that 
``As far as I can make out, a more rapid decrease of $\overline{a_k^2}$ with increasing $\underline{k}$ would require 
a `spotty' distribution of the regions in which the velocity varies rapidly between neighboring points.'' 
Onsager then notes that all velocity increments supported in discontinuities at vortex sheets would produce 
$S_2(r)\propto r,$ corresponding to a $k^{-2}$ energy spectrum as for Burgers equation. Such sheets would also be consistent 
with $S_3(r)\propto r$ but Onsager remarks that they would not agree well with experimental observations on velocity traces. 
It is worth remarking that vortex sheets cannot, in fact, be the origin of anomalous energy dissipation \citep{derosa2024dissipation}.
For more discussion of Onsager's insights on inertial-range intermittency, see \cite{eyink2006onsager}, Section IV.C. 

The local kinetic energy balance \eqref{Ebal} derived by \cite{duchon2000inertial} for weak Euler solutions is the 
most succinct formulation of Onsager's proposal for turbulent dissipation ``without the final assistance by viscosity."
This result has a very familiar appearance to a modern field-theorist, because it is reminiscent of
anomalies in classical conservation laws due to UV divergences in the quantum field theory.
This analogy seems to have been first pointed out by \cite{polyakov1992conformal,polyakov1993theory}, who 
did not know about Onsager's work, of course, but who based his discussion on the statistical theory of Kolmogorov. 
Polyakov pointed out the essential similarity between turbulent cascades and chiral anomalies in 
quantum Yang-Mills theory, as both involve a constant flux through wavenumbers which vitiates a na\"{\i}ve 
conservation law. He pointed out also that the first derivation of the chiral anomaly in quantum electrodynamics  
by \cite{schwinger1951gauge} using a point-splitting regularization was very similar to the derivation of the 
4/5th-law by \cite{kolmogorov1941dissipation}. It is interesting that Schwinger seemingly did not fully appreciate 
the significance of his own calculation and did not point out its implication that conservation of axial charge is 
violated \citep{adler2005anomalies}. On the other hand, we now see that Onsager had used a similar 
point-splitting regularization already in the 1940's and had fully appreciated its consequence that na\"{\i}ve 
conservation of kinetic energy in the inviscid limit is broken by the ``violet catastrophes'' in turbulent  flows.  
This phenomenon is thus now commonly called a turbulent {\it dissipative anomaly}, because of the close 
connection with quantum-field theory anomalies. For more detailed discussion of this analogy, see
\cite{eyink2006onsager}, Section IV.B. 

I would like to make just a brief remark why the term ``anomaly'' is apt for the turbulent phenomenon,
independent of the connection with quantum field-theory. It should be emphasized that no fundamental
microscopic conservations laws for total energy, total linear momentum, total angular momentum, etc. 
are ever violated by such anomalies. However, there is obviously no microscopic law of ``conservation 
of kinetic energy''! This conservation law is an {\it emergent property} of inviscid hydrodynamics due to its  
Hamiltonian character, in which the kinetic energy of the fluid is itself the conserved Hamiltonian 
\citep{salmon1988hamiltonian,morrison2006hamiltonian}. It is because of this Hamiltonian structure 
of the Euler equations that kinetic energy conservation is formally expected and thus the breakdown 
of this conservation is ``anomalous''. The same is true more generally for turbulent 
dissipative anomalies. For example, conservation of helicity and conservation of circulation on arbitrary 
loops are not sacrosanct microscopic laws but are instead emergent laws connected with ``relabelling symmetry'',
an infinite-dimensional symmetry group of the action for three-dimensional ideal Euler equations. It is such 
emergent conservation laws of the ideal fluid which may be afflicted with dissipative anomalies in 
turbulent flows. It is interesting that some of these symmetries are preserved in a formulation of the 
viscous Navier-Stokes via a stochastic least-action principle \citep{eyink2010stochastic}, suggesting 
that not all of these emergent conservation laws will be explicitly broken by dissipative anomalies 
but may instead be realized in an unconventional stochastic form. See further discussion in the 
following section \ref{sec:open} {\it (b)}. 

Finally, I  point out that the work of  \cite{duchon2000inertial} makes a connection between 
the ``ideal turbulence'' theory and the multifractal model of energy dissipation. After 
\cite{kolmogorov1962refinement} had proposed his lognormal model of turbulent energy dissipation 
to account for intermittency corrections, \cite{mandelbrot1974intermittent,mandelbrot1989multifractal}
introduced a more general description of the energy dissipation as a {\it multifractal measure} with a spectrum 
of singularities. Subsequent experimental studies of \cite{meneveau1991multifractal} supported these 
predicted scaling properties of turbulent energy dissipation. The result \eqref{match} of \cite{duchon2000inertial}
shows that the dissipative anomaly term $D(\bu)$ in the inviscid energy balance coincides exactly 
with this multifractal measure. In fact, the analysis of \cite{duchon2000inertial} implies that 
the inviscid limit of the viscous energy dissipation $\varepsilon^\nu=2\nu |\bS^\nu|^2$ exists and coincides 
with $D(\bu),$ which is a non-negative distribution and, thus, a measure. Recent rigorous results
relate the fractal dimension of the support of this measure to the inertial-range intermittency  
of the velocity increments \citep{derosa2022intermittency}. 

\subsubsection{Implications and Open Questions}\lb{sec:open} 

There are many questions raised by the original analysis of \cite{onsager1945distribution,onsager1949statistical} 
and also many subsequent developments extending and elaborating his ideas. Here I  shall briefly review 
and discuss such further implications and open issues. 

\vspace{10pt} 
\noindent 
{\it (a) Origin of singularities}. Onsager's result and its extensions by \cite{eyink1994energy,constantin1994onsager,duchon2000inertial} 
and others show that singularities of the velocity field are required to explain anomalous dissipation of kinetic energy in the inviscid limit 
of incompressible hydrodynamic turbulence. However, this theory does not explain the origin of such singularities. The traditional 
view has been that these singularities arise from finite-time blow-up of smooth Euler solutions, e.g. see \cite{frisch1995turbulence}, 
section 9.3. The most significant argument in favor of this view is based on a set of results in mathematical theory of PDE's which 
goes by the name {\it weak-strong uniqueness}. Such results state that a weak Euler solution (or an Euler solution in even a 
more general sense: see section \ref{sec:ConjIntro}) which has total kinetic energy non-increasing in time must coincide with a 
strong (i.e. smooth) solution with the same initial data, over the entire time-interval for which the latter exists. E.g. see 
\cite{lions1996mathematical,brenier2011weak,bardos2013mathematics,wiedemann2018weak}. These results can be used to infer that a weak Euler solution obtained by a 
zero-viscosity limit of a Navier-Stokes solution, even in a weaker sense of limit than discussed in section \ref{sec:DissAnom}, must coincide 
with any such smooth Euler solution. In particular, these weak-strong uniqueness results rule out the appearance of anomalous 
energy dissipation over any finite time-interval, if the Navier-Stokes equation is solved with initial data that is a smooth velocity field 
or that even converges (say in $L^2$) to a smooth field, and if the Euler solution with that initial data does not blow up. These arguments
seem to suggest that anomalous dissipation requires finite-time Euler singularities. There has also been recent progress 
in showing blow-up of strong Euler solutions, both by numerical simulations \citep{luo2014toward,hou2022potential} and by rigorous mathematical 
analysis \citep{elgindi2019finite,elgindi2021finite,chen2021finite,chen2022stable}.  

On the other hand, there are significant reasons to doubt that finite-time blow-up of smooth Euler solutions 
has anything at all to do with empirical observations on fluid turbulence. First, weak-strong uniqueness is 
known to break down in the presence of solid boundaries \citep{bardos2019onsager}, requiring in that case 
additional assumptions that may well be violated. This point will be discussed at length in section \ref{sec:wkstrng}. 
Furthermore, the hypothesis that dissipative singularities in incompressible turbulence form from finite-time Euler singularities
does not correlate well with observations. For example, it does not account for the dichotomy demonstrated by \cite{cadot1997energy} 
between flows with hydraulically smooth and rough walls, with vanishing dissipation for hydraulically smooth 
walls and a dissipative anomaly for hydraulically rough walls. In fact, the best numerical and mathematical evidence for a 
finite-time blow-up is in a cylindrical domain with smooth boundaries \citep{luo2014toward,chen2021finite,chen2022stable}, 
where empirical evidence suggests anomalous dissipation does not exist! I personally am of the opinion that solid walls must play 
a crucial role in the appearance of anomalous energy dissipation in incompressible fluid turbulence, as I  shall discuss in more 
detail in section \ref{walls}. I  just note here that {\it all} of the laboratory experiments and natural observations that support a turbulent 
anomaly involve fluid-solid interactions, e.g. the interactions with the solid grid for decaying turbulence in a wind-tunnel. 

The obvious objection to these remarks is the evidence for a dissipative anomaly arising from numerical simulations
in a periodic box, with no walls. However, the evidence for anomaly via blow-up from simulations is weak.  
For example, the simulation of \cite{kaneda2003energy} does not start with smooth initial data. As already discussed by 
\cite{drivas2019onsager}, Remark \#4, the simulation reported in \cite{kaneda2003energy} uses an iterative initialization 
procedure which builds in an initial quasi-singularity, corresponding to an increasing range of Kolmogorov-type spectrum.
This common device is widely regarded as a numerical short-cut to accelerate the convergence to a steady state, but it 
renders the simulations irrelevant to the issue of finite-time singularity. There are simulations that do employ 
smooth initial data, e.g. the Taylor-Green vortex as discussed recently by \cite{fehn2022numerical}. However, here 
the evidence for anomalous dissipation in a finite time is weaker, with no completely compelling convergence 
to a non-zero limit within computational limitations. 

\vspace{10pt} 
\noindent 
{\it (b) Lagrangian spontaneous stochasticity}. One of the most remarkable consequences of Onsager's H\"older 
singularity result was first discovered in the work of \cite{bernard1998slow}, who studied Lagrangian fluid 
particle trajectories that solve the initial-value problem:
\be \dot{\bx} =\bu(\bx,t), \quad \bx(0)=\bx_0. \lb{part-ivp} \ee
When $\bu$ is the limiting Euler solution at infinite-$Re$ then, as recalled by \cite{bernard1998slow}, 
the maximal H\"older regularity derived by \cite{onsager1949statistical} is insufficient to guarantee a 
unique solution of \eqref{part-ivp} and there is generally a continuous infinity of solutions with exactly 
the same initial data. It was realized by  \cite{bernard1998slow} that this non-uniqueness permits 
stochasticity to persist in the high-Reynolds limit. For example, the position of a Brownian particle such 
as a small colloid or a dye molecule advected by a turbulent flow will satisfy instead a stochastic ODE:
\be d\tilde{\bx} =\bu(\tilde{\bx},t)dt+\sqrt{2D}\, d\tilde{\bW}(t) , \quad \bx(0)=\bx_0, \lb{spart-ivp} \ee
where $D$ is the molecular diffusivity of the particle and $\tilde{{\bf W}}(t)$ is a Wiener process modeling the 
Brownian motion. To study the high-$Re$ limit, one may again non-dimensionalize by defining $\hat{\bu}=\bu/u',$
$\hat{\bx}=\bx/L,$ $\hat{t}=t/(L/u')$ and then $D$ is replaced by $1/Pe,$ where $Pe=u'L/D$ 
is the P\'eclet number. In the joint limit $Re\gg 1,$ $Pe\gg 1,$ the Lagrangian particle trajectories 
$\tilde{\bx}(t)$ solve the limiting deterministic ODE \eqref{part-ivp}, but  \cite{bernard1998slow} showed that 
they may remain random in the limit with a non-trivial transition probability density $p_\bu(\bx,t|\bx_0,0)$ 
to arrive at position $\bx$ at time $t.$ See Fig.~\ref{fig_SS}. The essential physics was implicit already 
in the work of \cite{richardson1926atmospheric} on 2-particle turbulent dispersion, according to which initial 
particle separations are ``forgotten'' at sufficiently long times. It is important to note however that there is 
no averaging over velocities in the prediction of \cite{bernard1998slow} and that the transition probability
represents stochastic particle evolution in a fixed realization $\bu$ of the turbulent velocity field.
\cite{bernard1998slow} furthermore showed that this ``spontaneous stochasticity'' of Lagrangian particles 
provides a mechanism for anomalous dissipation of a concentration field $c$ of such tracers, 
which satisfies the advection-diffusion equation 
\be \partial_t c(\bx,t)+\bu(\bx,t)\bdot\grad c(\bx,t)= D\Delta c(\bx,t), \quad c(\bx,0)=c_0(\bx). \lb{adv-diff} \ee
Since the solution of this equation (in non-dimensionalized form) has the exact Feynman-Kac 
representation 
\be c(\bx,t) =  \int p_\bu^{Re,Pe}(\bx,t|\bx_0,0)\, c_0(\bx_0) \, d^dx_0, \lb{cFKrep} \ee 
the existence of a nontrivial limit $\lim_{Re,Pe\to\infty} p_\bu^{Re,Pe}(\bx,t|\bx_0,0)
=p_\bu(\bx,t|\bx_0,0)$ implies by convexity that 
$\int \frac{1}{2} c^2(\bx,t)\, d^dx < \int \frac{1}{2} c^2_0(\bx_0)\, d^dx_0 $
in the limit $Re,Pe\to\infty.$

{\begin{figure}
\centering
\includegraphics[width=.5\textwidth,height=.44\textwidth]{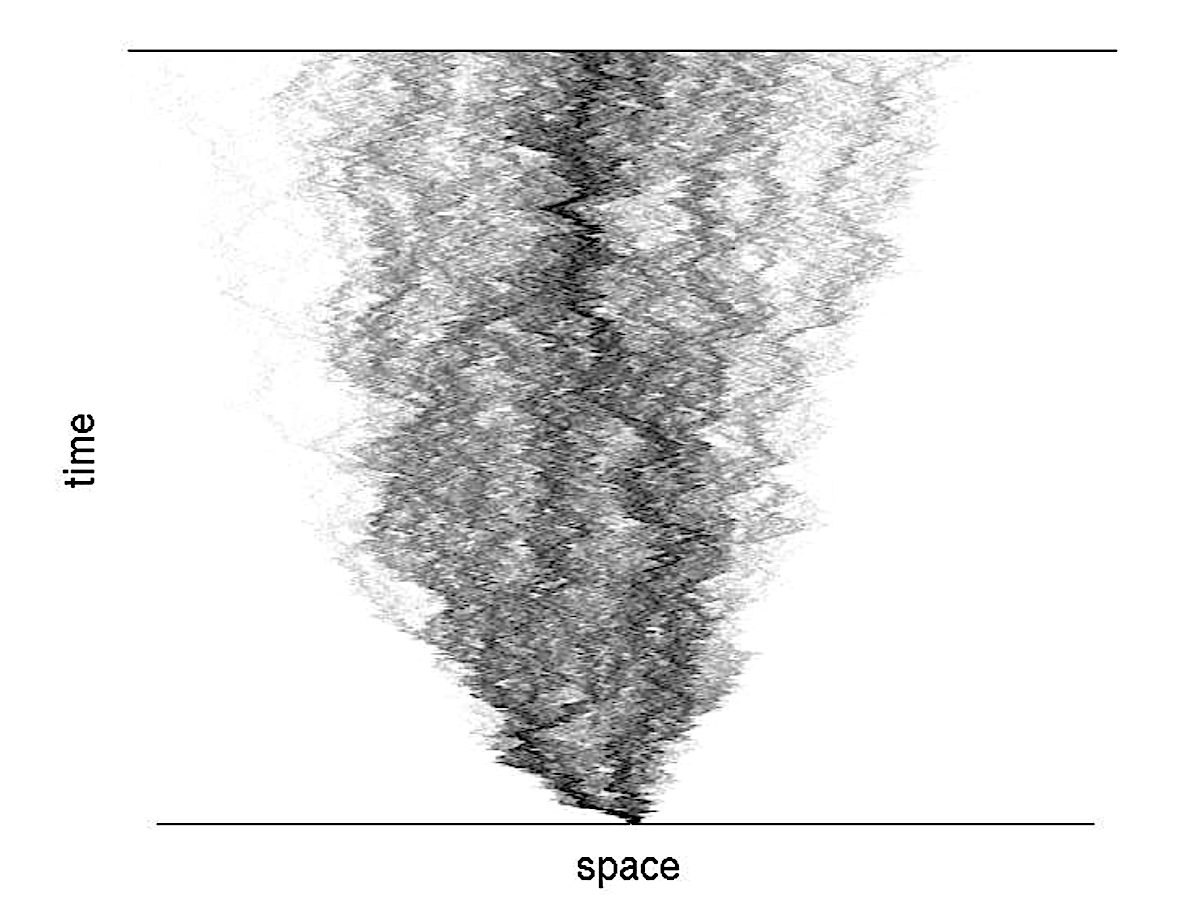}
\caption{Sketch of the solutions of a deterministic ODE $d\bx/dt=\bu(\bx,t)$ for deterministic initial data $\bx_0$ 
but with singular velocity $\bu$. Unlike traditional unique solutions, the trajectories spread randomly, like a plume of smoke. 
Reproduced with permission from G. Falkovich, K. Gaw\c{e}dzki, \& M. Vergassola, 
``Particles and fields in fluid turbulence,'' Rev. Mod. Phys. {\bf 73}, 913--975 (2001). 
Copyright (2001) by the American Physical Society.}
\label{fig_SS}
\end{figure}} 

The original work of \cite{bernard1998slow} was carried out for the synthetic turbulence model of 
\cite{kraichnan1968small}, who considered advection by Gaussian random velocity fields that are 
delta-correlated in time. Note that for suitable power-law spatial correlations to mimic inertial-range 
turbulent velocity fields, the realizations of the Kraichnan velocity ensemble are H\"older continuous 
with probability one. A very important feature of this model is that the limiting probability distributions
\be p_\bu(\bx,t|\bx_0,0)=\lim_{Re,Pe\to\infty} p_\bu^{Re,Pe}(\bx,t|\bx_0,0) \ee
are {\it independent} of the particular subsequences $Re_k,$ $Pe_k\to \infty$ and, in fact, are 
independent of the particular form of regularization and noise, e.g. the Brownian motion $\tilde{\bW}(t)$
in \eqref{spart-ivp} might be replaced with a fractional Brownian motion and the limit would be the same.
This strong result for the Kraichnan model follows from a rigorous study by 
\cite{lejan2002integration} who gave a direct construction of a Markov transition probability 
$p_\bu(\bx,t|\bx_0,0)$ for the zero-regularization, zero-noise problem by a Weiner chaos expansion 
in the white-noise advecting velocity field $\bu(\bx,t).$ It is a consequence of this construction 
that any reasonable spatial regularization and noisy perturbation will converge to this same transition
probability in the zero-regularization, zero-noise limit. It was also shown by \cite{lejan2002integration} 
that the realizations $\tilde{\bx}(t)$ of the limiting Markov process with transition densities $p_\bu(\bx,t|\bx_0,0)$
are solutions (in a generalized sense) of the deterministic initial-value problem \eqref{part-ivp}
for each fixed realization $\bu$ of the Kraichnan velocity ensemble. Thus, in this case, the Markov process 
$\tilde{\bx}(t)$ should be regarded as the proper solution of the above initial-value problem 
\eqref{part-ivp}, that is then well-posed in the sense of Hadamard. Spontaneous stochasticity in the strong 
sense should be understood to include such universality of the limiting random process, that is, the 
robustness of the stochastic solution to different regularizations and noisy perturbations. 

Extensions of the work of \cite{bernard1998slow} and further developments 
on the Kraichnan model are described by \cite{falkovich2001particles}. This comprehensive review makes 
clear that many of the results have validity extending beyond the Kraichnan model and, in fact, 
\cite{drivas2017lagrangianI} showed that spontaneous stochasticity is the only possible mechanism 
of scalar anomalous dissipation away from solid walls, for both passive and active scalars advected 
by a general incompressible velocity field. Note that universality of the limiting spontaneous statistics
is not required in this case, so that different limiting random processes yielding anomalous dissipation may result 
from different subsequences and/or different regularizations.There are, however, only a handful of model 
velocity fields for which Lagrangian spontaneous stochasticity can be demonstrated from first principles 
\citep{drivas2021life,drivas2020statistical}. Renormalization group methods  from statistical physics 
apply to velocity fields with suitable scaling properties and yield universality of the limit 
\citep{eyink2020renormalization}, but the underlying ergodic properties of the dynamical flows 
that are required for this method are difficult to establish in general. One interesting  
example of a velocity field solving a PDE is the solution of inviscid Burgers equation, where Lagrangian spontaneous 
stochasticity holds backward in time at shock locations and explains anomalous energy dissipation
\citep{eyink2015spontaneous}. There is no such spontaneous stochasticity for particles advected 
forward in time  by the Burgers velocity field and it has been speculated that similar time-asymmetry 
may be a more general feature of Lagrangian spontaneous stochasticity in dissipative weak solutions 
of PDE's obtained as inviscid limits. 

As far as Navier-Stokes turbulence is concerned, there is so far no compelling evidence of Lagrangian spontaneous
stochasticity from laboratory experiments \citep{bourgoin2006role,tan2022universality}.  On the other hand, 
numerical simulations of homogeneous, isotropic turbulence in a periodic domain provide reasonable 
evidence for the ``forgetting'' of initial separations of deterministic Lagrangian particles
\citep{bitane2013geometry,buaria2015characteristics} and ``forgetting'' of both initial separations and 
molecular diffusivity of stochastic Lagrangian trajectories \citep{eyink2011stochastic,buaria2016lagrangian}. 
Setting aside the important issue of empirical evidence, another crucial question is what implications 
Lagrangian spontaneous stochasticity might have for incompressible Navier-Stokes turbulence. 
A Feynman-Kac representation of Navier-Stokes solutions was derived by \cite{constantin2008stochastic} which 
is similar to \eqref{cFKrep} for the advection-diffusion equation but more non-trivial, because of the nonlinearity
of the equations. It was pointed out by \cite{eyink2010stochastic} that the Constantin-Iyer representation 
corresponds to a stochastic principle of least-action, thus making connection with Hamiltonian fluid mechanics. 
In particular, the stochastic action functional for incompressible Navier-Stokes is invariant under the infinite-dimensional
symmetry group of particle-relabelling just as is the action for deterministic Euler. The remarkable properties
of vorticity under Euler dynamics that follow from that symmetry therefore carry over to Navier-Stokes in a stochastic 
fashion. For example, the Kelvin Theorem of conservation of circulation on an arbitrary Lagrangian loop for 
Euler holds also for Navier-Stokes in the sense that the circulation on stochastically advected loops is a 
backward-in-time martingale \citep{constantin2008stochastic,eyink2010stochastic}. This property prescribes 
the ``arrow-of-time'' of the irreversible Navier-Stokes dynamics and it is plausible that a similar property
carries over to dissipative weak Euler solutions obtained in the inviscid limit \citep{eyink2006turbulent}.  
The dissipative anomaly for Navier-Stokes turbulence may be characterized also in terms of the time-asymmetry
of separation of stochastic Lagrangian particle \citep{drivas2019turbulent,cheminet2022eulerian}, which in 3D 
separate faster backward in time than they do forward in time \citep{eyink2011stochastic,buaria2016lagrangian}. 


\noindent 
{\it (c) Finite Reynolds numbers}. A frequently-made criticism of Onsager's ``ideal turbulence'' theory is that it 
applies only to the unrealistic limit $Re=\infty$ and is thus inapplicable to real-world turbulence which is, 
of course, always at a finite Reynolds number. As an example, I  may quote from the monograph of 
\cite{tsinober2009informal}, \S 10.3.2: ``... it is not clear why results for finite $Re$ (i.e., for NSE having no 
singularities or extremely `intermittent' ones) are relevant for the limit (if such exists) $Re\to\infty$ 
(e.g., for Euler equation with space-filling singularities)''. In some respects, 
such criticisms are a simple misunderstanding of the elementary concept of limit. Indeed, the predictions of a 
theory for $Re\to\infty$ are experimentally falsifiable, as they must be valid to arbitrary accuracy by taking 
Reynolds numbers larger (but finite). There is, however, a legitimate question regarding any such predictions of an 
``ultimate regime''  concerning how large $Re$ must be chosen to observe the predictions of the theory. 
This is especially important for Onsager's theory since, as has often been observed, the strict requirement for 
its validity is that the number of cascade steps must be large, that is, $\log_2 Re\gg 1$ and this condition is hard to 
satisfy in even the highest Reynolds-number terrestrial turbulent flows. There are two responses to this very important question. 

First, there are explicit error estimates in the mathematics of the Onsager theory which provide bounds 
on the correction terms at large but finite $Re.$ For example, in the coarse-grained energy balance 
equation \eqref{II15} there is at finite Reynolds number a viscous dissipation term $\nu |\grad\overline{\bu}_\ell|^2$
in addition to the inertial flux term $\Pi_\ell.$ It is not difficult to derive bounds of the form
\be  \nu |\grad\overline{\bu}_\ell|^2 = O(\nu \delta u^2(\ell)/\ell^2) \ee 
locally in space-time with $\delta u(\ell)=\sup_{|\br|<\ell} |\delta\bu(\br)|,$ or similar bounds in terms of Besov 
norms for space-integrated dissipation. These estimates provide concrete upper estimates on the finite-$Re$
corrections, which can explain departures from the $Re=\infty$ theory. 

Second, Onsager's RG-type arguments can also be applied directly to the Navier-Stokes equations 
at large but finite-$Re$ and do not require any assumptions about existence of weak Euler solutions 
in the limit $Re\to\infty.$ For example, \cite{drivas2019onsager} study the balance of the subscale 
kinetic energy $k_\ell:=(1/2){\rm Tr}\,\btau_\ell$ for the forced Navier-Stokes equation, which takes the form 
\be \partial_t k_\ell + \grad\bdot {\bf J}_\ell = -\overline{(\nu|\grad\bu|^2)}_\ell +\nu |\grad\overline{\bu}_\ell|^2 
+\Pi_\ell + \tau_\ell({\bf f};\bu) \ee
where ${\bf J}_\ell$ is a suitable spatial flux of the subscale kinetic energy and $\tau_\ell({\bf f};\bu):=\overline{(\bu\bdot{\bf f})}_\ell
-\overline{\bu}_\ell\bdot\overline{\bf f}_\ell$ is the direct power input by the force into unresolved scales. If one assumes
suitable Besov regularity of the Navier-Stokes solutions, $\bu^\nu\in B^{\sigma,\infty}_p,$ $p\geq 3,$ uniform in the Reynold number,
then one may derive bounds of the form  
\be D=(1/Re)|\widehat{\grad}\widehat{\bu}|^2=O(Re^{\frac{1-3\sigma}{1+\sigma}}). \lb{gen-On49} \ee  
These bounds follow by the same basic principle of independence of the physics on the arbitrary coarse-graining 
scale $\ell,$ which allows one to choose $\ell/L\propto Re^{-1/(1+\sigma)}$ to optimize the estimates. Following  
the same ideas of Onsager's argument, one can then deduce the existence of ``quasi-singularities'' in the Navier-Stokes 
solutions from the observation of energy dissipation vanishing slowly with $Re:$
\be D=(1/Re)|\widehat{\grad}\widehat{\bu}|^2\propto Re^{-\alpha} \lb{weak-anom} \ee  
which will require $\sigma\leq \sigma_{\alpha}:=\frac{1+\alpha}{3-\alpha}.$ Note that when $\alpha=0,$ one recovers 
Onsager's critical value $\sigma=\frac{1}{3},$ but when $\alpha>0$ instead $\sigma_\alpha>\frac{1}{3}.$ 
This is a significant strengthening of the Onsager's original result, because empirical observations alone 
could never allow one to distinguish $\alpha=0$ from a very tiny value of $\alpha.$ Nevertheless, Onsager's 
conclusions about ``quasi-singularities'' are robust, remaining valid even under the weaker condition \eqref{weak-anom}.

The condition \eqref{weak-anom} on the dimensionless dissipation rate for $\alpha>0$ has been evocatively termed a {\it weak dissipative anomaly} 
by \cite{bedrossian2019sufficient}. Note that whenever $\alpha<1$ in \eqref{weak-anom}, then 
$|\widehat{\grad}\widehat{\bu}|^2\propto Re^{1-\alpha}\to\infty$ as $Re\to\infty$  
and thus classical solutions of the PDE's can no longer exist in the limit of infinite-$Re$. Although no assumption 
about existence of weak solutions is necessary to derive the bound \eqref{gen-On49}, such weak  Euler solutions 
nevertheless emerge under the natural assumption of some Besov regularity with $\sigma<\sigma_\alpha$
that is uniform in $Re$ \citep{drivas2019onsager}. Note furthermore that \cite{bedrossian2019sufficient} have 
proved that the Kolmorogov 4/5th-law remains valid under the assumption of a weak dissipative anomaly only, 
unlike the original derivation of \cite{kolmogorov1941dissipation} which assumed a {\it strong dissipative anomaly} ($\alpha=0$). 
It is easy to check that the Taylor microscale defined by $\lambda^2=15\nu u^{\prime 2}/\varepsilon$
satisfies $\lim_{Re\to\infty} \lambda/L=0$ precisely when $\alpha<1$ in \eqref{weak-anom} and then it is proved 
in \cite{bedrossian2019sufficient} that the maximum difference between $\frac{\langle \delta u_L^3(r)\rangle}{r}$ and $-\frac{4}{5}\varepsilon^\nu$ 
becomes vanishingly small over an increasing range of length-scales $\lambda\ll r \ll L.$ This important result shows that 
validity of the 4/5th law cannot be taken as evidence for a strong anomaly. The proof in \cite{bedrossian2019sufficient} 
considers the statistical steady-state of turbulence in a periodic domain driven by a body force which is a Gaussian random field,
delta-correlated in time and takes advantage of simplifications of that stochastic forcing. More recently, analogous 
results have been derived by \cite{novack2023scaling} for incompressible Navier-Stokes equations with a deterministic 
forcing and valid for individual flow realizations. 

These results have gained some added significance by a recent surprising observation that the scaling 
exponent $\zeta_3$ of absolute 3rd-order structure functions defined by \eqref{Besov} or \eqref{zetap} 
for $p=3$ in fact satisfies $\zeta_3>1$ in some high-$Re$ numerical simulations of forced turbulence in a 
periodic domain \citep{iyer2024zeroth}. This observation is, of course, inconsistent with the bound
$\zeta_3\leq 1$ derived by \cite{constantin1994onsager} under the assumption of a strong dissipative
anomaly. \cite{iyer2024zeroth} appear to have found instead that the normalized dissipation rate
$D$ very slowly decays at high $Re$ according to the bound \eqref{gen-On49} derived by \cite{drivas2019onsager}.
Therefore, only a weak dissipative anomaly appears to occur in this particular simulation of forced homogeneous 
and isotropic turbulence in a periodic box. Consistent with the results of \cite{bedrossian2019sufficient} 
and \cite{novack2023scaling}, however, the Kolmogorov 4/5th law is still observed to hold in this simulation, in agreement with earlier observations 
of \cite{iyer2020scaling}. These observations need to be confirmed with other forcing schemes and extended 
to even higher Reynolds numbers, but they raise additional doubts about the simplification of forced turbulence 
in a periodic box as a valid paradigm for incompressible fluid turbulence more generally. 

\vspace{10pt} 
\noindent 
{\it (d) Extensions.} One of the great virtues of Onsager's ``ideal turbulence'' theory is that it extends readily to turbulent 
flows in other physical systems than incompressible fluids. For example, consequences have been worked out for turbulence in 
compressible fluids, both barotropic \citep{feireisl2017regularity} and with a general thermodynamic equation of state  
\citep{eyink2018cascades,drivas2018onsager}, in quantum superfluids \citep{tanogami2021theoretical}, 
in relativistic fluids \citep{eyink2018cascadesR}, in magnetohydrodynamics, both incompressible 
\citep{caflisch1997remarks,galtier2018origin,faraco2022rigorous} and compressible (\cite{lazarian20203d}, \S IV.B),    
and in kinetic plasmas at low collisionality \citep{eyink2015turbulent,eyink2018cascades,bardos2020onsager}. In many of these examples,
the dimensional and statistical reasoning of \cite{kolmogorov1941local,kolmogorov1941dissipation} does not obviously apply 
and the Onsager theory provides the only first-principles formulation. Almost all of the just-listed examples have 
applications in astrophysics and space science and, in particular, most astrophysical fluids and plasmas are 
compressible. Astrophysics provides, in many ways, the optimal arena for Onsager's theory. For one thing, 
the Reynolds numbers in astrophysics are often much larger than the highest Reynolds numbers 
attainable in terrestrial turbulence. In addition, problems such as the physical origin of fluid singularities are much 
easier to address in compressible fluids. A well-advanced mathematical theory already exists for generic development
of shocks in smooth solutions of compressible Euler equations and the consequent production of vorticity  
\citep{buckmaster2023shock}. Shocks and baroclinic generation of vorticity are expected to be a common 
source of astrophysical turbulence, e.g. supernovae-driven shocks in the interstellar medium \citep{gatto2015modelling}.  

Other recent results on the $Re\to \infty$ limit involve incompressible fluids, but fully 
incorporating the usually neglected effect of thermal fluctuations.  Starting with the fluctuating hydrodynamic 
equations \eqref{LLNS} of  \cite{landau1959fluid} for a low-Mach incompressible fluid (see also \cite{forster1977large,donev2014low}),
\cite{eyink2024emergence} have studied the limiting dynamics in the inertial range. The earlier work of  
\cite{bandak2024spontaneous}, which shall be discussed more below, already considered the standard large-scale 
non-dimensionalization of the equations via $\hat{\bu}=\bu/u',$ $\hat{\bx}=\bx/L,$ $\hat{t}=u't/L,$ which takes the form 
\begin{equation}
\partial_{\hat{t}}\hat{\bu} + (\hat{\bu}\cdot\hat{\grad})\hat{\bu} = -\hat{\grad}\hat{p} + \frac{1}{Re}\hat{\Delta}\hat{\bu} 
+ \sqrt{\frac{2\theta_\eta}{Re^{15/4}}} \hat{\grad}\cdot \hat{\boxi} + \digamma \hat{{\mathbf f}}. \lb{FNS3} 
\end{equation}
where $\theta_\eta=k_BT/\rho u_\eta^2\eta^3$ is the thermal energy relative to the energy of an 
eddy of Kolmogorov scale $\eta$ and with Kolmogorov velocity $u_\eta=(\nu\varepsilon)^{1/4},$
a dimensionless quantity generally of order $10^{-6}$ or even smaller.  In deriving \eqref{FNS3}, 
\cite{bandak2024spontaneous} assumed the Taylor relation $\varepsilon\sim u^{\prime 3}/L$
and an external body force with dimensionless magnitude $\digamma=f_{rms}L/u^{\prime 2}$ 
has also been included to generate turbulence. It is na\"{\i}vely clear that, on inertial-range scales, the 
direct effect of the thermal noise term must be negligible, even smaller than the viscous diffusion. 
In fact,  the noise term leads to a mathematically ill-defined dynamics, unless 
the stochastic noise field $\tilde{\boxi}$ and velocity field $\tilde{\bu}$ are both cut off at some high wave-number 
$\Lambda,$ or $\hat{\Lambda}=\Lambda L$ after non-dimensionalization. In the statistical physics 
literature, such a cut-off is standard and represents physically a coarse-graining length $\ell=\Lambda^{-1}$
of the measured velocity field, which is chosen somewhere between $\eta$ and the molecular mean-free 
path length $\ell_{mfp}$. In the limit $Re\to\infty$ and $\hat{\Lambda}\to \infty,$ 
\cite{eyink2024emergence} prove under natural conditions that are observed in experiment 
(see section \ref{sec:InfRe} below) that the limiting velocity fields are weak solutions of incompressible Euler, 
as assumed in the \cite{onsager1949statistical} theory. 

It is important to stress that Onsager's ``ideal turbulence'' description does not apply in the dissipation 
range and, indeed, Onsager never discussed dissipation-range turbulence in any of his published or 
unpublished writings, to my knowledge. This is perhaps one reason that he never discussed the possible 
interactions of thermal fluctuations and turbulence. This issue was raised about a decade later, however,  
by \cite{betchov1957fine,betchov1961thermal} who pointed out that the kinetic energy spectrum of thermal fluid 
fluctuations 
\be E(k) \sim \frac{k_BT}{\rho} \frac{4\pi k^2}{(2\pi)^3} \lb{thermal} \ee 
must surpass the spectrum of turbulent fluctuations at a wavenumber of order $k_\eta\sim 1/\eta.$ This same idea 
was rediscovered much later by \cite{bandak2022dissipation} who provided numerical evidence for the conjecture 
in a shell model, before full verification by \cite{bell2022thermal} in a numerical simulation of the Landau-Lischitz 
fluctuating hydrodynamic equations \eqref{LLNS}. Confirmation by laboratory experiment is crucial but appears 
much more difficult at this time. In principle, all turbulent processes at sub-Kolmogorov scales should be 
strongly affected by thermal noise, including droplet formation, combustion, biolocomotion, etc. but it is unclear 
whether such probes of the small-scale physics can show the clear signature of thermal noise at large scales. 
A prototypical example is high Schmidt-number turbulent advection, where the classical theories of   
\cite{batchelor1959small} and \cite{kraichnan1968small} that ignore thermal fluctuations miss power-law 
scaling of the concentration spectrum below the Batchelor dissipation scale \citep{eyink2022high}. 
On the other hand, the classical prediction of a $k^{-1}$ spectrum in the viscous-convective range 
remains intact, because the strong effect of thermal hydrodynamic fluctuations in renormalizing the 
molecular diffusivity in that range is exactly the same as in laminar flows \citep{onsager1945theories} and thus 
hidden phenomenologically. There are effects of thermal noise even in the turbulent inertial range 
\citep{bandak2024spontaneous}, but these are much more subtle and will be discussed further below.  

\vspace{10pt} 
\noindent 
{\it (e) Physicality of Weak Solutions.} A frequent objection made by physicists and fluid mechanicians 
to the ideal turbulence theory is that the mathematical concept of ``weak solution'' is unphysical. 
In fact, such criticisms are directed not only at Onsager's proposed Euler solutions, but also against the 
weak solutions of incompressible Navier-Stokes equation constructed by \cite{leray1934mouvement}. 
To quote again from \cite{tsinober2009informal}, \S 10.3.1, regarding Leray's work:  ``An important point is 
that if one looks at real turbulence at finite Reynolds numbers (however large) there seems to be no need 
for weak solutions at all.'' Here the presumption is that smooth, strong solutions of Navier-Stokes must exist, 
as there is no obvious empirical evidence for the severe singularities required to overcome the viscous 
regularization. This issue of the physical meaning of weak solutions is quite important, since it is at the center 
of the empirical testability of Onsager's theory. Thus, I  must discuss the matter briefly here. Furthermore,
my own views may not coincide with those of most mathematicians who work on Onsager's theory,
and these differences must be carefully explained. The issues are subtle and intimately related with the famous 
Sixth Problem of \cite{hilbert1900mathematische}, which was ``the problem of developing mathematically the 
limiting processes ... which lead from the atomistic view to the laws of motion of continua'' and which remains 
still largely unresolved. 

As I  have already emphasized, ``weak solutions'' are mathematically equivalent to ``coarse-grained solutions'' 
such as \eqref{II9} for Navier-Stokes and \eqref{wEuler} for Euler, when those equations are imposed 
for all $\ell>0$ \citep{drivas2018onsager}. In reality, the ``coarse-grained solution'' of Euler equations \eqref{wEuler} 
is an accurate physical description of a turbulent flow only in the inertial range of scales for $\ell\gg \eta.$ The most 
important fact about the formulation in \eqref{wEuler} of ``weak Euler solutions'', or its equivalent \eqref{FEuler}, is 
that it involves only the coarse-grained fields $\overline{\bu}_\ell$ and $\overline{p}_\ell$ for $\ell\gg \eta,$
since the subscale stress $\btau_\ell(\bu,\bu)$ depends as well only upon $\overline{\bu}_{\ell'}$ with $\ell'\lesssim\ell$
by the fundamental property of scale-locality. To state my view succinctly, it is the coarse-grained fields 
such as $\overline{\bu}_\ell$ and $\overline{p}_\ell$ which are physical, because they correspond to what is 
experimentally measurable. In fact, every experiment has some spatial resolution $\ell,$ such that only averaged 
properties for length-scales $>\ell$ are obtained. It is instead the fine-grained/bare fields such as $\bu(\bx)$ which are unphysical, 
because they are unobservable objects corresponding to a mathematical idealization $\overline{\bu}_\ell\to\bu$ as $\ell\to 0,$ 
which goes beyond the validity of a hydrodynamic description and is physically unachievable. 

The above views are probably not the same as those held currently by the majority of mathematicians and fluid mechanicians who work 
on turbulence, who mostly have followed \cite{vonneumann1949recent} in believing that  
\begin{quotation} \noindent 
``$\,$`Turbulent' and `molecular' disorder 
are in general distinct. A macroscopic and yet turbulent $\bu$ (and, more generally, a complete system of system of fluid 
dynamics with such characteristics) can be defined.'' --- \cite{vonneumann1949recent}, \S 3.2, p.446. 
\end{quotation} 
According to this common view,  the deterministic Navier-Stokes equations 
are assumed valid for scales $\ell\gtrsim \lambda_{mfp},$ the molecular mean free path length, with $\overline{\bu}_\ell\approx \bu,$
the Navier-Stokes solution, for all $\ell$ in the range $\eta\gtrsim \ell\gtrsim\lambda_{mfp}.$ There is, however, no rigorous 
mathematical theory that justifies this assumption and there are no experimental measurements available at scales $\ell<\eta$
to confirm it.  A major problem with this view is that it omits the physical effects of thermal fluctuations, which leads to deviations 
from the predictions of deterministic Navier-Stokes at scales $\ell\gg \lambda_{mfp}$ that are experimentally well-observed 
in laminar flows \citep{dezarate2006hydrodynamic}. In fact,  deterministic Navier-Stokes is a physically inconsistent set of equations, 
because it incorporates molecular dissipation without including the corresponding molecular fluctuations. It thus violates 
the fundamental fluctuation-dissipation theorem of statistical physics \citep{onsager1953fluctuations}. I  therefore 
do not regard solutions of deterministic Navier-Stokes, either weak or strong, as having fundamental physical significance. 
Deterministic Navier-Stokes is instead just a reasonably accurate  ``large-eddy simulation'' model of turbulence for 
resolved scales down to $\ell\sim \eta$. 

A more fundamental model of molecular fluids are the fluctuating hydrodynamic equations of \cite{landau1959fluid} 
which incorporate the fluctuation-dissipation relation and which explain the experiments on thermal fluctuations 
in laminar flows \citep{dezarate2006hydrodynamic}. It was first pointed out by \cite{betchov1961thermal} that 
fluctuating hydrodynamics is an appropriate model of the turbulent dissipation range and, in fact, the incompressible,
low-Mach equations \eqref{LLNS} were proposed in his work independently of  \cite{landau1959fluid}. However,
as observed in the previous subsection, fluctuating hydrodynamics is not a continuum theory but instead involves
an explicit cutoff $\Lambda=\ell^{-1}$ and it is believed to describe the evolution of the coarse-grained fields $\overline{\bu}_\ell$ 
for all scales $\ell\gtrsim \lambda_{mfp}$ (or, in fact, for the incompressible version  \eqref{LLNS}, only at 
scales where $|\overline{\bu}_\ell|\lesssim c_s,$ the sound speed \citep{donev2014low}). Furthermore, the viscosity 
$\nu_\ell$ in the model depends upon the scale $\ell,$ as shown, for example, by the renormalization group analysis of 
\cite{forster1977large}. Thus, even in laminar flows, the observed viscosity $\nu_\ell$ at length scale $\ell$ is an 
``eddy-viscosity'', although the fluctuating eddies arise there from thermal fluctuations rather than turbulent fluctuations. 
In the language of modern physics, the Landau-Lifschitz fluctuating hydrodynamics equations 
are low-wavenumber ``effective field theories" \citep{schwenk2012renormalization,liu2018lectures}. 

An important consequence of these considerations is that velocity-gradients are $\ell$-dependent 
at {\it all} length-scales $\ell$ in a turbulent flow and not only for $\ell>\eta.$ However, it should be true 
to a good approximation that, for typical thermal fluctuations, 
\be \grad\overline{\bu}_\ell\sim \grad\bu+O\left(\left(\frac{k_BT}{\rho \ell^5}\right)^{1/2}\right), \quad \ell\lesssim \eta, \ee 
where $\bu$ is again the solution of deterministic Navier-Stokes. It then easily follows that 
\be \nu_\ell\langle |\grad\overline{\bu}_\ell|^2\rangle \simeq \nu \langle |\grad\bu|^2\rangle, \quad 
\nu=\nu_\eta \ee 
where $\langle \cdot\rangle$ denotes a space-time average and $\ell/\eta$ spans about 2-3 decades of scales. 
For more details, see \cite{eyink2007turbulenceI}, section II.E.  
In this sense, typical thermal fluctuations should matter little for mean energy dissipation. On the other 
hand, this statement is presumably not true for large, rare thermal fluctuations. Indeed, in a simple particle model
of hydrodynamics, the ``totally asymmetric exclusion process'', dissipative weak solutions of inviscid Burgers 
equation arise with overwhelming probability (law of large numbers) but rare fluctuations (large deviations) 
can lead to weak solutions with local energy production in some space-time regions rather than dissipation 
everywhere \citep{jensen2000large,varadhan2004large}. Furthermore, higher-order gradients $\grad^k\overline{\bu}_\ell$
with $k>1$ will be much more strongly affected by thermal noise and, indeed, the numerical simulation of \cite{bell2022thermal} 
found that such higher-order gradients are completely dominated by thermal noise already for $\ell\lesssim \eta.$
  
I therefore fundamentally disagree with a common view, clearly expressed by \cite{tsinober2009informal}, 
that hypothetical fine-grained fields $\bu$ and their gradients are physically more ``objective'' than coarse-grained fields 
$\overline{\bu}_\ell$: 
\begin{quotation}\noindent 
``There is a generic ambiguity in defining the meaning of the term {\it small scales} (or more generally scales) 
and consequently the meaning of the term {\it cascade} in turbulence research. As mentioned in chapter 5, 
the specific meaning of this term and associated inter-scale energy exchange/‘cascade’ (e.g., spectral energy 
transfer) is essentially decomposition/representation dependent. Perhaps, the only common element in all 
decompositions/ representations (D/R) is that the small scales are associated with the field of velocity derivatives. 
Therefore, it is natural to look at this field as the one {\it objectively} (i.e., D/R independent) representing the small 
scales. Indeed, the dissipation is associated precisely with the strain field, $s_{ij}$, both in Newtonian and 
non-Newtonian fluids.'' --- \cite{tsinober2009informal}, \S 6.2, p.127. 
\end{quotation} 
This point of view is based on the belief, without evidence, that there is some hypothetical fine-grained field $\bu
=\lim_{\ell\to 0} \overline{\bu}_\ell$ and that  objectivity is achieved by ``convergence'' as $\ell\to 0.$
The existence of such a fine-grained velocity in turbulent flow was probably first questioned by the 
visionary scientist, Lewis Fry Richardson, whose famous paper on turbulent pair-dispersion contained a section 
entitled ``{\it Does the Wind possess a Velocity?}'' He wrote there
\begin{quotation}\noindent 
``This question, at first sight foolish, improves on acquaintance. A velocity is defined, for example, in Lamb's 
`Dynamics'  to this effect : Let $\Delta x$ be the distance in the $x$ direction passed over in a time $\Delta t,$ 
then the $x$-component of velocity is the limit of $\Delta x/\Delta t$ as $\Delta t\to 0$. But for an air particle 
it is not obvious that $\Delta x/\Delta t$ attains a limit as $\Delta t\to 0.$ --- \cite{richardson1926atmospheric}, \S 1.2, p.709. 
\end{quotation} 
Indeed, each individual air molecule is moving at about 346 m/sec, the speed of sound in that fluid, with the mean wind 
velocity a small correction, and the local measured velocity will depend entirely upon the resolution scale $\ell$ which is adopted.  

I certainly agree that science should be concerned with objective facts. However, 
in contrast to the 19th-century continuum mechanics perspective that objectivity is achieved by 
convergence, the modern tool to achieve objectivity is the renormalization group. I  may cite 
here the physics Nobel laureate Kenneth G. Wilson: 
\begin{quotation}\noindent 
``A procedure is now being developed to understand the statistical continuum limit. The procedure is called
the renormalization group. It is the tool that one uses to study the statistical continuum limit in the same 
way that the derivative is the basic procedure for studying the ordinary continuum limit.'' --- 
\cite{wilson1975renormalization}, p.774. 
\end{quotation} 
The fundamental understanding that arose in physics was that for systems with strong fluctuations at all length
scales --- a class which includes both quantum field theories and turbulent flows --- the effective description varies with 
the length scale $\ell$ of resolution. In that case, the proper goal is not to seek for some idealized and unobservable  
``truth'' at $\ell\to 0$ but instead to attain objectivity by understanding how the description changes as $\ell$ is varied. 
This application of this approach to hydrodynamics has been mostly been restricted to flows much simpler 
than turbulence, c.f. \cite{forster1977large}, where weak-coupling perturbation expansions are applicable,
and only recently has some progress been made on non-perturbative application to turbulence by functional 
renormalization group methods \citep{canet2022functional}.   
It is conceptually important, however, to understand that LES models resolved at length scale $\ell$ are not really 
continuum models. Indeed, it is a truism in the LES community that the true model is not the ``continuum'' LES model  
but instead the discretization of the model used to solve it numerically. It therefore makes sense to combine the steps 
of coarse-graining and numerical discretization, as in some approaches to modeling thermal fluctuations
\citep{espanol2009microscopic}. I  shall return to this theme later in discussing the relation of Onsager's 
``ideal turbulence'' theory with LES modeling.  

\subsection{Onsager's Conjecture}\lb{sec:ConjIntro}  

The quote from \cite{onsager1949statistical} on which I  base this essay began with a provocative suggestion that 
``turbulent dissipation as described could take place just as readily without the final assistance by viscosity.'' 
It is important to emphasize that none of the mathematical results that I  have reviewed so far, including those of Onsager himself, 
establish this hypothesis. Such arguments give only necessary conditions for anomalous dissipation, but not sufficient conditions.
An example of an instantaneous velocity field in $C^h$ was found by \cite{eyink1994energy} showing that these proofs cannot 
be extended to $h\leq 1/3,$ but such examples also fall far short of proving Onsager's conjecture, as already stated there: 
\begin{quotation}
\noindent 
``It must not, of course, be concluded that, simply because our argument fails when $h\leq 1/3$, that non-conservation 
is actually possible for $h\leq 1/3$. We emphasize that to demonstrate this it is necessary to construct an appropriate 
solution ${\bf v}(.,.)$ with ${\bf v}(.,t)\in C^h$, $0< h < 1/3$ for $t\in [0,T]$, for which the energy indeed decreases or 
increases in the interval.'' --- \cite{eyink1994energy}, p.235. 
\end{quotation}
There are several elements involved in Onsager's conjecture which are highly non-trivial, e.g. that weak Euler solutions exist in the 
limit $Re\to \infty$ and that dissipation remains positive in that limit. As to the first matter, I  note that inviscid 
limits of Navier-Stokes solutions can be proved to exist without any further assumptions, at least for flows in infinite Euclidean 
space or in a periodic domain. However, the limiting Euler solutions so obtained are much weaker even than those postulated 
by Onsager, consisting of ``measure-valued solutions'' with a distribution $p_{\bx,t}(d\bu)$ of possible velocity values
at each space-time point $(\bx,t)$ \citep{diperna1987oscillations,wiedemann2018weak} or satisfying cumbersome 
energy inequalities \citep{lions1996mathematical}.  In addition, such generalized weak Euler solutions are proved to exist 
only for subsequences of viscosity $\nu_k\to 0$ and may differ from subsequence to subsequence.

Recently, however, considerable progress has been made on these problems, in particular a rigorous proof  has been found 
that dissipative Euler solutions as conjectured by  \cite{onsager1949statistical} do exist with $\bu\in C^{1/3-\epsilon}$
for any $\epsilon>0$ \citep{isett2018proof,buckmaster2018onsager}. Furthermore, advances have been made on deriving 
such results from the incompressible Navier-Stokes equations in the limit $Re\to\infty$ \citep{brue2023anomalous} and 
even more progress has been made on related problems in passive scalar turbulence \citep{bedrossian2022batchelor}. 
In this section I  shall attempt to give a broad overview of these developments as appropriate to this essay, 
but certainly no comprehensive review, which would be difficult in any case for such a rapidly moving and growing literature. 
See \cite{delellis2019turbulence,buckmaster2020convex} for some recent reviews by lead researchers in this area.  

It must be emphasized at the outset that, as spectacular as these recent mathematical developments might be, they 
cannot in principle justify the correctness of Onsager's ``ideal turbulence'' as a physical theory. In fact, the final 
judge of the truth of any physical theory is experiment, not mathematics. I  whole-heartedly agree with the famous dictum 
of Einstein: 
\begin{quotation} \noindent 
``Insofern sich die S\"atze der Mathematik auf die Wirklichkeit beziehen, sind sie nicht
sicher, und insofern sie sicher sind, beziehen sie sich nicht auf die Wirklichkeit. [Insofar as the propositions 
of mathematics refer to reality, they are not certain, and insofar as they are certain, they do not refer to reality.]'' --\cite{einstein1921geometrie}, p.3. 
\end{quotation} 
Even if Onsager's conjectures could be rigorously derived from a fundamental model such as the  Landau-Lifschitz fluctuating 
hydrodynamic equations \eqref{LLNS}, serious questions arise whether those equations themselves might break down locally in spacetime 
in the vicinity of extreme intermittent events \citep{bandak2022dissipation}. Instead, the novel predictions of Onsager's theory must 
be carefully compared with experiments to test their validity. In fact, we know already from the brief review of experimental 
observations in section \ref{empiric} that Onsager's $1/3$ H\"older result cannot be the whole story, because it makes no reference 
to solid walls and, in particular, it does not elucidate the difference in the wall geometries where anomalous dissipation occurs  
and where it does not. I  discuss this crucial issue more in section \ref{walls}.  

\subsubsection{Existence of Dissipative Euler Solutions}\lb{sec:exist} 

The program which has successfully led to a mathematical construction of the weak Euler solutions conjectured 
by \cite{onsager1949statistical} in breakthrough work by \cite{isett2018proof} was initiated 
\cite{delellis2009euler} and \cite{delellis2010admissibility}, which both appeared as preprints already in 2007. 
A very good review of this early work is given by \cite{delellis2013continuous} 
and an updated review is to be found in \cite{delellis2019turbulence}. These authors pointed out a remarkable connection of 
Onsager's conjecture with famous work of the mathematician John \cite{nash1954c1} on $C^1$ isometric embeddings. 
I give here a very succinct review, following closely the discussions in the previous references.

Nash (1954) addressed a classical problem of differential geometry, whether a smooth
manifold $M$ of dimension $n\geq 2$ with Riemannian metric $g$ may be isometrically
imbedded in $m$-dimensional Euclidean space ${\mathbb R}^m$, i.e. whether a $C^1$ embedding 
map ${\bf u}: M \to {\mathbb R}^m$ exists so that the Riemannian metric induced by the embedding 
agrees with $g,$ or 
\be \partial_i{\bf u}\cdot\partial_j{\bf u}= g_{ij}. \lb{isometry} \ee 
To answer this question, Nash considered a more general problem of {\it short embeddings} which do not 
preserve lengths of curves on $M$ but can only {\it decrease} lengths, so that 
\be \partial_i  \overline{{\bf u}}\cdot\partial_j\overline{{\bf u}}\leq g_{ij} \lb{short} \ee
in the matrix sense. The startling result obtained by Nash, with some improvement due to \cite{kuiper1955c1},
is the following:

\begin{quotation}
\noindent
\underline{Nash-Kuiper Theorem}: {\it Let $(M, g)$ be a smooth closed $n$-dimensional Riemannian manifold, 
and let $\overline{{\bf u}}:M\to {\mathbb R}^m$  be a $C^\infty$ strictly short embedding with $m\geq n+1$. For any 
$\epsilon>0$ there exists a $C^1$ isometric embedding ${\bf u}: M\to  {\mathbb R}^m$ 
with $|{\bf u}-\overline{{\bf u}}|_{C^0}<\epsilon.$}  
\end{quotation} 

\noindent 
This result is surprising for two reasons. First, the condition \eqref{isometry} is a set of $n(n+1)/2$ equations in $m$ unknowns.
A reasonable guess would be that the system is solvable, at least locally, when $m\geq  n(n+1)/2$ 
and this indeed was a classical conjecture of Schl\"afli. However, for $n \geq  3$ and $m = n + 1$, 
the system \eqref{isometry} is hugely overdetermined! It is not obvious that there should be any solutions at all, but the 
Nash-Kuiper Theorem shows that there exists an enormous set of $C^1$-solutions that are $C^0$-dense in the set of 
short embeddings. I  shall not discuss here the details of Nash's proof of his astonishing result, but just remark that his construction of the isometry ${\bf u}$
was in a series of stages, by adding at each stage a new small, high-frequency perturbation. The construction is not an 
``abstract nonsense'' argument that requires the Axiom of Choice or some other non-constructive element, 
and, in fact, 
the entire procedure can be implemented to any finite stage of iteration, in principle, by a computer algorithm.
For example, see the work of \cite{borrelli2012flat} which calculates numerically a $C^1$-isometric embedding of the flat 
2-torus ${\mathbb T}^2$ into ${\mathbb R}^3$ and Fig.~\ref{fig_nash} for the first four stages of iteration. It interesting 
to note that Nash himself regarded his 1954 paper as a ``sidetrack'' on the problem of isometric embeddings  
\citep{raussen2016interview} and attached greater importance to his later work that constructed $C^\infty$ embeddings
in higher dimensions \citep{nash1956imbedding}. The latter paper is notable for its introduction of the Nash-Moser 
implicit function theorem and also for its exploitation of a smooth mollification scheme with a continuous parameter $\ell,$
involving a careful study of variations with $\ell$ similar to an RG approach (L. Sz\'ekelyhidi, Jr., private communication). 
A very intriguing recent stochastic formulation of \cite{menon2021information,inauen2023stochastic} makes this analogy more clear. 

{\begin{figure}
\centering
\includegraphics[width=.5\textwidth]{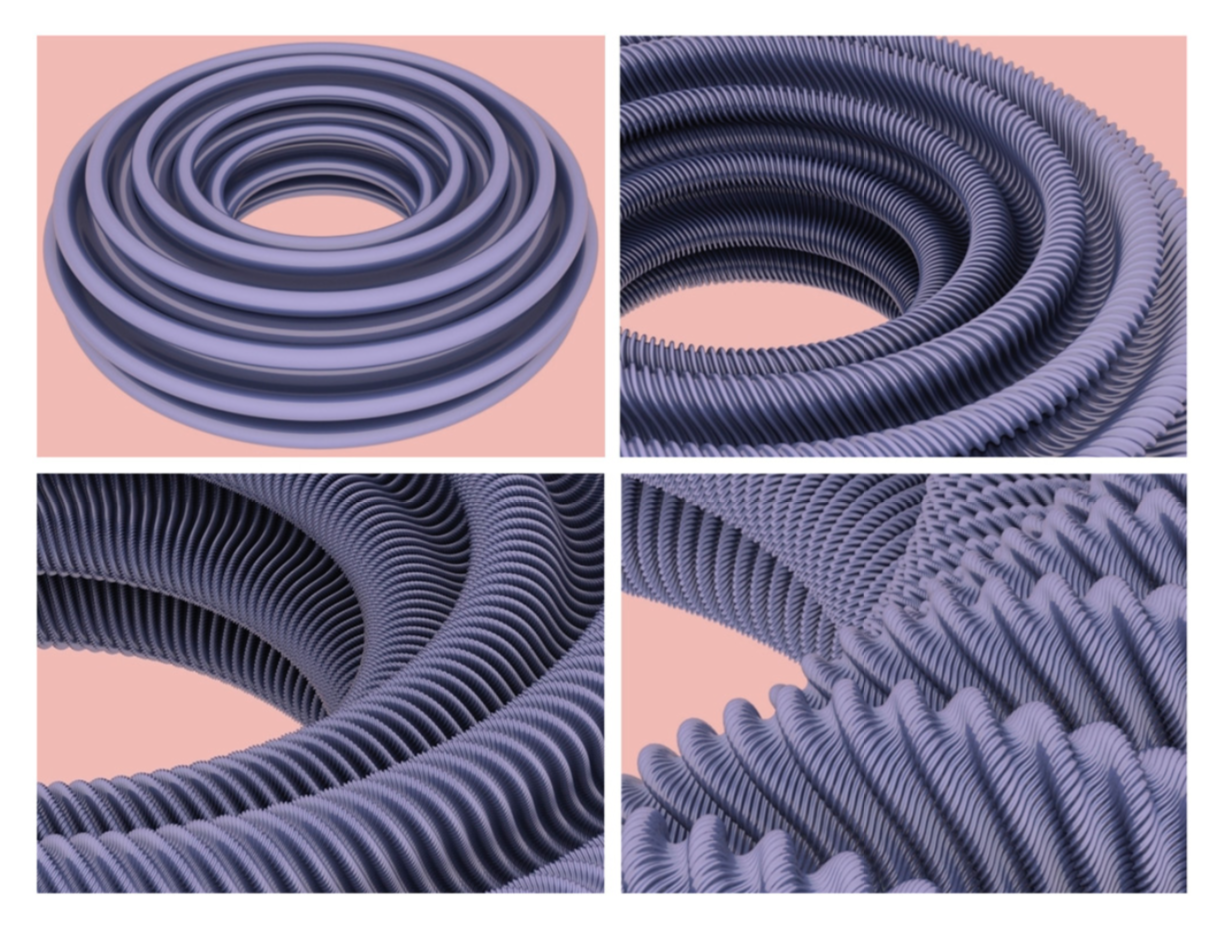}
\caption{The first four stages in the iterative construction of a $C^1$-isometric embedding of the 
flat 2-torus ${\mathbb T}^2$ into ${\mathbb R}^3.$  The initial map is corrugated along the meridians 
to increase their length. Corrugations are then applied repeatedly in various directions to produce a sequence of maps.
Each successive map is strictly short, with reduced isometric default. Reproduced from V. Borrelli et al.,
``Flat tori in three-dimensional space and convex integration,'' Proc. Natl. Acad. Sci. {\bf 109} 7218--7223 (2012), 
with permission of PNAS.}
\label{fig_nash}
\end{figure}} 

The fundamental contribution of De Lellis and Sz\'ekelyhidi Jr was to realize that there is a very close 
mathematical analogy between the problem of isometrically embedding a smooth manifold by a 
map of low regularity and the problem of solving the Cauchy initial-value problem for 
incompressible Euler equations by a velocity field of low regularity and that Nash's method 
of construction can be carried over to the latter.  The analog of a ``short mapping'' for the Euler system is what 
De Lellis and Sz\'ekelyhidi Jr call a smooth {\it subsolution}, i.e.  a smooth triple 
$(\overline{\bu}, \overline{p},\overline{\btau})$ with $\overline{\btau}$ a symmetric, positive-definite tensor such that 
$$ \partial_t\overline{\bu}+\grad\cdot(\overline{\bu}\ \overline{\bu}+\overline{\btau})=-\grad\overline{p}, 
\quad \grad\cdot\overline{\bu}=0. $$
Everyone from the turbulence modelling community will recognize at once that this has the form of an LES model equation 
incorporating a positive-definite ``turbulent stress'' tensor $\overline{\btau}.$ Approximation results can be obtained for subsolutions 
entirely analogous to the Nash-Kuiper theorem on approximation of short embeddings. As an example, a theorem of 
\cite{buckmaster2018onsager} states the following:
\begin{quotation}
\noindent 
\underline{Theorem} \citep{buckmaster2018onsager}: {\it Let $(\overline{{\bu}},\overline{p},\overline{\btau})$ be any smooth, strict subsolution of the Euler equations 
on ${\mathbb T}^3 \times [0,T]$ and let $h< 1/3$. Then there exists a sequence $(\bu_k,p_k)$ of weak Euler solutions 
such that $\bu_k\in C^h({\mathbb T}^3 \times [0,T])$ satisfy, as $k\to\infty,$
$$ \int_{{\mathbb T}^3} d^3x\ f \ \bu_k\rightarrow  \int_{{\mathbb T}^3} d^3x\ f \overline{\bu},
\quad  \int_{{\mathbb T}^3} d^3x\ f \ \bu_k\bu_k\rightarrow  \int_{{\mathbb T}^3} d^3x\ f (\overline{\bu}\ \overline{\bu}+\overline{\btau}) $$
for all $f\in L^1({\mathbb T}^3)$ uniformly in time, and furthermore for all $t\in [0, T ]$ and all $k$
$$  \int_{{\mathbb T}^3} d^3x\ \frac{1}{2}|\bu_k|^2 =\int_{{\mathbb T}^3} d^3x\ \frac{1}{2}(|\overline{\bu}|^2+{\rm Tr}\ \overline{\btau}). $$} 
\end{quotation} 
The notion of convergence in this theorem can be made more physically transparent by taking $f(\br)=\widetilde{G}_\delta(\bx+\br)$ 
for $\delta>0,$ in which case the approximation property implies that pointwise 
\be \widetilde{\bu}_{k,\delta}\to\widetilde{\overline{\bu}}_\delta, \quad \widetilde{\tau}_\delta(\bu_k,\bu_k)\to\widetilde{\tau}_\delta(\overline{\bu},\overline{\bu})
+\widetilde{\overline{\btau}}_\delta \ee
Thus, if the subsolution $(\overline{{\bu}},\overline{p},\overline{\btau})$ arose from a coarse-grained Euler solution in the sense of \cite{drivas2018onsager},
so that $\overline{{\bu}}=\overline{{\bu}}_\ell$, $\overline{p}=\overline{p}_\ell$, $\overline{\btau}=\overline{\btau}_\ell(\bu,\bu),$ then the simple 
identity of \cite{germano1992turbulence} for the convolution filter $\widetilde{\overline{G}}_{\ell,\delta}=\overline{G}_\ell*\widetilde{G}_\delta$
\be \widetilde{\overline{\btau}}_{\ell,\delta}(\bu,\bu)= \widetilde{\tau}_\delta(\overline{\bu}_\ell,\overline{\bu}_\ell)
      +  \widetilde{(\overline{\btau}_\ell(\bu,\bu))}_\delta
\ee 
implies that for any $\delta>0$ 
\be  \widetilde{\bu}_{k,\delta}\to\widetilde{\overline{\bu}}_{\ell,\delta}, \quad \widetilde{\tau}_\delta(\bu_k,\bu_k)\to\widetilde{\overline{\tau}}_{\ell,\delta}(\bu,\bu)
\ee 
pointwise in space. When $\delta\ll \ell,$ this yields the remarkable statement that a coarse-grained Euler solution $(\widetilde{\overline{\bu}}_{\ell,\delta},\widetilde{\overline{p}}_{\ell,\delta})
\simeq (\overline{\bu}_{\ell},\overline{p}_{\ell})$ at scale $\ell$ can be well approximated pointwise in space by a sequence $(\widetilde{\bu}_{k,\delta},\widetilde{p}_{k,\delta})$
of coarse-grained Euler solutions at the much smaller scale $\delta.$

The constructions of such weak Euler solutions follow a strategy similar to that of Nash, by a sequence
of stages $\bu_0,$ $\bu_1,$ $\bu_2,$ .... At stage $n$ one has, after coarse-graining, a subsolution  
$$ \partial_t \bu_n +\grad\cdot(\bu_n\bu_n+\btau_n)=-\grad p_n $$
which is supported on wavenumbers $<\Lambda_n.$ By adding a small-scale carefully chosen perturbation 
one can succeed to cancel a large part of the stress $\btau_n$ so that, 
in the limit, $\btau_n\to {\bf 0}$ weakly and one obtains a weak limit $\bu$ which is a distributional Euler 
solution. A number of different models of the small scales have been employed, such as Beltrami flows 
\citep{delellis2013dissipative} and Mikado flows \citep{isett2018proof,buckmaster2018onsager}, together with other operations, 
such as evolving under smooth Euler dynamics locally in time and ``gluing'' the different time-segments.
In the language of LES modelling, the constructions can be regarded as a sort of iterative ``defiltering'',
in which unresolved scales are successively restored. A physicist might prefer to call this an ``inverse renormalization 
group'', the reverse of the successive coarse-graining employed by \cite{wilson1975renormalization}. 
Clearly, there is a huge number of subsolutions, since one may adopt any positive definite tensor $\overline{\btau}$ 
whatsoever. A consequence is therefore results such as the following: 
\begin{quotation}
\noindent 
\underline{Theorem} \citep{buckmaster2018onsager}: {\it Let $e:[0,T]\to {\mathbb R}^+$ be any strictly positive, 
smooth  function. Then for any $0<h<1/3$ there exists a weak Euler solution $\bu\in C^h({\mathbb T}^3 \times [0,T])$ 
such that 
$$  \int_{{\mathbb T}^3} d^3x\ \frac{1}{2}|\bu|^2 =e(t). $$} 
\end{quotation} 
In particular, one may take $e(t)$ to be any function strictly decreasing in time and then the Euler solution 
of the above theorem (globally) dissipates kinetic energy. 

The details of the constructions are complex and rapidly evolving, so that they may be best gleaned 
from the journal articles themselves. I  therefore make here just a few remarks. First, 
in addition to solutions that satisfy physical constraints, one can construct by the same methods solutions 
with obviously unphysical behaviors (e.g. kinetic energy {\it increasing} in time for unforced 3D incompressible flows!)  
Thus, it is clear that the methods suffice to answer some very fundamental PDE questions, but they still miss 
much essential physics. Another important point is that the existence proofs of the weak Euler solutions are 
constructive, exactly as are the Nash constructions of isometric embeddings. \cite{frisch2018mathematical} 
have attempted to compute numerically the weak Euler solutions constructed by Mikado flows 
\citep{isett2018proof,buckmaster2018onsager}. Visualization, however, appears difficult with current constructions 
because of the very tiny scales of the added eddies. These weak Euler solutions can be regarded as 
a sophisticated sort of ``synthetic turbulence'' \citep{juneja1994synthetic}, manufactured spacetime fields 
with many of the same properties as a physical turbulent flow velocity, but in addition solving exactly the 
inviscid equations of motion. Such solutions have been constructed which satisfy additional observed properties 
such as spatial intermittency \citep{novack2023intermittent,derosa2022intermittency}. We might hope 
in coming years to see weak Euler solutions with features in ever closer agreement with physical turbulent flows. 
As in the field of artificial intelligence and the Turing test, it would then be crucial to design tests of increasing sensitivity 
that could distinguish a true turbulent velocity field and such artificially constructed flows. This exercise 
could help to clarify better the essential properties that characterize physical turbulence. 

\subsubsection{Non-Uniqueness for the Initial-Value Problem}\label{sec:nonuniq} 

One very fundamental feature of the \cite{nash1954c1} result which I  have not yet duly emphasized is the 
profligate non-uniqueness of the $C^1$ isometric embeddings produced by his construction. This stands 
in stark contrast to embeddings with higher smoothness $C^n,$ $n\geq 2$ for which the the Gauss-Codazzi 
equations relate the second fundamental form of the embedded surface to the Riemann curvature tensor. 
In fact, it is known that a compact Riemannian surface with positive Gauss curvature 
may be isometrically embedded into ${\mathbb R}^3$ by a $C^2$ map in just one way, up to a rigid-body motion.
Thus it is clear that isometric embeddings have very different qualitative behavior at low and high regularity, 
i.e. $C^1$ versus $C^2$, often referred to as a dichotomy between {\it rigidity} at high regularity versus 
{\it flexibility} at low regularity.  This type of wild non-uniqueness at low regularity is a central aspect of the {\it $h$-principle} 
introduced by mathematician Mikhail \cite{gromov1971topological,gromov1986partial}, with the isometric embedding 
problem as a primary example. The theorem of \cite{buckmaster2018onsager} that I  cited in the previous 
subsection, as emphasized by those authors, in an extension of the $h$-principle to low-regularity weak 
solutions of the Euler equations, which exhibit a similar ``flexibility''. The name {\it convex integration} for the discussed 
techniques to construct these Euler solutions, by the way, goes back also to the work of \cite{gromov1971topological,gromov1986partial}, 
who introduced it as one of several methods to prove the $h$-principle. The origin of the name has to do with 
differential relations of the type $Df\in A,$ which satisfy the $h$-principle when the convex hull of $A$
contains a small neighborhood of the origin. 

Convex integration methods have fundamental implications for non-uniqueness of weak solutions to the Cauchy initial-value 
problem for the Euler equations, as discussed already by \cite{delellis2010admissibility}. The Theorem 1 from that paper
shook many previous expectations: 
\begin{quotation}
\noindent 
\underline{Theorem} \citep{delellis2010admissibility}: {\it Let $d\geq 2$. There exist compactly-supported divergence-free vector fields 
$\bu_0\in L^\infty$ for which there are infinitely many weak Euler solutions with that initial data, satisfying both the global energy 
equality
$$ \int_{{\mathbb R}^d} |\bu(\bx,t)|^2\,d^nx = \int_{{\mathbb R}^d} |\bu(\bx,s)|^2\,d^dx, \quad \mbox{for all $t>s$} $$
and the local energy equality 
$$ \partial_t\left(\frac{1}{2}|\bu|^2\right)+\grad\cdot\left[\left(\frac{1}{2}|\bu|^2+p\right)\bu\right]=0.$$
Furthermore, there are infinitely many weak Euler solutions with that initial data satisfying the global energy inequality: 
$$ \int_{{\mathbb R}^d} |\bu(\bx,t)|^2\,d^dx < \int_{{\mathbb R}^n} |\bu(\bx,s)|^2\,d^dx, \quad \mbox{for all $t>s$} $$
}
\end{quotation} 
This result showed that one cannot add a local energy inequality  
\be \partial_t\left(\frac{1}{2}|\bu|^2\right)+\grad\cdot\left[\left(\frac{1}{2}|\bu|^2+p\right)\bu\right]\leq 0, \lb{leineq} \ee 
(or even an equality) and obtain a unique weak solution to the Euler equations for certain $L^\infty$ initial data. 
Furthermore, non-uniqueness occurs even if total energy is strictly decreasing. This is in stark contrast to simpler 
equations such as the inviscid Burgers model (or general hyperbolic scalar conservation laws) where it is known 
that insisting on such a local energy inequality selects a unique weak solution, which also coincides with the 
``viscosity solution'' obtained by incorporating viscosity and passing to the inviscid limit. The above theorem 
shows that such a simple selection principle for ``physical weak solutions'' cannot succeed for the incompressible Euler 
equations. 

Note that such initial data with non-unique solutions cannot have smoothness such as $C^{1+\epsilon}$ 
(velocity-gradients existing, with $C^\epsilon$ H\"older regularity) or higher smoothness, because this would violate the following 
important type of result:
\begin{quotation}
\noindent 
\underline{Theorem}: {\it Let $\bu\in L^\infty((0,T);L^2({\mathbb T}^d))$ be a weak Euler solution, $\bU\in C^1({\mathbb T}^d\times [0,T])$
a strong solution, and assume that $\bu$ and $\bU$ share the same initial datum $\bu_0$. Assume moreover that 
\be \int_{{\mathbb T}^d} |\bu(\bx,t)|^2\,d^dx \leq  \int_{{\mathbb T}^d} |\bu_0(\bx)|^2\,d^d x \lb{geineq} \ee 
for almost every $t\in (0,T)$. Then $\bu(\bx,t) =\bU(\bx,t)$ for almost every $(\bx,t).$}
\end{quotation} 
Results of this type go by the name of {\it strong-weak uniqueness}. For a clear and lucid review, see
\cite{wiedemann2018weak}. The conclusion of such results is that any ``admissible''  weak Euler solution, i.e. satisfying the 
global kinetic energy inequality \eqref{geineq} must coincide with a classical Euler solution, as long as that exists. Since 
local-in-time existence of a classical solution is guaranteed for $\bu_0\in C^{1+\epsilon},$ then 
the results in the cited theorem of \citep{delellis2010admissibility} cannot hold for such initial data, at least 
for some finite time interval. Note, by the way, that strong-weak uniqueness applies even to ``measure-valued weak Euler 
solutions" such as those constructed by \cite{diperna1987oscillations}. See \cite{wiedemann2018weak}. 

These non-uniqueness results have since been considerably extended and the question is still currently under active 
investigation. Some very important results are contained in the more recent paper of \cite{daneri2021non}, who prove that  
\begin{quotation}
\noindent 
\underline{Theorem} \citep{daneri2021non}: {\it For any $h\in(0,1/3)$,  there is a set of divergence-free vector fields $\bu_0\in C^h({\mathbb T}^3)$
which is a dense subset of the divergence-free vector fields in $L^2({\mathbb T}^3)$ such that infinitely many 
Euler solutions exist with that initial data  for which $\bu(t) \in C^{h'}({\mathbb T}^3)$ for all $h'<h,$ $t\in (0,T)$ 
and for which the global energy inequality \eqref{geineq} holds.} 
\end{quotation} 
This theorem shows that the non-uniqueness holds right up to the critical Onsager 1/3 exponent and 
for a dense set of initial data. Thus, non-uniqueness is in some sense ``typical''. In recent remarkable 
work of \cite{giri2023L3} it has been shown by an $L^3$-generalization of convex integration methods 
that weak Euler solutions exist, with $\bu\in C([0,T],B^\sigma_3({\mathbb T}^3)$ for any $\sigma<1/3,$ 
which are even locally dissipative, satisfying the local energy inequality \eqref{leineq} and not merely the 
global inequality \eqref{geineq}. This result suggests that non-uniqueness may hold also for locally 
dissipative Euler solutions up to Onsager's $1/3$ exponent. To my knowledge, such results 
have been proved in the $C^h$-setting so far only for $h<1/15$ by \cite{isett2022nonuniqueness} and 
subsequently for $h<1/7$ by \cite{delellis2022non}.  Very intriguingly, \cite{delellis2022weak} have argued 
heuristically that $h=1/3$ should be the critical H\"older exponent not only for conservation of kinetic energy 
but also for convex integration constructions. For a physicist, this remarkable coincidence smacks of a 
turbulent ``fluctuation-dissipation relation.'' 

These startling results on the ``flexibility'' or non-uniqueness of the dissipative weak Euler solutions that were 
conjectured by Onsager pose a fascinating problem for their physical interpretation. One possible attitude 
is that additional conditions should be added to select the ``physically correct'' weak solution obtained 
in the inviscid limit, presumed also to be unique. It is certainly true that the Euler solutions constructed by 
current convex integration methods have unphysical features and thus solutions obtained in the inviscid limit 
should plausibly have additional properties, e.g. the ``martingale property'' of fluid circulations conjectured by 
\cite{eyink2006turbulent}. However, I believe myself that the non-uniqueness or ``flexibility'' uncovered 
by the convex integration theory is probably physically real and that, even if sufficient conditions are identified 
to select physical solutions, nevertheless infinitely many, non-unique solutions for the same initial-data will remain. 
One motivation for this conjecture is the ``spontaneous stochasticity'' phenomenon discovered by \cite{bernard1998slow}, 
which was proposed by \cite{eyink2014mathematical} to manifest not only in Lagrangian particle trajectories 
but also in the Eulerian evolution of the turbulent velocity field itself. Independently and much more concretely, 
\cite{mailybaev2016spontaneously} presented strong evidence from numerical simulations that such Eulerian spontaneous 
stochasticity appears in the Sabra shell model in the high-$Re$ limit. \cite{mailybaev2016spontaneously} 
used another evocative term ``stochastic anomaly'' for this phenomenon, stressing the close formal relation with 
the ``dissipative anomaly'', and pointed out also connections with the classic paper of \cite{lorenz1969predictability} 
on turbulence predictability. Indeed, the pioneering work of \cite{lorenz1969predictability} nearly anticipated the modern 
concept of Eulerian spontaneous stochasticity, as is clear from the first sentences in the abstract of that paper:
\begin{quotation}\noindent 
``It is proposed that certain formally deterministic fluid systems which possess many scales of motion are 
observationally indistinguishable from indeterministic systems; specifically, that two states of the system differing 
initially by a small `observational error' will evolve into two states differing as greatly as randomly chosen states 
of the system within a finite time interval, which cannot be lengthened by reducing the amplitude of the initial error. '' 
--\cite{lorenz1969predictability}, p. 289
\end{quotation} 
Stated differently, the claim of \cite{lorenz1969predictability} was that deterministic fluid systems at high Reynolds numbers
with ``many scales of motion'' may exhibit stochastic solutions, even as random errors in the initial data are taken to zero.
This is the essence of spontaneous stochasticity.  The only thing missing from the concept of \cite{bernard1998slow}
is the clear understanding that such stochastic behavior is made possible by non-uniqueness of 
the solutions of the initial-value problem for the limiting ideal evolution equation as $Re\to \infty.$

Mailybaev and his collaborators have gone on to provide numerical evidence for Eulerian spontaneous 
stochasticity in some prototypical fluid instabilities, such as the Rayleigh-Taylor mixing layer 
\citep{biferale2018rayleigh} and a Kelvin-Helmholtz unstable vortex sheet \citep{thalabard2020butterfly}.
It is noteworthy that for both of these problems, Rayleigh-Taylor \citep{gebhard2021new} and 
Kelvin-Helmholtz \citep{szekelyhidi2011weak,mengual2023dissipative}, non-unique solutions 
of the Cauchy problem for the limiting inviscid equations have been demonstrated by convex 
integration methods. These examples suggest that spontaneous stochasticity must be a 
fairly widespread phenomenon in geophysics and astrophysics. \cite{mailybaev2023spontaneously}
discuss also some simple multi-scale models with known ergodic mixing properties, based on 
the Arnold cat map or irrational rotations of the circle, where Eulerian spontaneous stochasticity 
can be rigorously proved {\it a priori}. Furthermore, \cite{mailybaev2023spontaneous} have set up  
a general renormalization group approach which can demonstrate Eulerian spontaneous stochasticity  
by identifying a suitable RG fixed point and establish also universality properties of the limiting spontaneous 
statistics. Finally, \cite{bandak2024spontaneous} provide evidence that the Landau-Lifschitz fluctuating Navier-Stokes 
equations \eqref{LLNS} should exhibit spontaneously stochastic solutions in the limit $Re\to\infty$ for 
turbulent initial data, triggered just by thermal noise, by means of shell-model simulations. 

I therefore believe that the non-uniqueness or ``flexibility'' phenomenon uncovered by convex integration  
theory is manifested by the essential unpredictability of individual turbulent flow realizations, but whose 
statistics are universal and predictable. The ground-breaking work of \cite{lorenz1969predictability} identified 
a physical mechanism for such  turbulent unpredictability through an ``inverse error cascade'', in which 
dissipation-scale errors propagate up to the largest scales of the flow in a time on the order of the large-eddy turnover time 
 \citep{bandak2024spontaneous}.  The ideas of \cite{lorenz1969predictability} are paradigm-altering,
fundamentally changing the methods and aims of scientific prediction. I  believe that the revolutionary
nature of the  \cite{lorenz1969predictability} work was obscured by his use of a quasi-normal closure
to enable numerical calculations, but which hid the underlying mathematical foundations of his ideas.   
The paper of \cite{palmer2014real} has recalled attention to the ``real butterfly effect'' of \cite{lorenz1969predictability},
but in my opinion mis-identified the origin of the phenomenon in the non-uniqueness of solutions of the viscous 
Navier-Stokes dynamics. However, their proposal does not explain the observations of \cite{thalabard2020butterfly}
on the 2D vortex sheet, for example, since the 2D Navier-Stokes equation has unique solutions. I  believe 
that the fundamental unpredictability identified by \cite{lorenz1969predictability} has its origin instead in the 
non-uniqueness of solutions of the ideal Euler equations obtained in the limit $Re\to 0$ and the Eulerian spontaneous 
stochasticity associated to statistical distributions over those non-unique solutions \citep{bandak2024spontaneous}. 
I hope that proper understanding of the mathematical foundations will help to accelerate progress on this important problem. 

We see furthermore that probability re-enters Onsager's ``ideal turbulence'' theory, as an essential aspect. 
Although your stirred morning cup of coffee, or similar turbulent flows at higher Reynolds numbers, are 
described by a single velocity realization, the same is not true if one considers the ensemble of flows that 
arise from all of the cups prepared over a long sequence of mornings. No matter how carefully one arranges 
to stir the coffee in the same manner each day, the individual flows in the cup and the patterns and whorls 
of cream will differ on each individual day. In fact, the coarse-grained Euler equation \eqref{FEuler} becomes
a stochastic equation when considered as an initial-value problem for $\overline{\bu}_\ell$ alone, since the initial
data for the unresolved eddies $\bu_\ell':=\bu-\overline{\bu}_\ell$ are unknown. In that case, the turbulent
stress term $\btau_\ell$ becomes a random quantity \citep{eyink1996turbulence}, rationalizing the introduction
of ``eddy noise'' \citep{rose1977eddy} or ``stochastic backscatter'' \citep{leith1990stochastic} in large-eddy 
simulation modeling. What spontaneous stochasticity tells us is that such randomness persists even in the 
idealized limit $\ell\to 0.$  It is in the prediction of the future from present initial data that 
probability enters turbulence theory intrinsically.  

\subsubsection{The Infinite Reynolds Number Limit}\lb{sec:InfRe} 

None of the weak Euler solutions discussed so far are obtained by taking infinite Reynolds-number limits of the solutions 
of fluid equations valid at dissipation-range scales, such as the incompressible Navier-Stokes equation \eqref{NS} 
(valid down to about the Kolmogorov scale $\eta$) or the incompressible Landau-Lifschitz equations \eqref{LLNS} 
(valid down to about the mean-free-path length $\ell_{mfp}$). The Euler solutions of physical relevance are just 
approximations to these more basic equations in the inertial-range of length-scales ($\ell\gg \eta$) and such solutions 
must possess very special properties not shared by general weak Euler solutions, including those manufactured 
by convex integration methods. Several very challenging problems remain in the investigation of the infinite-$Re$
limits, which include the existence and uniqueness of the limits themselves. Another important issue is the 
{\it selection problem}, which is the question whether suitable constraints or conditions can be formulated 
which fully characterize {\it a priori} the Euler solutions obtained in the inviscid limit, without actually passing to the limit. 
The latter problem is important for potential practical applications, since one of the great hopes of Onsager's ``ideal turbulence'' 
theory is that it might provide a shortcut to describe and compute the infinite-$Re$ limit directly, avoiding the great 
expense to resolve the small dissipation-range scales. I  provide here a very brief review of a few significant 
mathematical results that bear on these questions.    

A basic problem in the theory of the inviscid limit is that the principal {\it a priori} bounds on the velocity field 
follow from the decay of kinetic energy and such ``$L^2$-bounds'' alone provide insufficient
compactness to guarantee that inviscid limits are standard weak Euler solutions, even along 
suitable subsequences. Without additional conditions, the inviscid limits are more general objects 
such as ``measure-valued Euler solutions'' \citep{diperna1987oscillations,wiedemann2018weak} or
clumsy ``dissipative Euler solutions'' in the sense of \cite{lions1996mathematical}. There are some 
researchers who take measure-valued solutions as the proper physical description and who 
investigate numerical schemes converging to such solutions \citep{fjordholm2017construction}.  
However, an important observation was made by \cite{chen2012kolmogorov}  and 
\cite{isett2022nonuniqueness} (see also \cite{drivas2019onsager,drivas2019remarks}) who observed 
that empirical observations on scaling laws of energy spectra and velocity structure functions guarantee 
strong inviscid limits that are standard weak Euler solutions. The criterion can be stated in terms 
of instantaneous structure functions of absolute velocity increments, where, as in \eqref{Besov}, strict power-law 
scaling is not required but only a power-law upper bound of the form 
\be S_p(\br,t)\leq C_p(t)U^p |\br/L|^{\zeta_p}, \quad \eta_p\leq |\br|\leq L, \lb{Besov2} \ee
where $\eta_p/L\to 0$ as $Re\to\infty.$ If the $p$th-order moment condition \eqref{Lp} holds and if also 
\be \overline{C}_p:=\frac{1}{T}\int_0^T C_p(t)\, dt <\infty, \ee
for some $p\geq 2$ and for some $Re$-independent values $\zeta_p>0$ and $\overline{C}_p>0,$
then it follows that strong limits of Navier-Stokes solutions exist in $L^p([0,T]\times {\mathbb T}^d)$ along 
subsequences $Re_k\to\infty$ and the limiting velocities are standard weak Euler solutions. Of course, these 
solutions may not be unique for the same initial data $\bu_0$ and different subsequences $Re_k\to \infty$ may 
yield different limiting Euler solutions. These results have recently been extended to the Landau-Lifschitz equations 
\eqref{LLNS} in a probabilistic setting \citep{eyink2024emergence}, where the argument now yields a 
probability distribution ${\mathbb P}$ on space-time velocity fields in the limit $Re_k\to\infty$ whose realizations, 
with probability one, are standard weak Euler solutions with the specified initial data $\bu_0$
(either deterministic or random). 

Perhaps the most impressive results that have been achieved to date are on the problem of {\it scalar turbulence},
in particular for the viscous-convective regime described by the theories of \cite{batchelor1959small} and 
\cite{kraichnan1968small} for the limit of high Schmidt number. In the breakthrough work of  \cite{bedrossian2022batchelor},
the Onsager ``ideal turbulence'' predictions have been derived from first principles for the vanishing diffusivity limit 
of a passive scalar advected by a 2D Navier-Stokes flow (or a 3D hyperviscous Navier-Stokes flow) at fixed Reynolds number,
when both the scalar advection-diffusion equation \eqref{adv-diff} and the Navier-Stokes equation \eqref{NS} are forced by 
low-wavenumber random sources, Gaussian and white-in-time.  \cite{bedrossian2022batchelor} study the unique stationary 
measures $\mu^{Re,Sc}$ of the joint velocity-scalar fields $(\bu,c)$ in the limit as $Sc=\nu/D\to \infty$ and establish the existence 
of limiting measures  $\mu^{Re,\infty}$ for which the scalar concentration field $c$ has the Batchelor $k^{-1}$ spectrum 
and a constant scalar flux $\chi$ to infinite wavenumbers. Furthermore, \cite{bedrossian2022batchelor} 
show that the scalar advection equation obtained from \eqref{adv-diff} in the formal limit $D\to 0$ still has 
unique weak solutions in a natural integral form and the limiting measures $\mu^{Re,\infty}$ are invariant 
under this limiting joint dynamics of $(\bu,c)$ for $Sc\to\infty.$ The analogue of Onsager's $1/3$ criterion 
for dissipative weak solutions of the scalar in the Batchelor range advected by a smooth velocity field $\bu$ is 
the statement that $c\notin B^{\sigma,\infty}_2$ for any $\sigma>0$ and \cite{bedrossian2022batchelor} 
prove that the scalar fields $c$ for the limiting measures $\mu^{Re,\infty}$ indeed cannot possess such regularity. 
This is the only example I  know where, in a quite physical setting, the analogue of Onsager's conjectures 
can be rigorously derived from first principles. Although probabilistic methods are not ``intrinsic'' to this problem 
in the same sense as for spontaneous stochasticity, this is a beautiful example where a probabilistic approach is  
able to achieve physical results where purely deterministic techniques are still inadequate. 
Extending such results appears quite difficult. The enstrophy cascade of two-dimensional turbulence 
\citep{kraichnan1967inertial,batchelor1969computation} is physically very similar to the Batchelor regime,
but here the vorticity $\omega$ is an active scalar with $Sc=1$. The analogue of the Onsager theory for the 
2D enstrophy cascade \citep{eyink2001dissipation,lopes2006weak} closely resembles the results derived 
by \cite{bedrossian2022batchelor} for the Batchelor regime, but now the limit $Re\to\infty$ must be tackled
directly in order to prove anything rigorously. 

It seems very difficult to extend the results of \cite{bedrossian2022batchelor} even to the passive scalar in the 
inertial-convective range, where likewise $Sc$ is held fixed and $Re\to\infty$ or, equivalently, P\'{e}clet number 
$Pe=Sc Re\to \infty.$  However, there has been interesting progress on this problem by \cite{colombo2023anomalous} 
and especially by \cite{armstrong2023anomalous} and \cite{burczak2023anomalous}. None of these authors 
consider an advecting velocity field which satisfies the Navier-Stokes equations, but instead they construct 
``synthetic turbulence'' velocity fields which are divergence-free and such that the passive scalar satisfying 
the advection-diffusion equation \eqref{adv-diff} exhibits anomalous dissipation. The velocity fields in these 
constructions all have (in a suitable sense) the spatial H\"older regularity $C^h$ with $0<h<1$
that is expected for an inertial-range velocity field and the advected scalar exhibits anomalous dissipation
in the sense that, with dimensionless diffusivity $\hat{D}=1/Pe,$ 
\be \limsup_{\hat{D}\to 0} \int_0^{\hat{T}} d\hat{t} \int_{{\mathbb T}^2} d^2\hat{x}\ \hat{D}|\hat{\grad}\hat{c}|^2 >0, \ee 
where $\widehat{(\cdot)}$ denotes outer-scale non-dimensionalization. A striking result of these constructions 
is that there are distinct limit points $\hat{c}_*$ for the scalar field obtained along different subsequences 
$\hat{D}_k\to 0,$ which give different weak solutions of the ideal advection equation for the same initial data $\hat{c}_0.$
Similar non-uniqueness was found also by \cite{huysmans2023nonuniqueness} for a passive scalar advected 
by a rougher divergence-free velocity field that is only $L^\infty$ in space, and here the non-unique solutions 
of the ideal advection equation conserve the scalar energy. Such non-uniqueness is somewhat surprising for 
passive scalar advection. Recall that the scalar advection equation for the Batchelor regime at $Sc\to\infty$ was 
shown by \cite{bedrossian2022batchelor} to have unique solutions and likewise the white-noise advection model 
of \cite{kraichnan1968small} in the inertial-convective regime is known to have unique weak solutions of the 
scalar advection equations (\cite{lototskii2004passive}; note that the latter solutions are ``weak'' in the PDE sense 
but strong in the stochastic sense, with exactly one solution for each realization of the advecting 
random velocity field). The recent results raise the intriguing possibility of Eulerian spontaneous 
stochasticity also for passive scalar advection in the inertial-convective range. 

There are important differences between the methods in the above papers which may be appreciated 
by reading the originals, but which deserve to be emphasized briefly here. See also the extensive review
in \cite{armstrong2023anomalous}, section 1.1.  The construction of \cite{colombo2023anomalous} 
combines alternating shear flows that focus in a singular manner at the final time-step. A drawback
of this approach is that the anomalous scalar dissipation occurs only at the final time and the velocity 
field has only one active scale at each time, both features very uncharacteristic of physical turbulence.
\cite{armstrong2023anomalous} overcome these problems, synthesizing a 
true multiscale velocity field for $h<1/3$ that produces anomalous scalar dissipation at all times. Their approach is an iterative 
homogenization technique which has much in common with renormalization group methods in physics, 
generating an eddy-diffusivity at each scale by integrating over the smaller scales. 
Furthermore, the synthetic turbulent field which is constructed by \cite{armstrong2023anomalous} has
a very fundamental feature of true turbulent flows that small eddies are advected by larger eddies, which turns 
out to be essential in obtaining anomalous dissipation for the scalar at all times. To my knowledge, this is the first 
example of a synthetic turbulent velocity which captures the sweeping properties of a physical Navier-Stokes
or Euler solution.  \cite{burczak2023anomalous} have extended this approach by combining it with 
convex integration techniques to ensure that the advecting velocity field is in fact a weak solution of the Euler equations. 

The methods of \cite{colombo2023anomalous} have recently been applied to prove the existence 
of kinetic energy dissipation anomaly for the forced 3D Navier-Stokes equation \citep{brue2023anomalous,brue2022onsager}
For a sequence of body forces ${\bf f}^{Re}$ that satisfy some modest uniform smoothness properties, such as 
\be \sup_{Re>0} \|{\bf f}^{Re}\|_{C([0,T],C^h({\mathbb T}^3))}<\infty \ee 
for any $h\in (0,1),$ it has been proved that the viscous dissipation by the Navier-Stokes solution is anomalous 
for $\hat{\nu}=1/Re\to 0$ in the sense that 
\be \limsup_{\hat{\nu}\to 0}  \int_0^{\hat{T}} d\hat{t} \int_{{\mathbb T}^3} d^3\hat{x}\ \hat{\nu} |\hat{\grad}\hat{\bu}|^2 >0. \ee
Although the body force  is not a smooth, large-scale forcing as usually assumed in turbulence theory, it is 
regular enough that the previous lim-sup vanishes for the linear Stokes equation, which lacks nonlinear energy cascade.
Even more interestingly, the inviscid limits are found to be non-unique exactly as in the passive scalar case, with 
different subsequences of Reynolds numbers $Re_k\to \infty$ yielding distinct weak solutions of the Euler equations 
with the same initial data $\bu_0$ and the same body force ${\bf f}=\lim_{Re\to\infty} {\bf f}^{Re}$  \citep{brue2022onsager}. 

\section{Turbulence Interactions With Solid Walls}\label{walls}

\subsection{Overture on Turbulence and Solid Surfaces}\lb{sec:overture} 

All of the previous theoretical results that I  have discussed involve turbulence away from walls 
and solid boundaries, although terrestrial turbulence arises most frequently at fluid-solid interfaces
(and somewhat less often at gas-liquid interfaces). Surprisingly, it is only recently that the methods 
of \cite{onsager1949statistical} have been applied to such problems, in what may be called the 
``Onsager theory of wall-bounded turbulence''. Here I shall briefly review this recent work following 
\cite{eyink2023turbulenceII}, where more details can be found. I  just note that it is not entirely 
ahistorical to refer to an ``Onsager theory of wall-bounded turbulence'', since \cite{onsager1949statistical} mentioned
turbulent pipe flow as an example of this type:							
\begin{quotation} \noindent 
Such a familiar type of turbulence as exists in a liquid flowing through a cylindrical tube is neither homogeneous 
nor isotropic. The mean fluctuations of the velocity vary over the cross-section of the tube, the local macroscale 
is generally comparable to the distance from the wall, and fluctuations as well as correlations are more or less anisotropic.
--L. \cite{onsager1949statistical}, p.283. 
\end{quotation} 
citing the paper of \cite{montgomery1943generalization} on smooth- and rough-walled pipes for the reported observations.  
Onsager was also a friend from their time in Trondheim together of Theodore Theodorsen, the originator of the concept of 
``hairpin vortices''  in wall-bounded turbulence \citep{theodorsen1952mechanism}, and they reportedly discussed turbulence,
among other scientific topics. I shall unfortunately not have space in this essay to explore how Onsager's ideas 
relate to the modern descendent of Theodorsen's ideas, the ``Attached Eddy Model'' \citep{marusic2019attached}, although 
both attempt to describe the high Reynolds limit of wall-bounded flows.  Connections can be glimpsed from 
the results presented in sections \ref{sec:moment}-\ref{sec:vorticity}. In fact, we shall see that Onsager pursued his interest 
in wall-bounded turbulence further in his unpublished notes and that his calculations anticipate some modern 
research directions.   

On the other hand, it is clear already from the brief summary of the experimental evidence presented
in section \ref{empiric} that Onsager's $1/3$ H\"older result and the various analyses that follow it, ignoring the 
crucial role of the solid walls, cannot be a complete theory of the turbulent dissipative anomaly for incompressible fluid
flows.  Onsager's result applies indeed to wall-bounded turbulence, as we shall see, but it gives only a necessary condition 
for anomalies, without explaining the sufficient conditions that may generate them. In particular, the literature reports
a sharp dichotomy in certain flows which have been characterized as ``internal'' or ``closed'' \citep{cadot1997energy},
but which I believe are better described as ``wall-parallel'', with the mean flow everywhere tangent 
to the solid boundary. In these flows --- which include such common examples as flows through straight pipes 
and flat-walled channels, flat-plate boundary layers, and Taylor-Couette flows --- there is no empirical evidence for 
energy dissipative anomaly with walls that are hydraulically smooth, but only with walls that are hydraulically rough.
Furthermore, this dichotomy does not apply to turbulent wakes behind solid bodies, either bluff or streamlined, 
where there is solid evidence for a dissipative anomaly in the wake independent of the wall roughness. No theory 
of the turbulent dissipative anomaly can be complete which does not account for these crucial dependences on the nature 
of the wall. It is these conundrums, among others, which makes the problem of high-$Re$ wall-bounded turbulence 
such a fascinating science problem. 

It should be emphasized at the outset that the very meaning of ``infinite Reynolds number limit'' 
becomes ambiguous for flows with wall roughness, a feature always present in any 
physical surface to some degree \citep{jimenez2004turbulent}. The difficulty is that the typical roughness 
height $k$ introduces a new length scale besides the overall dimension $L$ of the flow and thus Reynolds' 
law of hydrodynamic similarity breaks down. Dimensional analysis yields that, in addition to the usual 
Reynolds number $Re=UL/\nu,$ there is another dimensionless number group, which may be taken to 
be either the length ratio $\hat{k}=k/L$ or else the ``roughness Reynolds number'' $Re_k=Uk/\nu.$
It now matters greatly how the parameter regime $Re\gg 1$ is reached. In a fixed laboratory 
apparatus of dimension $L,$ it is most common to vary $Re$ by varying the flow speed $U,$
but this has the effect also of varying $Re_k.$ When $Re_k\simeq 1,$ the transitionally rough regime,
then roughness effects begin to be felt and these effects become dominant in the fully rough regime for $Re_k\gg 1.$
The high-$Re$ limit with smooth walls that is most popular with mathematicians instead corresponds 
to flow around bodies of very large size $L$ so that $Re=UL/\nu\gg 1$ while simultaneously $Re_k\lesssim 1$
and $\hat{k}\ll 1.$ 
  
Another crucial difference with turbulence away from solid walls is that kinetic energy is not 
the only inviscid invariant of smooth Euler solutions that is subject to possible dissipative 
anomalies. For example, fluid linear momentum is strictly conserved in smooth wall-parallel Euler 
flows, where pressure forces act only normal to the surface and cannot transfer the momentum 
component parallel to the wall. Instead in viscous flow, the skin friction vector 
\be \btau_w=2\nu \bS\bdot\bn, \ee 
with $\bn$ the surface unit normal pointing into the fluid, can transfer parallel momentum, possibly even in the 
limit as $Re\to \infty.$ This possibility seems to have been first raised by G.~I. \cite{taylor1915eddy}
in a paper on eddy motion in the atmosphere:
\begin{quotation}\noindent
``...a very large amount of momentum is communicated by means of eddies from the atmosphere to the ground. 
This momentum must ultimately pass from the eddies to the 
ground by means of the almost infinitesimal viscosity of the air. The actual value 
of the viscosity of the air does not affect the rate at which momentum is communicated 
to the ground, although it is the agent by means of which the transference is effected. 

\noindent
$\cdots$

\noindent 
The finite loss of momentum at the walls due to an infinitesimal viscosity may be 
compared with the finite loss of energy due to infinitesimal viscosity at a 
surface of discontinuity in a gas.${\,\!}^*$'' --\cite{taylor1915eddy}, pp.25-26. 
\end{quotation}
It is remarkable that Taylor in this early paper not only recognized that there 
could be a finite loss of momentum due to an ``infinitesimal viscosity'', but also compared 
this phenomenon with discontinuous shock solutions which we now understand
to be described, in modern language, by weak solutions of inviscid fluid equations. 
This proposal goes well beyond the laminar boundary-layer theory of \cite{prandtl1905ueber}. 
In addition to momentum, another inviscid conserved quantity is the total volume-integrated 
vorticity, which vanishes identically according to the theorem of \cite{foppl1913wirbelbewegung}
and is thus, trivially, conserved. However, it has been emphasized by \cite{lighthill1963introduction} and 
\cite{morton1984generation} that this ``trivial'' conservation law imposes very important constraints 
on the flow and, in particular, has a local source at solid surfaces of the form 
\be \bsigma=-\bn\btimes\,\grad p=\bn\btimes \nu(\grad\btimes\bomega). \lb{sigma} \ee
The middle expression in \eqref{sigma} describes the creation of vorticity by tangential pressure 
gradients (which occurs even in  smooth Euler flows) and the final expression describes the 
viscous diffusion of vorticity from the surface, with both terms instantaneously in balance for stick boundary 
conditions $\bu=\bzed$ on the velocity field. In a turbulent flow, the source \eqref{sigma} drives an 
``inverse cascade of vorticity'' away from the solid surface \citep{eyink2008turbulent,kumar2023vorticity} 
and one naturally wonders if this cascade may persist in the infinite-$Re$ limit. 

Before undertaking any mathematical analysis, it is crucial to review briefly some representative 
results from experiments and numerical simulations, providing a few more details than in section \ref{empiric},
so that we have some clear idea of the observations that require explanation. \cite{mckeon2004friction} report 
on experimental measurements of the friction factor \eqref{lambda} 
in a smooth-wall pipe with $Re_k\lesssim 1$, finding good agreement with the classical Prandtl-K\'arm\'an law 
for appropriate constants. Extrapolated to very high Reynolds numbers, this result gives $\lambda\sim C/(\ln Re)^2$ 
for $Re\gg 1.$ Although there is considerable ongoing debate about the precise asymptotics, all observations 
are consistent with $\lambda$ tending to zero as $Re\to\infty,$ although much more slowly than the rate 
$\lambda\sim 64/Re$ for laminar pipe flow. 
Noting the simple relation $\partial \overline{p}/\partial x=
2\overline{\tau}_{w,x}/R$ from mean momentum balance, with $R$ the pipe radius, we see that 
$\overline{\tau}_{w,x}/\bar{U}^2 \sim C'/(\ln Re)^2$ according to the results of \cite{mckeon2004friction}. 
There is thus no ``strong momentum anomaly'' in this flow of the type conjectured by \cite{taylor1915eddy},
but only a ``weak anomaly'' in the terminology of \cite{bedrossian2019sufficient}. 
On the other hand, the experiment of \cite{nikuradse1933laws} with ``sand-grain'' roughness,  
which varied $Re$ by changing $\bar{U}$ for fixed values of $\hat{k}=k/R,$ found that $\lambda \sim 
C \hat{k}^{1/3}$ for $Re\gg 1.$ Although this experiment indicates a ``strong anomaly'', it does not 
mean that $\overline{\tau}_{w,x}/\bar{U}^2$ tends to a non-vanishing value. In fact, various numerical 
simulations of rough-walled pipe and channel flow such as that of  \cite{busse2017reynolds} indicate 
that indeed $\overline{\tau}_{w,x}/\bar{U}^2\to 0$ as $Re\to \infty$ and the non-vanishing drag is due 
to the net pressure force acting on the roughness elements, or the ``form drag.''  There is therefore 
a close correspondence between the wall-parallel flows with hydraulically rough walls and the wake flows
past bodies, where likewise the time-averaged skin friction tends to zero with increasing $Re$ and 
non-vanishing drag is due to the asymmetric pressure forces acting on the body. A good example of the 
latter type is the flow past the sphere studied at high-$Re$ by \cite{achenbach1972experiments}, who measured 
both skin friction and pressure profiles on the surface of the sphere. Experimental visualizations 
of flow around individual ``sand-grain'' roughness elements in turbulent duct flow \citep{gao2021experimental} 
show remarkable similarity to flow around bluff bodies such as a sphere, with separating boundary 
layers and recirculation zones in the wake. The apparently non-vanishing surface pressure gradients 
in these flows suggest also a ``strong vorticity anomaly'', due to persistent vorticity flux into the flow.

Most of the attempts to understand these observations have been based on the incompressible 
Navier-Stokes equation \eqref{NS} in the flow domain $\Omega$, with stick b.c. $\bu=\bzed$ of the velocity 
on the boundary $\partial\Omega$ of the domain and kinematic pressure obtained from the Poisson 
problem with Neumann boundary conditions:
\be -\Delta p = {\rm tr}(\grad\bu)^2 \ \ \mbox{ in $\Omega$}, 
\quad \frac{\partial p}{\partial n}=\bn\bdot\nu\Delta\bu \ \ \mbox{ on $\partial\Omega$}. \lb{press} \ee
Of course, this mathematical model is just an approximation to reality for molecular fluids in nature
and various refinements have been considered. For example, many analyses have considered also 
the {\it slip conditions} of  Navier that determine a velocity slip $\delta\bu_s$ at the wall 
\be \delta \bu_s = \ell_s \frac{\partial \bu}{\partial n} \ \ \mbox{ on $\partial\Omega$} \ee 
rather than stick b.c. for the velocity field, with slip length $\ell_s$ estimated by Maxwell to be of the
order of the molecular mean free path length $\ell_{mfp}.$ One also cannot rule out effects 
of thermal fluctuations, since these generally appear at lengths much larger than $\ell_{mfp}.$ 
Here it should be noted that the correct formulation of the Landau-Lifschitz fluctuating hydrodynamics 
equations in the vicinity of a fluid-solid interface is still a hotly debated issue; see \cite{reichelsdorfer2016foundations} 
for a very readable summary. I leave such issues mostly to side in my review below, where 
I shall discuss how far the ``standard model'' of incompressible Navier-Stokes with stick b.c. 
can be used to explain the empirical observations. I  just note finally that the various conditions
for convergence of Navier-Stokes solutions to weak Euler solutions in the limit $Re\to\infty,$ which 
were discussed in section \ref{sec:InfRe}, carry over also to wall-bounded flows. See 
\cite{drivas2019remarks} for an excellent overview of such results. 

\subsection{Onsager RG Analysis}\lb{wall:RG} 

\subsubsection{Regularization of Ultraviolet Divergences}\lb{wall:UV}  

A key conclusion that may be be drawn from the brief summary of empirical observations in the 
previous section is that new UV divergences appear at the solid interfaces in wall-bounded 
turbulence for $Re\gg 1$. In fact, whenever the non-dimensionalized wall stress 
$\hat{\btau}_w=\btau_w/U^2=\frac{1}{Re} \frac{\partial \hat{\bu}}{\partial \hat{n}}$ 
vanishes more slowly than the laminar rate $\propto 1/Re,$ then the non-dimensionalized 
velocity-gradient at the wall diverges: 
\be  \lim_{Re\to\infty} \frac{\partial \hat{\bu}}{\partial \hat{n}}= \infty \ee
Note that such divergences appear at the wall even if there is only a ``weak anomaly'' in the turbulent 
momentum balance. In order to obtain a dynamical description that remains valid in the limit $Re\to\infty$, 
these new divergences at the wall must be regularized along with the divergences of the
non-dimensionalized gradients $\hat{\grad}\hat{\bu}$ in the bulk of the flow. 

A recent paper of \cite{bardos2018onsager} has initiated the mathematical investigation 
of Onsager's ``ideal turbulence'' theory for wall-bounded flows, followed already by several works with improvements  
\citep{drivasN2018onsager,bardos2019onsager,chen2020kato}. These works employ as regularizer 
a modification of the spatial coarse-graining of the velocity defined in \eqref{II5}. It is convenient to assume that the filter kernel 
$G$ is supported in a ball of radius 1, so that the definition of $\bar{{\mathbf u}}_\ell({\mathbf x},t)$ 
makes sense for points ${\mathbf x}\in \Omega$ with distance at least $\ell$ from the boundary $\partial\Omega.$
To eliminate also the divergences of the velocity-gradients at the wall and to obtain a well-defined coarse-grained
velocity, one may also smoothly ``window out'' eddies at distances $<h$ to the wall, with $h>\ell.$ This can be accomplished 
by taking a smooth windowing function $\theta_{h,\ell}(\delta)$ with the properties that
\begin{align}  \theta_{h,\ell}(\delta)=\left\{\begin{array}{ll}
                           0   & \delta<h \cr
                           1   & \delta>h+\ell 
                           \end{array}\right. \end{align} 
and $\theta_{h,\ell}(\delta)$ monotone increasing on the interval $[h,h+\ell]$. See Fig.~\ref{fig_window}. 
Finally, one defines $\eta_{\ell,h}({\mathbf x}):=\theta_{\ell,h}(d({\mathbf x}))$ and for all ${\mathbf x}\in \Omega$
\begin{align}
    \widetilde{\mathbf{u}}_{\ell,h}({\mathbf x},t)=\eta_{\ell,h}({\mathbf x})
    \bar{\mathbf{u}}_\ell({\mathbf x},t) \label{utilde} 
\end{align}
where $d({\mathbf x})$ measures the distance of ${\mathbf x}\in \Omega$ to the boundary: 
\begin{align} d({\mathbf x})=\inf_{{\mathbf y}\in\partial\Omega}|{\mathbf x}-{\mathbf y}|. 
\label{d-def} \end{align} 
With this definition 
$\nabla d({\mathbf x}) = {\mathbf n}({\mathbf y}_{\mathbf x})
:={\mathbf n}({\mathbf x}),$ 
where ${\mathbf n}({\mathbf y})$ is the inward-pointing unit normal vector at a point 
${\mathbf y}\in\partial\Omega$ and ${\mathbf y}_{\mathbf x}\in\partial\Omega$
is the point at which the infimum in \eqref{d-def} is achieved for each ${\mathbf x}\in\Omega.$ 
See \cite{bardos2018onsager}.  The coarse-grained velocity defined by \eqref{utilde} may 
be described picturesquely as the fluid velocity seen by an observer who is myopic and who also 
has tunnel vision, with parameter $\ell$ characterizing the blurriness of their eyesight and 
$h$ their loss of peripheral vision. 

{\begin{figure}
\centering
\includegraphics[width=.4\textwidth]{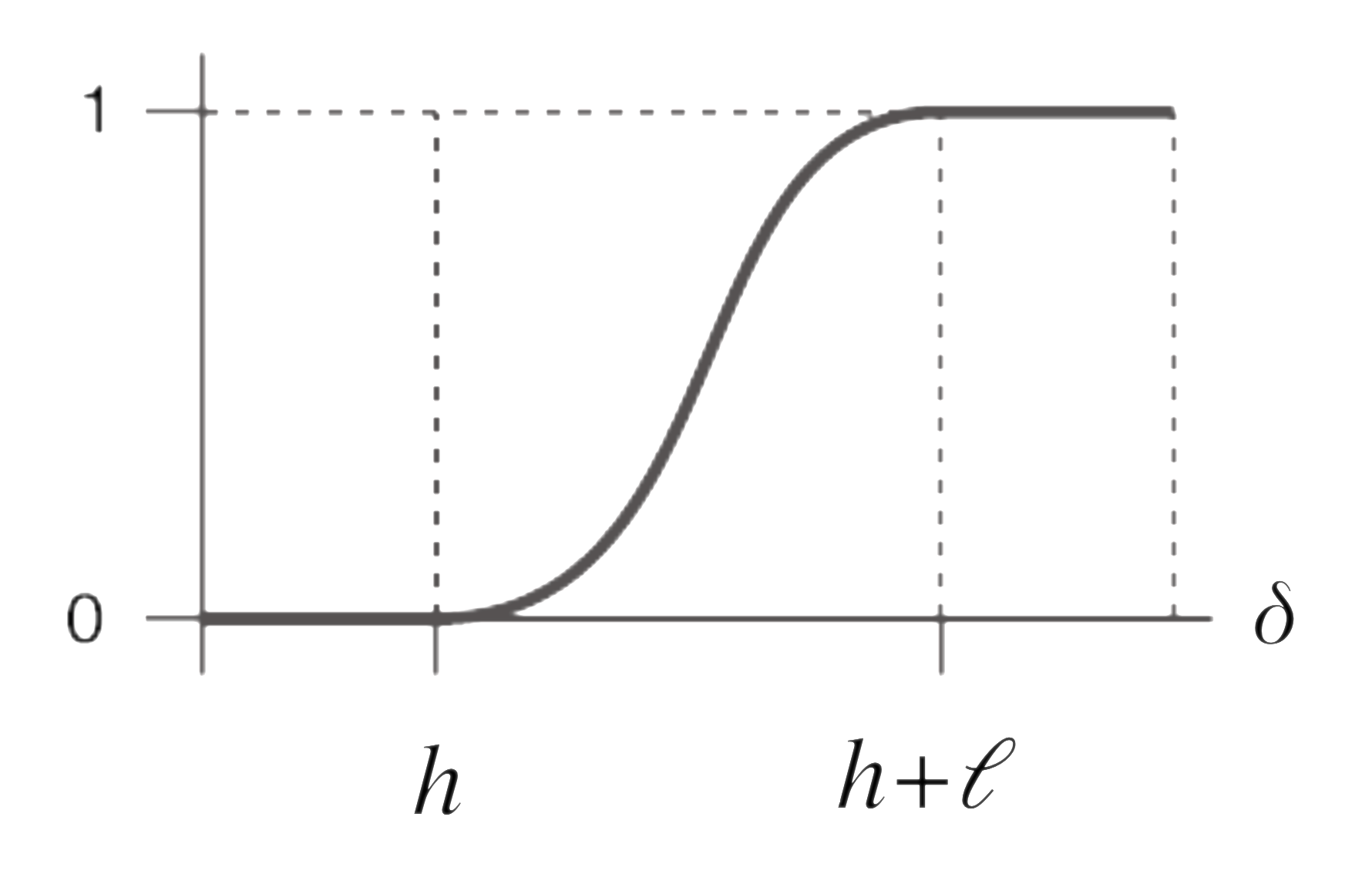}
\caption{Window function $\theta_{h,\ell}$ used to screen from observation fluid eddies near the wall.}
\label{fig_window}
\end{figure}} 

It is important to emphasize here that this specific choice of regularization is not essential 
to the theory. Just as in quantum field theory, there are many possible regularizers 
to eliminate UV divergences \citep{gross1976applications} and it is ultimately a matter 
of convenience and ease of application which to choose for a particular problem. Onsager in his unpublished 
research notes in fact explored a different regularization method for plane-parallel channel
flow based on expansion of the velocity field in eigenmodes of the linear Stokes operator. 
See the folder of notes in the online Onsager Archive at \url{https://ntnu.tind.io/record/121186},
 pp.4-8, where Onsager solves the eigenproblem for the Stokes operator 
 ${\mathit A}_D\bw:={\mathit P}(-\Delta)\bw$ with Dirichlet b.c. $\bw=\bzed,$ where
 ${\mathit P}$ is the Leray projection onto divergence-free fields: 
 \be {\mathit A}_D\bw_n = \lambda_n \bw_n, \quad n=0,1,2,....\ee 
Identical results were published later by \cite{rummler1997eigenfunctions}. Onsager clearly had the 
idea to expand the turbulent velocity field in the channel into such eigenmodes, thus regularizing 
all UV divergences. With a smooth exponential cut-off, this would correspond to taking
\be \overline{\bu}_\ell = \sum_{n=0}^\infty e^{-\lambda_n\ell^2} c_n \bw_n \ee
for some regularization scale $\ell,$ where $\bu=\sum_{n=0}^\infty c_n\bw_n.$ With this approach, 
$\grad\bdot\overline{\bu}_\ell=0$ in $\Omega$ exactly and $\overline{\bu}_\ell=\bzed$ on $\partial\Omega.$
Note that the above expansion is equivalent to defining the regularized velocity field by the 
auxilliary initial-value problem for the Stokes equation:
\be \frac{\partial\overline{\bu}_\ell}{\partial\ell^2} = -{\mathit A}_D\overline{\bu}_\ell, 
\quad \left.\overline{\bu}_\ell\right|_{\ell=0}=\bu. \ee 
Exactly this PDE approach to spatial filtering in wall-bounded domains was proposed recently by 
\cite{johnson2022physics}, which can be regarded as a version of the heat-kernel regularization
of \cite{isett2016heat}, adapted to manifolds with boundaries. This PDE approach is more flexible 
than the original eigenmode-expansion of Onsager, since it is difficult for general domains to 
calculate the eigenfunctions explicitly. Note that one may apply this Stokes regularization also 
with the operator ${\mathit A}_N$ for Neumann b.c. $\partial \bw/\partial n=\bzed$ on $\partial \Omega,$
so that $\int_\Omega \overline{\bu}_\ell \, dV=\int_\Omega \bu\, dV.$ 

In general, the necessity of some spatial filtering or coarse-graining closely connects the Onsager 
theory with the turbulence modeling method of Large-Eddy Simulation (LES). Note that 
wall-bounded turbulence is currently one of the most challenging areas of LES research and 
the subject of intense effort \citep{bose2018wall}. Many other types of spatial filtering or 
regularizations have been explored in that context, including regularizations specified by 
elliptic PDE problems such as 
\be \overline{\bu}_\ell 
- \frac{\partial}{\partial x_k} \left(\ell^2 \frac{\partial\overline{\bu}_\ell}{\partial x_k}\right)=\bu, 
\quad \mbox{ in $\Omega$} \lb{elliptic} \ee 
for some chosen length-scale $\ell(\bx),$ which may now be chosen to depend on position. 
See \cite{germano1986differentialA,germano1986differentialB,bose2014dynamic,bae2019dynamic}. 
As stated in the quote of \cite{onsager1949statistical}, the integral length in wall-bounded 
turbulence is generally proportional to the distance from the wall, or $L(\bx)\sim \kappa d(\bx)$
with $\kappa$ the so-called K\'arm\'an constant. 
Thus, one might be tempted to take $\ell(\bx)\propto L(\bx),$ which would correspond to integrating   
out all eddies smaller than the local integral length. This so-called ``wall-resolved LES'' is, however, 
almost as computationally expensive as direct numerical simulation (DNS) of the full 
Navier-Stokes equation \citep{yang2021grid}, so that one is generally interested in taking $\ell(\bx)\gg L(\bx)$
for practical applications. This problem of coarse-graining and LES modelling in the presence of solid 
walls is an important application area where mathematicians and engineers can very productively interact. 

\subsubsection{Coarse-Grained Equations}\lb{wall:coarse} 

The filtering procedures employed in LES lead to effective coarse-grained equations which contain
quantities unclosed in terms of the coarse-grained velocity itself, which may then be modeled.
The same is true also of the regularization procedures used by mathematicians to study the 
infinite-Reynolds limit. For example,  the combined spatial filtering and windowing introduced by
\cite{bardos2018onsager} when applied to the Navier-Stokes equation \eqref{NS} leads to an exact 
regularized equation for the coarse-grained field $\widetilde{\mathbf{u}}_{\ell,h}$ of the form
\begin{align}
    \frac{\partial \widetilde{\mathbf{u}}_{\ell,h} }{\partial t}  +\nabla \cdot \bigg[\widetilde{\tau}_{\ell,h}(\mathbf{u},\mathbf{u})
    + \widetilde{\mathbf{u}}_{\ell,h}\widetilde{\mathbf{u}}_{\ell,h} + \widetilde{p}_{\ell,h} \mathbf{I}  -  \nu\widetilde{\mathbf{\nabla u}}_{\ell,h}\bigg]=
    -{\mathbf f}_{\ell,h}. 
   \label{Windowed}
    \end{align}
Here, the {\it turbulent (or subgrid) stress} due to eddies of size $<\ell$ may be defined as usual by     
\begin{align}
    \widetilde{\tau}_{\ell,h}(\mathbf{u},\mathbf{u}) = (\widetilde{\mathbf{uu}})_{\ell,h}-\widetilde{\mathbf{u}}_{\ell,h}\widetilde{\mathbf{u}}_{\ell,h}.
\end{align}    
In addition, a new {\it inertial drag force} appears associated to the eliminated 
near-wall eddies
\begin{align} 
{\mathbf f}_{\ell,h}= 
-\theta_{\ell,h}'(d({\mathbf x}))\,{\mathbf n}({\mathbf x}) \cdot 
\bigg[ (\overline{\mathbf{uu}})_\ell + \overline {p}_\ell \mathbf{I}-\nu \nabla\bar{\mathbf{u}}_\ell  \bigg],  
\lb{drag} \end{align} 
and represents momentum transfer to the unresolved near-wall eddies. This force includes resolved viscous 
momentum diffusion, but nonlinear transfer dominates for $Re\gg 1$ at fixed $\ell,$ $h.$
Mathematically, the regularized equation \eqref{Windowed} when considered for all possible choices of $h>\ell$
is equivalent to the standard weak formulation of the incompressible Navier-Stokes equation in a domain with boundary
(see \cite{boyer2012mathematical}, Ch. V.1.2). 

If \eqref{Windowed} were considered for the purpose of LES modeling, both the subgrid stress 
$\widetilde{{\boldmath \mbox{$\tau$}}}_{\ell,h}$ and the inertial drag force ${\mathbf f}_{\ell,h}$ would need 
to be somehow closed or parameterized. Indeed, one can expect that these quantities have universal 
statistical properties independent of the small-scale dissipation and of the detailed properties of the wall,
if the distance $h$ is chosen in the inertial sublayer and length-scale $\ell$ is chosen also sufficiently small. 
In addition, however, another quantity must be modelled which appears in the coarse-grained 
mass balance: 
\begin{align}
\nabla\cdot \tilde{{\mathbf u}}_{\ell,h}=\tilde{\sigma}_{\ell,h} \label{mass-bal}  \end{align}
which I  call the {\it inertial mass source} 
\begin{align} 
\tilde{\sigma}_{\ell,h}:=\theta_{h,\ell}'(d({\mathbf x})) {\mathbf n}({\mathbf x})
\cdot\bar{\mathbf u}_\ell. \end{align}
This quantity measures the mass-exchange with the unresolved near-wall eddies and it is 
non-vanishing only for $h<d({\mathbf x})<h+\ell,$ with furthermore a vanishing space integral
$\int_{h<d<h+\ell} \tilde{\sigma}_{\ell,h} \, dV=0.$ The windowing operation has thus introduced 
effective ``compressibility'' so that the Poisson equation for coarse-grained pressure becomes 
\begin{align}  -\triangle\tilde{p}_{\ell,h}=
\partial_t \tilde{\sigma}_{\ell,h}
+\nabla\nabla:\left[\widetilde{\mathbf{u}}_{\ell,h}\widetilde{\mathbf{u}}_{\ell,h}
+\widetilde{{\boldmath \mbox{$\tau$}}}_{\ell,h}
-\nu\widetilde{\mathbf{\nabla u}}_{\ell,h}\right] 
-\nabla\cdot \mathbf{f}_{h,\ell} \end{align}
which involves derivatives of all three modelled quantities but which can yield the resolved
pressure by applying standard Poisson solvers. Similar effects appear with other coarse-graining
methods, e.g. the velocity field from the elliptic filtering \eqref{elliptic} satisfies an equation 
$\grad\bdot\overline{\bu}_\ell=\overline{\sigma}_\ell$ analogous to \eqref{mass-bal} with 
\be \overline{\sigma}_\ell = 
\overline{\left(\frac{\partial}{\partial x_k} \left[\grad\ell^2 \bdot\frac{\partial\overline{\bu}_\ell}{\partial x_k}\right]\right)}_\ell\ee
In this case, however, it is not true that $\int_{\Omega} \overline{\sigma}_{\ell} \, dV=0$ unless $\partial\overline{\bu}_\ell/\partial n=0$
at $\partial\Omega.$ 

The regularized equation \eqref{Windowed} involves two arbitrary lengths $\ell$ and $h$
and, as well, three arbitrary functions $G,$ $\eta,$ and $d.$ Other choices of distance function 
might be more useful for some purposes, e.g. $d({\mathbf x})=\min\{|y-H|,|y+H|\}$ could be useful
in an iterative RG analysis of a rough-wall channel flow, with $H$ the channel half-width, in order to 
establish universal statistics in the ``inertial sublayer''. A similar type of ``time windowing'' 
was employed in the recent RG analysis of Lagrangian spontaneous stochasticity 
\citep{eyink2020renormalization}, where it corresponds to ignoring non-universal initial times
of the particle position histories. The ``principle of 
renormalization group invariance'' is that no objective physics can depend upon these 
arbitrary quantities introduced for the purpose of regularization 
\citep{gross1976applications,eyink2018review}.
The present example is a case of a several-parameter renormalization group involving changes 
of the entire regularization scheme, which was encountered already in quantum field theory 
\citep{stueckelberg1953normalisation,stevenson1981optimized,peterman1982several} 
and which has been applied since to PDE's, including boundary-value problems
\citep{chen1996renormalization,kovalev1999functional}. A key idea in RG methods  
is that arbitrariness in regularization parameters may be exploited by 
choosing them in some optimal way to deduce non-trivial consequences. I  
describe some applications of that principle in the following section. 

\subsubsection{Momentum Cascade in Space}\lb{sec:moment} 

Most of the prior work on the Onsager theory for wall-bounded flows has considered energy cascade 
and energy dissipation \citep{bardos2018onsager,bardos2019onsager,drivasN2018onsager,chen2020kato},
but I  prefer to begin first with the momentum anomaly suggested by \cite{taylor1915eddy}. I  shall
employ the regularization by spatial filtering and windowing introduced  by \cite{bardos2018onsager}, 
which leads to the important concept of {\it spatial momentum cascade}. To avoid any confusion, I  
emphasize at the outset that the notion of momentum cascade arising in the Onsager theory does {\it not} coincide 
with the one which is now standard in the literature on wall-bounded turbulence \citep{tennekes1972first,jimenez2012cascades}.
I shall discuss the essential differences below. My analysis shall follow closely the work of  \cite{quan2022inertial},
where momentum balance in the limit $Re\to\infty$ is treated. It is also possible to extend this analysis to 
finite Reynolds numbers and some beginning steps in this direction have been taken by \cite{eyink2022onsager}, 
following the ideas of \cite{drivas2019onsager} for energy cascade, but I  consider here only the regime $Re\gg 1.$ 

The work of \cite{quan2022inertial} followed the same RG strategy as in prior works on the Onsager theory
such as \cite{duchon2000inertial}, by considering the limit $Re\to\infty$ for both the fine-grained description 
and the coarse-grained description with regularization scales $\ell,$ $h.$ Nontrivial consequences can be 
deduced simply by requiring that both descriptions lead to the same observable conclusions and exploiting 
the freedom to vary $h,\ell$ by taking $h,$ $\ell\to 0.$ The specific example analyzed by \cite{quan2022inertial}
was external flow around a finite, smooth body $B,$ as considered in the famous paradox of 
\cite{dalembert1749theoria,dalembert1768paradoxe}. As seen from the empirical results reviewed 
in section \ref{sec:overture}, this is probably the simplest flow which provides evidence for a 
strong dissipation anomaly. However, the analysis of \cite{quan2022inertial} carries over straightforwardly 
to related more complex flows, such as turbulent flows through pipes with hydraulically rough (but 
mathematically smooth) walls. Since only mathematically smooth surfaces are considered, we are 
investigating the limit $Re=UL/\nu\gg 1$ as achieved by taking $L$ very large (where $L$ would be a 
length such as the diameter of the body or the radius of the pipe). In the case of a hydraulically rough wall
we keep the ratio $\hat{k}=k/L$ fixed, so that the body remains geometrically similar as $Re$ increases.  

The analysis of \cite{quan2022inertial} (with some pedagogical simplifications introduced in \cite{eyink2023turbulenceII})
begins with the fine-grained momentum balance/Navier-Stokes equation smeared against a smooth space-time
test function $\barphi(\bx,t):$ 
\begin{equation}\lb{NSphi} 
    \begin{aligned}
       \int_0^T\int_{\Omega} \left[ (\partial_t\barphi+\nu \Delta \barphi) \bdot \mathbf{u}^{\nu}\right. & \left.+\, \grad\barphi\bdots(\mathbf{u}^{\nu}\otimes\mathbf{u}^{\nu} 
        + p^{\nu}\mathbf{I}) \right] \, dV dt  \\
        & \vspace{50pt} = \int_0^T\int_{\partial\Omega} \left[\btau_w^\nu\bdot\barphi - p_w^\nu(\bn\bdot\barphi) \right] \,dS\,dt. 
    \end{aligned}
\end{equation}
Note that I am using the standard non-dimensionalization with $\hat{\nu}=1/Re,$ but I  drop hats $\hat{(\cdot)}$
to keep notations simple. In contrast to usual discussions of weak solutions in domains $\Omega$ with boundaries, however, 
the test functions adopted in this study need {\it not} vanish near the surface $\partial\Omega,$  and thus the smeared 
balance equation gets surface contributions both from the skin friction $\btau_w$ and also from $p_w,$ the pressure 
at the wall. Under the assumption of strong $L^2$-convergence in the spacetime domain $[0,T]\times \Omega$ as 
$\nu\to 0$ (see section \ref{sec:InfRe}), we see that the lefthand side converges and thus also the limits exist,  
$\btau_w=\lim_{\nu\to 0} \btau_w^\nu,$ $p_w=\lim_{\nu\to 0} p_w^\nu,$ as distributions on the surface $\partial \Omega,$
so that 
\begin{equation}\lb{Eulerphi} 
    \begin{aligned}
       \int_0^T\int_{\Omega} \left[ \partial_t\barphi \bdot \mathbf{u} \right. & \left.+\, \grad\barphi\bdots(\mathbf{u}\otimes\mathbf{u}
        + p\mathbf{I}) \right] \, dV dt  \\
        & \vspace{50pt} = \int_0^T\int_{\partial\Omega} \left[\btau_w \bdot\barphi - p_w (\bn\bdot\barphi) \right] \,dS\,dt. 
    \end{aligned}
\end{equation}
The limiting fields $(\bu,p)$ in $[0,T]\times \Omega$ thus constitute a weak Euler solution, as seen by restricting 
\eqref{Eulerphi} to test functions $\barphi$ whose supports vanish on $\partial\Omega,$ but the equation 
\eqref{Eulerphi} gives additional information at the surface inherited from the Navier-Stokes solutions. 

The key step is now to consider the corresponding smeared version of the coarse-grained momentum balance
\eqref{Windowed} 
\begin{align}\label{cgEuler}
    \int_0^T\int_{\Omega}\partial_t\barphi\bdot(\eta_{h,\ell}\Bar{\mathbf{u}}_{\ell}^\nu)\,dV\,dt &
    +\int_0^T\int_{\Omega}\grad\barphi\bdots\eta_{h,\ell}(\Bar{\mathbf{T}}_{\ell}^\nu+\Bar{p}_{\ell}^\nu\mathbf{I}-\grad\Bar{\mathbf{u}}_{\ell}^\nu)
    \,dV\,dt \\\nonumber 
    &=-\int_0^T\int_{\Omega}\barphi\bdot(\Bar{\mathbf{T}}_{\ell}^\nu \bdot \grad\eta_{h,\ell}
    + \Bar{p}_{\ell}^\nu\grad\eta_{h,\ell})\,dV\,dt
\end{align}
where I  have defined $\Bar{\mathbf{T}}_{\ell}^\nu:=\overline{(\bu^\nu\bu^\nu)}_\ell$ and also I  have recalled the 
definition $\widetilde{a}_{h,\ell}=\eta_{h,\ell}\overline{a}_\ell$ for any quantity $a.$ Taking the double limits first $\nu\to 0$
and subsequently $h,\ell\to 0,$ then it is easy to see that the lefthand side of \eqref{cgEuler} converges exactly to the 
lefthand side of \eqref{Eulerphi}. In that case, however, the righthand side of  \eqref{cgEuler} must converge also to 
the righthand side of \eqref{Eulerphi}. We thereby get results on {\it spatial cascade of wall-parallel momentum}
\be \lim_{h,\ell\to 0} -\bn\times \Bar{\mathbf{T}}_{\ell}\bdot\grad\eta_{h,\ell} =\bn\btimes \btau_w \lb{parallel} \ee 
with $\Bar{\mathbf{T}}_{\ell}:=\overline{(\bu\bu)}_\ell$, and on {\it spatial cascade of wall-normal momentum}
\be \lim_{h,\ell\to 0} \bn\bdot \left[ \Bar{\mathbf{T}}_{\ell}\bdot\grad\eta_{h,\ell}+ \Bar{p}_{\ell}\grad\eta_{h,\ell}\right]=p_w, \lb{perp} \ee 
both convergence statements interpreted in the sense of distributions on $[0,T]\times \partial\Omega.$ To obtain these results I  had to
make suitable choices for test functions $\barphi$ and, in particular, this allowed us to neglect
$\lim_{h,\ell\to 0} -\bn\times \overline{p}_{\ell}\grad\eta_{h,\ell} =0$ in the first statement. The physical meaning of these results 
are straightforward. The first result \eqref{parallel} simply states that the spatial cascade 
of tranverse momentum to the wall via turbulent nonlinear advection in the limiting Euler solution must match onto 
the direct infinite-$Re$ limit of the viscous skin friction at the wall.  Recall that $\grad\eta_{h,\ell}(\bx)=\bn(\bx) \theta_{h,\ell}'(d(\bx))
\simeq \bn(\bx) \delta(d(\bx)-h)$ so that the quantity on the lefthand side of \eqref{parallel} indeed represents
advective momentum flux toward the wall at distance $h$ from the wall. Likewise, the second result \eqref{perp} states that 
the spatial cascade of wall-normal momentum away from the wall (note the change of sign) in the limiting Euler solution must 
match onto the direct infinite-$Re$ limit of the kinematic pressure at the wall. These two results are exact analogues of that 
derived for scale-cascade by \cite{duchon2000inertial}, i.e. the matching of the expressions \eqref{DR}, \eqref{DPi} for nonlinear 
energy cascade  in the limiting Euler solution and the direct infinite-$Re$ limit of viscous energy dissipation \eqref{match}.  

Unlike the scale-cascade of kinetic energy studied by \cite{duchon2000inertial}, which can be understood very roughly 
as a deterministic version of the cascade of \cite{kolmogorov1941local,kolmogorov1941dissipation}, in our case the meaning 
of ``momentum cascade'' is distinctly different from the traditional one in the literature on wall-bounded turbulence.
I may recall that \cite{tennekes1972first,jimenez2012cascades}, and many others, posit the existence in canonical 
wall-bounded flows (pipe, channel, boundary-layer) of an ``inertial layer'' where the Reynolds stress or mean momentum flux
$-\langle u_x u_y\rangle$ ($x$ wall-parallel, $y$ wall-normal) is approximately constant and equal to $u_\tau^2=\langle \tau_{w,xy}\rangle,$
the mean skin friction.  This looks superficially similar to \eqref{parallel}, except that the latter is a deterministic result for the weak 
Euler solutions obtained in the infinite-$Re$ limit. However, the result \eqref{parallel} apparently becomes trivial in all of these canonical flows 
where there is only a weak anomaly and $\btau_w=\bzed!$ Conversely, both the traditional and the deterministic cascades occur 
in a turbulent pipe flow with a hydraulically rough wall, but at completely different wall normal distances. Indeed, the standard mean 
momentum cascade then holds in the traditional inertial layer $k\ll y \ll H,$ whereas result \eqref{parallel} holds in the range 
$\nu/u_\tau\ll h\ll k,$ i.e. within the roughness sublayer! Recall that our limit is one in which $Re_k\to\infty$ so that our flows 
are in the fully rough regime and completely inertial-dominated turbulence exists between the roughness elements. 

So far I  have not established a result equivalent to Onsager's $1/3$ H\"older exponent result. As glimpsed already 
by \cite{taylor1915eddy}, a corresponding condition has to do with continuity of the velocity at the wall. In fact, 
\cite{bardos2018onsager,bardos2019onsager} understood that the essential condition has to do with the vanishing 
of the {\it normal velocity component} approaching the wall, since it is the normal velocity which transports momentum 
and other conserved quantities to and from the wall. The specific condition that I  consider is that of \cite{drivasN2018onsager}:
\be \lim_{\delta\to 0} \|\bn\bdot\bu\|_{L^2([0,T],L^\infty(\Omega_\delta))} = 0, \lb{DN-cond} \ee
which means that the {\it maximum} wall-normal velocity in the $\delta$-neighborhood $\Omega_\delta=\{\bx\in \Omega: d(\bx)<\delta\}$
of the boundary $\partial\Omega$ must vanish as $\delta\to 0.$ 
With this assumption the advective transport of momentum to the wall can  be shown without much difficulty to vanish:
\be \lim_{h,\ell\to 0} \Bar{\mathbf{T}}_{\ell}\bdot\grad\eta_{h,\ell}=\bzed,  \ee
see \cite{quan2022inertial,eyink2023turbulenceII} for details. In that case, the previous results simplify drastically.
The first result \eqref{parallel} reduces to
\be \btau_w=\bzed,  \lb{tauw0} \ee 
which is the statement that condition \eqref{DN-cond} rules out a strong momentum anomaly. The second result \eqref{perp}
furthermore becomes
\be \lim_{h,\ell\to 0} \overline{p}_\ell\theta_{h,l}'(d(\bx)) = p_w. \lb{p-cont} \ee 
This is essentially a statement about commutation of two limits: the same pressure field at the wall is obtained either 
by taking the $Re\to\infty$ limit $p_w$ of the pressure $p_w^{Re}$ directly at the wall or instead by taking $Re\to\infty$ first 
and then taking the limit of the pressure $p$ of the Euler solution in the flow interior at points approaching to the wall. 
This commutation of limits may be loosely termed ``continuity of pressure at the wall.'' 

The last two results seem very satisfactory. As I  have already discussed in section \ref{sec:overture},
the mean skin friction $\langle \btau\rangle$ (e.g. a long-time average) seems to vanish quite generally.
Thus, absence of a strong momentum anomaly \eqref{tauw0} looks consistent with experiment. The second 
result \eqref{p-cont} on the continuity of pressure at the wall is very good news for LES modeling. As I  have 
discussed in section \ref{sec:overture}, the limiting drag force on a moving body at high-$Re$ seems 
to arise entirely from pressure forces (form drag) with negligible drag arising from viscous skin friction.  
Thus, the result \eqref{p-cont} means that the asymptotic drag on a body can be successfully calculated 
from the weak Euler solution, with the limiting pressure on the body correctly calculated from the Euler pressure 
in the flow interior after taking the limit approaching the body. However, it is a bit premature to take this as good news!
First, there is currently no consensus how to determine the pressure uniquely for the weak Euler solution 
and in particular what boundary condition should be imposed to replace the Neumann condition 
in \eqref{press} for the Navier-Stokes pressure
\citep{bardos2022c,derosa2023double,bardos2023hoelder,derosa2023full}.
Second, and more seriously, we shall see in section \ref{sec:wkstrng} that the assumption 
that $\btau_w\equiv \bzed$ identically leads to the paradoxical conclusion that all drag vanishes 
in the limit $Re\to 0$ and that the Navier-Stokes solution converges to the smooth potential 
Euler solution of \cite{dalembert1749theoria,dalembert1768paradoxe} with no drag. 

\subsubsection{Energy Cascade in Space and in Scale}\lb{sec:energy} 

Kinetic energy balance can be analyzed in a very similar manner. \cite{quan2022onsager} considered 
external flow past a body $B$ with fluid velocity constant at infinity, so that the total kinetic energy was infinite. 
The flow was thus divided into a potential part $\bu_\phi$ and a rotational part $\bu_\omega=\bu-\bu_\phi,$
so that the ``relative energy''  $E_{rel}(t)=\int_\Omega \left[\bu_\phi\bdot\bu_\omega+\frac{1}{2}|\bu_\omega|^2\right] \, dV$
(heuristically, total energy minus energy of the potential flow) remains finite. To keep notations simple, however, I  consider 
here internal flow in a bounded domain $\Omega$ as did  
\cite{bardos2018onsager,bardos2019onsager,drivasN2018onsager,chen2020kato}. 

The fine-grained kinetic energy balance for incompressible Navier-Stokes when smeared with a scalar test function $\varphi$ 
is easily calculated to be 
\begin{eqnarray} 
        &&   \int_0^T\int_{\Omega}(\partial_t+\nu\Delta) \varphi \cdot \frac{1}{2}|\mathbf{u}^\nu|^2\,dV\, dt
               + \int_0^T\int_{\Omega}\grad\varphi\bdot \left(\frac{1}{2}|\mathbf{u}^\nu|^2+p^\nu\right)\mathbf{u}^\nu \, dV\,dt\cr 
        && \hspace{30pt}  + \int_0^T\int_{\Omega} \nu(\grad\grad\varphi)\bdots \bu^\nu\bu^\nu \, dV\, dt \ = \ \int_0^T\int_{\Omega}\varphi \, \varepsilon^\nu\,dV\,dt
        \label{Enu}
    \end{eqnarray} 
with $\varepsilon^\nu=2\nu |\bS^\nu|^2$ the viscous energy dissipation. As in the previous section, I  allow the smooth test function $\varphi$
to be non-vanishing on the boundary $\partial\Omega$ but no boundary terms arise after integration by parts because of the stick b.c. 
Assuming strong convergence of $\bu^\nu,p^\nu$ in the bulk to $\bu,p$ as $\nu\to 0,$ 
it is easy just as in a periodic domain to obtain the inviscid limit 
 \begin{eqnarray} 
        &&   \int_0^T\int_{\Omega} \partial_t\varphi \cdot \frac{1}{2}|\mathbf{u}|^2\,dV\, dt
               + \int_0^T\int_{\Omega}\grad\varphi\bdot \left(\frac{1}{2}|\mathbf{u}|^2+p\right)\mathbf{u} \, dV\,dt\cr 
        && \hspace{70pt}   = \ \int_0^T\int_{\Omega}\varphi \, D \,dV\,dt
        \label{E0}
    \end{eqnarray} 
where 
\be \lim_{\nu\to 0} \int_0^T\int_{\Omega}\varphi \, \varepsilon^\nu\,dV\,dt =  \int_0^T\int_{\Omega}\varphi \, D \,dV\,dt. \ee 

One can also write the inviscid coarse-grained kinetic energy balance \eqref{II15}, window out the near-wall region, and then 
smear with the test function $\varphi$ to obtain
 \begin{eqnarray} 
        &&   \int_0^T\int_{\Omega} \partial_t\varphi \cdot \eta_{h,\ell} \frac{1}{2}|\overline{\bu}_\ell|^2\,dV\, dt
               + \int_0^T\int_{\Omega}\grad\varphi\bdot \eta_{h,\ell} \overline{\bJ}_e \, dV\,dt\cr 
        && \hspace{70pt}   = \ \int_0^T\int_{\Omega}\varphi \, \eta_{h,\ell} \Pi_\ell\,dV\,dt
              -  \int_0^T\int_{\Omega}\varphi \, \overline{\bJ}_\ell \bdot\grad\eta_{h,\ell} \,dV\,dt
        \label{Ehl}
    \end{eqnarray} 
where I  have introduced the spatial flux of coarse-grained kinetic energy: 
\be  \overline{\bJ}_e =\left(\frac{1}{2}|\overline{\bu}_\ell|^2+\overline{p}_\ell\right)\overline{\bu}_\ell + \btau_\ell\bdot\overline{\bu}_\ell.  \ee
Taking the limit $h,\ell\to 0,$ the lefthand side of the coarse-grained balance \eqref{Ehl} coincides with the lefthand side 
of the fine-grained balance \eqref{E0}. Thus, the righthand sides must also coincide and one obtains 
\be \lim_{h,\ell\to 0} \left( \eta_{h,\ell} \Pi_\ell - \overline{\bJ}_\ell \bdot\grad\eta_{h,\ell}\right)= D. \lb{E-cont} \ee 
In contrast to the result \eqref{DPi} obtained in periodic domains where anomalous dissipation $D$ matches
to inertial energy flux $\Pi_\ell$ alone, in wall-bounded domains there is an additional contribution from 
$- \overline{\bJ}_\ell \bdot\grad\eta_{h,\ell}$ which represents spatial energy flux to the wall. I mention 
in passing that, at any finite distance from the wall, the expression \eqref{DR} of \cite{duchon2000inertial} can also 
be derived here for $D$, as well as the 4/5th law \eqref{45th} and 4/15th law \eqref{415th}.  
   
A condition on vanishing of the wall-normal velocity analogous to \eqref{DN-cond} (but $L^3$ in time) suffices 
to show that the additional spatial flux contribution must vanish. However, the potential presence of this additional 
term raises the question whether the anomalous dissipation measure might have some contribution from the 
boundary, so that $D(\partial\Omega)>0.$ This question recalls the famous theorem of \cite{kato1984remarks} 
(see also \cite{sueur2012kato,bardos2013mathematics}, among others), according to which vanishing of the
viscous dissipation in the boundary strip $\Omega_{c\nu}=\{\bx\in \Omega: d(\bx)<c\nu\}$ for any constant $c>0,$ or
\be \lim_{\nu\to 0}\int_0^T \int_{\Omega_{c\nu}} \varepsilon^\nu \, dV \, dt = 0 \ee 
is necessary and sufficient that Navier-Stokes solution $\bu^\nu$ converges strongly in $L^2([0,T]\times\Omega)$
to the {\it smooth} Euler solution $\bu$ with the same initial data $\bu_0,$ over any time period $[0,T]$ for which 
this smooth Euler solution exists. This type of result is closely allied to the strong-weak uniqueness theory which I  
shall discuss in section \ref{sec:wkstrng}. A consequence is that vanishing of the energy dissipation in the ``Kato layer'' 
$\Omega_{c\nu}$  implies vanishing energy dissipation everywhere in $\Omega,$ unless the smooth Euler 
solution $\bu$ develops a singularity before time $T.$ \cite{bardos2013mathematics} have shown furthermore that 
the vanishing of energy dissipation in $\Omega_{c\nu}$ implies also that $\btau_w=\bzed.$
Here I  may recall the numerical study by 
\cite{nguyenvanyen2018energy} of a compact quadrupole vortex in 2D impacting on a flat, solid wall. 
At very high Reynolds numbers these authors find evidence for anomalous energy dissipation arising in the Kato 
layer from maximum vorticity values scaling as $\omega^\nu\sim 1/\nu.$ These high values at the wall imply that
for some points on the boundary, at least, skin friction $\btau_w=\nu\bomega^\nu\btimes\bn\not\to \bzed$ as $\nu\to 0.$
I return to these issues again in section \ref{sec:wkstrng}.  

\subsubsection{Vorticity Cascade in Space}\lb{sec:vorticity}  

The same methods as discussed previously can be applied also to vorticity and, in some 
respects, the analysis is much easier, because the vorticity balance is just the curl of the 
momentum balance already treated. I  describe here only a few key results, following the  
more extended treatment of \cite{eyink2023turbulenceII}.  

The fine-grained balance  of vorticity for the incompressible Navier-Stokes equation is, of course, 
the viscous Helmholtz equation:  
\be \partial_t\bomega^\nu=\grad\btimes(\bu^\nu\btimes\bomega^\nu-\nu\grad\btimes\bomega^\nu):=
-\grad\bdot\bSigma^\nu, \ee
where $\bSigma^\nu$ is a suitable anti-symmetric tensor representing vorticity flux through the flow volume
\citep{huggins1971dynamical,huggins1994vortex}. We might adopt this as the starting point to derive
the fine-grained vorticity balance in the limit $\nu\to 0$, but it is easier just to take $\barphi\to\grad\btimes\barphi$
in the ideal momentum balance equation \eqref{Eulerphi}. This yields directly:
 \begin{equation}\lb{Omegaphi} 
    \begin{aligned}
       \int_0^T\int_{\Omega} \left[ \partial_t (\grad\btimes\barphi) \bdot \mathbf{u} \right. & \left.+\, \grad(\grad\btimes\barphi)\bdots \mathbf{u}\otimes\mathbf{u}\right] \, dV dt  \\
        & \vspace{50pt} = \int_0^T\int_{\partial\Omega} \left[\btau_w \bdot(\grad\btimes\barphi) - \bsigma_w\bdot\barphi \right] \,dA\,dt, 
    \end{aligned}
\end{equation}
where surface integration by parts $\int_{\partial\Omega} p_w (\bn\btimes\grad)\bdot\barphi  \,dA=
-\int_{\partial\Omega} (\bn\btimes\grad)p_w \bdot\barphi \,dA$ was used in the last line to introduce  
the vorticity source $\bsigma_w=-\bn\btimes\grad p_w$ of \cite{lighthill1963introduction}. The terms in this 
equation have mostly transparent physical meaning. Note in particular that the $\btau_w$-term arises from 
viscous diffusion of vorticity away from the surface in the limit $\nu\to 0$, through the integration-by-parts 
identity
\be \int_\Omega(\grad\btimes\barphi)\bdot\nu\grad\btimes\bomega^\nu=
 \nu\int_\Omega \grad\btimes (\grad\btimes\barphi)\bdot\bomega^\nu+\int_{\partial\Omega}(\grad\btimes\barphi)\bdot\btau_w^\nu\, dA.  \ee 
We thus see that anomalous viscous diffusion of vorticity may remain as $\nu\to 0$ if  $\btau_w\neq \bzed.$ 

The corresponding coarse-grained vorticity balance is obtained by taking the curl of the coarse-grained 
momentum balance equation \eqref{Windowed} and can be written most simply as 
\begin{align}
    \partial_t (\eta_{h,\ell} \overline{\bomega}_{\ell}+\grad\eta_{h,\ell}\btimes\overline{\bu}_\ell)  
    +\grad\btimes\left[\grad\bdot(\eta_{h,\ell}\overline{\bT}_\ell)\right] \ = \ 
    -\grad\btimes{\mathbf f}_{\ell,h}. 
   \label{WindowOmega}
    \end{align}
In the time-derivative term I  used the identity 
$\grad\btimes\widetilde{\bu}_{h,\ell}=\eta_{h,\ell} \overline{\bomega}_{\ell}+\grad\eta_{h,\ell}\btimes\overline{\bu}_\ell,$ 
which makes clear that the balance equation is for the sum of a vorticity $\bomega=\grad\btimes\bu$ in the
flow interior and a vortex sheet $\bn\btimes\bu$ on the solid surface. Note that ${\mathbf f}_{\ell,h}$
is just the inviscid limit of the inertial-drag force defined in equation \eqref{drag}, which may also be written  
as ${\mathbf f}_{\ell,h}= -\left(\overline{\bT}_\ell + \overline {p}_\ell \mathbf{I} \right) \bdot \grad\eta_{h,\ell}.$
The term $ -\grad\btimes{\mathbf f}_{\ell,h}$ in the coarse-grained vorticity balance rationalizes the 
{\it force field method} used in aerodynamics and wake flows behind complex bodies, where the solid 
surface is replaced by a suitable body force \citep{vankuik2022non}. Smearing the coarse-grained vorticity
balance \eqref{WindowOmega} with a vector test function $\barphi$ and taking the limit $h,\ell\to 0$ recovers 
the same equation \eqref{Omegaphi} obtained from the fine-grained balance.  Note in particular that   
\be -\lim_{h,\ell\to 0} \int_0^T \int_\Omega \,  \barphi\bdot \grad\btimes{\mathbf f}_{\ell,h} \, dV\, dt 
= \int_0^T\int_{\partial\Omega} \left[-\btau_w \bdot(\grad\btimes\barphi) + \bsigma_w\bdot\barphi \right] \,dA\,dt. 
\lb{curlf-vort} \ee
which relates $-\grad\btimes{\mathbf f}_{\ell,h}$ to inviscid generation of vorticity at the body surface
and possible anomalous viscous diffusion out of the surface. \eqref{curlf-vort} follows directly 
from \eqref{parallel},\eqref{perp}.  


If the strong continuity condition for wall-normal velocity \eqref{DN-cond} holds so that $\btau_w\equiv \bzed,$ 
then one may derive another simple relation for the surface vorticity source of \cite{lighthill1963introduction}. 
In that case \citep{quan2022inertial}, the pressure-continuity result \eqref{p-cont} implies that
\begin{align}\lb{lighthill0} 
-\int_0^T dt\int_{\partial\Omega} dA\, \barphi\bdot(\bn\btimes\grad)p_w
&=\int_0^T dt\int_{\partial\Omega} dA\, (\bn\btimes\grad)\bdot\barphi p_w\\\nonumber
&=\lim_{h,\ell\to 0} \int_0^T dt\int_\Omega dV \, \grad\eta_{h,\ell}\btimes\grad\bdot\barphi \overline{p}_\ell \\\nonumber
&=\lim_{h,\ell\to 0} \int_0^T dt\int_\Omega dV \, -\barphi\bdot\grad\eta_{h,\ell}\btimes\grad\overline{p}_\ell \\
&=\lim_{h,\ell\to 0} \int_0^T dt\int_\Omega dV \,\barphi\bdot\left(\grad\eta_{h,\ell}\btimes\partial_t\overline{\bu}_\ell \nonumber
 +\grad\eta_{h,\ell}\bdot\bSigma_{\ell}\right)
 \end{align} 
where in the last line I  used the coarse-grained momentum balance in the flow interior, written as 
\be \partial_t\overline{\bu}_{\ell} = -\grad\bdot\overline{\bT}_\ell-\grad\overline{p}_\ell = \bSigma_\ell^*-\grad\overline{p}_\ell \ee 
where $\bSigma_\ell^*$ is the Hodge dual to anti-symmetric vorticity flux tensor $\bSigma_\ell,$ or in components
$\bSigma_{\ell,i}^*=\frac{1}{2}\epsilon_{ijk} \bSigma_{\ell,jk}.$ Relation \eqref{lighthill0} states that, under the condition
\eqref{DN-cond}, tangential pressure gradients at the body surface act in the inviscid limit as a local source of vorticity, 
which either  accumulates in the surface vortex sheet or else flows away via turbulent advection. 

Vorticity flux from the body surface can be directly related to drag. Relations of this type have been derived
by many authors, including Burgers, Lighthill and many others, as discussed in the very comprehensive 
review of \cite{biesheuvel2006force}. The paper of \cite{eyink2021josephson} derives such a relation
for the power consumed by the instantaneous drag force ${\bf F}_B(t)$ on a body $B$ held fixed in a flow with 
velocity ${\bf V},$ making connection with ideas in quantum superfluids. This {\it Josephson-Anderson relation}
\begin{eqnarray}\lb{JA} 
{\bf F}_B^\nu(t)\bdot {\bf V} &=& -\rho\int_\Omega \bu_\phi\bdot(\bu^\nu\btimes\bomega^\nu-\nu\grad\btimes\bomega^\nu)\, dV \cr
                                 &=& -\rho \int_{\Omega} \grad\bu_\phi\bdots\bu_\omega^\nu \bu_\omega^\nu \, dV 
                                 +\rho\int_{\partial\Omega} \bu_\phi\bdot\btau_w^\nu\, dA 
\end{eqnarray} 
is in fact a special case of a relation derived by \cite{howe1995force}. In the last line of \eqref{JA}, integration by 
parts has been used to rewrite the relation in a form which should remain valid in the limit $\nu\to 0,$ as pointed out by 
\cite{eyink2021josephson}. This inviscid limit has been rigorously derived by \cite{quan2022onsager} when the body 
surface $\partial B$ is mathematically smooth, i.e. when the body is very large. The Josephson-Anderson relation 
and the drag on the body are related to Onsager's dissipation anomaly by the balance equation of the kinetic energy 
$(1/2)|\bu_\omega|^2$ in the 
rotational flow $\bu_\omega=\bu-\bu_\phi,$ where $\bu_\phi$ is the stationary potential flow of 
\cite{dalembert1749theoria,dalembert1768paradoxe} with no drag. This balance takes the form 
\begin{eqnarray} 
        &&   -\int_0^T\int_{\Omega}\frac{1}{2}\partial_t\varphi|\mathbf{u}_{\omega}|^2\,dV\, dt
               -\int_0^T\int_{\Omega}\grad\varphi\bdot\left[\frac{1}{2}|\mathbf{u}_{\omega}|^2\mathbf{u}+p_{\omega}\mathbf{u}_{\omega}\right]
        dV\,dt\cr 
        && \hspace{30pt} =\int_0^T\int_{\Omega} \varphi \, \bu_\phi\bdot \btau_w \, dA\, dt 
         -\int_0^T\int_{\Omega}\varphi\grad\bu_{\phi}\bdots
        \mathbf{u}_{\omega}\otimes\mathbf{u}_{\omega}\,dV\,dt \cr
        && \hspace{100pt}  -\int_0^T\int_{\Omega} \varphi \, D(\bu) \, dV\, dt 
        \label{rotE}
    \end{eqnarray} 
with $\varphi$ a smooth test function that may not vanish at the boundary \citep{quan2022onsager}.  The 
Josephson-Anderson relation represents a transfer of kinetic energy into rotational fluid motions, 
which is then disposed by the dissipation $D$ into heat energy.   

\subsubsection{Weak-Strong Uniqueness and Extreme Near-Wall Events}\lb{sec:wkstrng}

A fact of fundamental importance for the theory of incompressible turbulence is that weak-strong 
uniqueness, as reviewed in section \ref{sec:open}, is {\it not} valid unconditionally in flows with solid walls, 
as it is for incompressible flows in periodic domains or in infinite space. \cite{bardos2014non} 
(and see also \cite{bardos2013mathematics,wiedemann2018weak}) have noted that additional 
conditions are required in wall-bounded flows to obtain weak-strong uniqueness and the first work
uses convex integration methods to construct a counterexample showing that uniqueness indeed fails
without such added hypotheses. One sufficient condition for weak-strong uniqueness is that the inviscid limit 
of the skin friction, $\btau_w=\lim_{\nu\to 0} \btau_w^\nu,$ should be vanishing as a distribution at all times 
on the solid surface, i.e. $\btau_w\equiv \bzed$ when smeared with any smooth space-time test function
\citep{bardos2013mathematics}. Another condition which guarantees the later is the uniform vanishing 
of wall-normal velocity \eqref{DN-cond}, which I  have already noted in \eqref{tauw0} implies 
vanishing skin friction \citep{quan2022onsager}. 

The gist of the weak-strong uniqueness result can be easily understood from the following sketch of a 
proof taken from \cite{quan2024weak}, which shows in the context of flow around a smooth body $B$
that even the weaker condition $\btau_w\bdot\bu_\phi\equiv 0$ on $[0,T]\times \partial B$ implies weak-strong 
uniqueness. The argument is based on the inviscid balance of global kinetic energy in the 
rotational flow, $E_\omega(\tau)=\frac{1}{2}\int_\Omega |\bu_\omega(\cdot,\tau)|^2\, dV,$ which is 
\be E_\omega(\tau) = E_\omega(0) + \int_0^\tau \int_{\partial\Omega}  \bu_\phi\bdot\btau_w \, dA\, dt
-\int_0^\tau \int_\Omega \grad\bu_\phi\bdots\bu_\omega\bu_\omega \, dV \, dt -\int_0^\tau \int_\Omega D\, dV\, dt. 
\lb{rotE-global} \ee 
This result can be obtained formally by taking $\varphi\equiv \chi_{[0,\tau]\times\Omega},$ the characteristic function of 
the set $[0,\tau]\times\Omega,$ in the local balance \eqref{rotE} and can be obtained rigorously by a careful limiting argument. 
Using the condition $\btau_w\bdot\bu_\phi\equiv 0$ and the fact that $D\geq 0,$ one obtains 
 \be E_\omega(\tau) \leq E_\omega(0) -\int_0^\tau \int_\Omega \grad\bu_\phi\bdots\bu_\omega\bu_\omega \, dV\, dt.  \ee 
Since $\bu_\phi$ is smooth and globally bounded, application of Cauchy-Schwartz inequality gives
\be E_\omega(\tau) \leq E_\omega(0) + C\|\grad\bu_\phi\|_\infty \int_0^\tau dt\, E_\omega(t) \ee 
and then the Gronwall inequality yields finally:
\be E_\omega(\tau) \leq E_\omega(0) \exp\left(C\|\grad\bu_\phi\|_\infty\cdot\tau\right) \lb{stable} \ee 
If $E_\omega(0)=0$ or equivalently $\bu(0)\equiv \bu_\phi,$ then $E_\omega(\tau)\equiv 0$
for all $\tau>0$ and thus $\bu\equiv \bu_\phi$ identically for all time. Careful examination of the proof
shows that this result remains valid even if $\|\bu^\nu(0)-\bu_\phi\|_{L^2(\Omega)}\to 0$ as $\nu\to 0,$
which allows for a viscous boundary layer in the Navier-Stokes initial data, as long as the energy 
$E_\omega^\nu(0)$ in that layer shrinks to zero as $\nu\to 0.$  Note that the estimate \eqref{stable} allows for 
instability of $\bu_\phi$ in the dynamical systems sense, so that a small perturbation may grow exponentially. 
However, the inequality \eqref{stable} shows that, if $\bu_\phi\bdot \btau_w\equiv 0,$ then the smooth 
potential Euler solution $\bu_\phi$ is {\it stable} in the sense of Hadamard well-posedness 
and that any Navier-Stokes solution whose initial data $\bu^\nu(0)$ converges to $\bu_\phi$ in $L^2$ as $\nu\to 0$
will converge globally to $\bu_\phi$ in the inviscid limit.  

In my opinion, a plausible conclusion is that weak-strong uniqueness is an inappropriate condition to 
select physically correct weak solutions of the Euler equations, unlike what has been often proposed in 
the mathematics literature. The clearest example is the constant flow past a smooth body,
the subject of the famous paradox of \cite{dalembert1749theoria,dalembert1768paradoxe},
where the potential Euler solution is smooth and stationary in time, thus existing globally without
blow-up. Nevertheless, drag does not seem to vanish in the $\nu\to 0$ limit, as shown by laboratory experiments 
like that of  \cite{achenbach1972experiments} for flow past a sphere and by numerical simulations such as 
that of \cite{chatzimanolakis2022vortex} for 2D flow past a disk of diameter $D$ impulsively accelerated to 
velocity $U.$ In the latter case, the initial conditions are exactly $\bu(0)=\bu_\phi$ at $t=0$ and, after a short period of a few 
mean-free-times when a kinetic description is required, the dynamics is well-described by incompressible Navier-Stokes. 
In principle, empirical observations like those above might be explained by instability of the potential Euler solution in a dynamical 
systems sense, so that small perturbations, e.g. due to external flow disturbances, molecular viscosity, or even thermal 
noise, grow exponentially in time. However, excluding for the moment external and thermal noise, there  
is an asymptotically exact solution of the 2D incompressible Navier-Stokes equation for the case of the impulsively 
accelerated disk as $\nu\to 0$ \citep{barlev1975initial} and although its time of validity is expected to be only 
$\sim D/U,$ this nevertheless suffices to show that $E_\omega^\nu(0+)\to 0$ where $t=0+$
is the initial time of validity of the Navier-Stokes description. In that case, the bound \eqref{stable} expressing
weak-strong uniqueness, if valid, would imply that $\bu=\lim_{\nu\to 0}\bu^\nu$ coincides with $\bu_\phi$ 
identically. Although I  have explained this result in the case of the impulsively accelerated body via the 
energy balance \eqref{rotE-global}, similar arguments apply when the body is accelerated over a finite 
time-interval. However, the empirical observations suggest instead that the limit $\bu$ is a dissipative 
weak Euler solution with initial condition $\bu(0)=\bu_\phi$ but distinct from $\bu_\phi$ and co-existing with that smooth,
stationary Euler solution. 

The latter conclusion can be correct only if all of the sufficient conditions guaranteeing the weak-strong
uniqueness result \eqref{stable} are in fact invalid. The possibility that $\btau_w\not\equiv\bzed$ is not in contradiction
with the general empirical observation that $\langle\btau_w\rangle=\bzed,$ since the long-time average could be zero 
if the condition $\btau_w\neq\bzed$ existed only in some early period of the acceleration. The condition
\eqref{DN-cond} on uniform vanishing of wall-normal velocity must also be violated if weak-strong uniqueness in fact
fails. It is worth noting that experimental and numerical investigations of wall-bounded turbulence at high-$Re$ have already 
observed very extreme events both of skin friction and of wall-normal velocity \citep{lenaers2012rare,orlu2020instantaneous}. 
Numerical simulations of turbulent channel flow show that, while $u_\tau^2=\langle\btau_w\rangle$ decreases slowly 
with $Re,$ the probability of large, rare fluctuations of the non-dimensionalized variable $\btau_w^+:=\btau_w/u_\tau^2$
increase with $Re$ \citep{lenaers2012rare}. The same study shows also very large fluctuations of the wall-normal
velocity $v=\bu\bdot/\bn$ in the viscous sublayer and buffer layer, with the probability of large values of $v^+=v/u_\tau$
increasing with $Re.$ These large $v^+$ fluctuations seem to be associated with explosive boundary-layer separation 
induced by strong near-wall vortices \citep{doligalski1994vortex}. The events predicted by the breakdown of the condition
\eqref{DN-cond}, however, are much more extreme and occur at very different wall distances in the flows with strong
dissipative anomaly, such as flow past a sphere or flow in a hydraulically rough pipe. In fact, restated in dimensional variables, 
the maximum value of $\bn\bdot\bu/U$ attained on the interval $\nu/u_\tau \ll h < \delta\cdot L$ must be independent 
of the choice of $\delta>0$ and also independent of Reynolds number for $Re$ sufficiently large. This is a striking prediction
of the breakdown of weak-strong uniqueness. 

There is some serious question, however, whether the high-$Re$ ranges considered are within the regime of validity 
of the standard Navier-Stokes description. Using the simple kinetic theory estimate for kinematic viscosity $\nu\sim c_s \ell_{mfp},$
where $c_s$ is the thermal velocity/sound speed and $\ell_{mfp}$ is the mean-free-path length,  one can see that the 
usual viscous unit of length at the wall $\delta_\nu:=\nu/u_\tau$  \citep{tennekes1972first} 
can be rewritten as $\delta_\nu\sim \ell_{mfp}/Ma_\tau$ where $Ma_\tau=u_\tau/c_s$ 
is the Mach number based on $u_\tau.$ In flows with a strong dissipative anomaly where
$u_\tau\sim U,$ this estimate becomes $\delta_\nu\sim \ell_{mfp}/Ma,$ which is only larger than $\ell_{mfp}$ by a factor 
of the inverse of the global Mach number $Ma=U/c_s.$ Note that this coincides with the estimate $\delta_\nu/L\sim 1/Re$
and thus corresponds to what in mathematics is called the ``Kato layer thickness''.  This distance is often just 
several microns in realistic flows. Furthermore, the value of $\btau_w^\nu$ is strongly fluctuating, as I have just discussed,
and the local fluctuating wall unit $\delta_\nu(\bx,t):=\nu /|\btau_w^\nu(\bx,t)|^{1/2}$ may therefore take values even a couple 
of orders of magnitude smaller. Such tiny lengths are dangerously close to $\ell_{mfp},$ where the validity of any 
hydrodynamic description must break down. Furthermore, thermal fluctuation effects may be expected to appear at
much larger length scales than $\ell_{mfp},$ just as in the bulk flow (sections \ref{empiric} and \ref{sec:open},({\it e})). 
Understanding the ultimate origin of drag on solid bodies at very high Reynolds numbers may therefore require an 
extension of Landau-Lifschitz fluctuating hydrodynamics to fluid-solid interfaces \citep{reichelsdorfer2016foundations}.

\subsection{Dissipative Euler Solutions and Zero-Viscosity Limit}\lb{wall:exist} 

Given the recent vintage of the work applying Onsager's ``ideal turbulence'' theory to wall-bounded flows, 
there is relatively little mathematical research attempting to construct corresponding weak Euler solutions. 
I thus remark only briefly about relevant works.  

I have already mentioned the application of convex integration methods by \cite{bardos2014non} to 
establish non-uniqueness of globally dissipative weak Euler solutions in a wall-bounded flow. A more
recent, interesting work by \cite{vasseur2023boundary} has applied the convex integration 
technique of \cite{szekelyhidi2011weak} for vortex sheets to the problem of turbulent channel flow,
constructing weak Euler solutions $\bu$ exhibiting boundary-layer separation,  although the initial data $\bu_0$
are given by the plug flow $\bU=U\hat{\bx},$ which is a smooth, stationary, potential Euler solution.
These weak solutions thus differ from plug flow, although the space-$L^2$ norm of the difference 
at each time is bounded as $\|\bu(,\cdot,t)-\bU\|^2_{L^2(\Omega)}\leq C H W U^3t,$ where $H$ and $W$
are the half-height and width of the channel, respectively.  Furthermore, \cite{vasseur2023boundary} study also 
Navier-Stokes solutions $\bu^\nu$ with stick b.c.  in the limit as $\nu\to 0$ and with initial data $\bu^\nu_0$ converging in
$L^2$ to the plug flow $\bU,$ proving that all weak limits $\bu^*$ as $\nu\to 0$ satisfy similar upper bounds as
the Euler solutions from convex integration, or $\|\bu^*(,\cdot,t)-\bU\|^2_{L^2(\Omega)}\leq C' H W U^3t.$
As noted already by \cite{vasseur2023boundary}, vanishing energy dissipation in the ``Kato layer'' near the wall 
would imply that all sequences $\bu^\nu$ of Navier-Stokes solutions must converge strongly in $L^2$ to $\bU$ as $\nu\to 0$. 
In standard terminology of wall-bounded flows, this ``Kato layer'' in turbulent channel flow corresponds to the viscous 
sublayer and buffer layer, over which volume the energy dissipation rate integrates to $u_\tau^3 W L $ 
for channel length $L$ \citep{tennekes1972first}.  This integrated dissipation tends to zero according to the standard 
Prandtl-K\'arm\'an theory  and then the theorem of \cite{kato1984remarks} implies that the inviscid limit  in this geometry
is just plug flow $\bU,$ in agreement with expectations in the fluid mechanics community
\citep{cantwell2019universal}. The methods of \cite{vasseur2023boundary} should yield more realistic 
results for bluff-body flows, such as flow past a sphere, with observable separation  and anomalous dissipation. 

The existing mathematical results already suggest that Eulerian spontaneous stochasticity as discussed
in section \ref{sec:nonuniq} should appear at very high Reynolds numbers also in wall-bounded turbulence. 
In particular, the non-uniqueness seen in convex integration constructions such as that of 
\cite{vasseur2023boundary} suggests that such stochasticity should be a general feature of 
separating turbulent boundary layers at high Reynolds numbers. The limits on predictability of 
turbulent wall-bounded flows due to spontaneous stochasticity will exacerbate the difficulties 
in data assimilation already arising due to ordinary chaos \citep{zaki2021limited,mons2021ensemble}.
Lagrangian spontaneous stochasticity should also appear in wall-bounded flows  
\citep{drivas2017lagrangianII} and this will have implications for turbulent vorticity dynamics
via stochastic Lagrangian representations 
\citep{constantin2011stochastic,eyink2020stochasticA,eyink2020stochasticB,wang2022origin}.
In particular Lagrangian spontaneous stochasticity should allow an alternative formulation of 
vorticity creation and flux from the boundary surface in the inviscid limit, alternative to the 
Eulerian description of turbulent vortex dynamics that was sketched in section \ref{sec:vorticity}.  

\section{Prospects}\lb{sec:conclude} 

The ideas proposed by \cite{onsager1949statistical} on an ``ideal turbulence'' at very high $Re$ have been 
the inspiration for extensive work by many researchers over the past three decades. I  have attempted 
to provide an overview of the great progress on this problem, emphasizing intuitive ideas rather  than rigorous 
mathematical details or exhaustive references to the literature. I  address finally here the key question: 
what is the promise for future advances? 

\noindent 

\subsection{How Do We Check If It's True?}\lb{sec:true} 

The works of the past decades have confirmed the remarkable insights of \cite{onsager1949statistical} 
on the mathematical description of Navier-Stokes solutions in the limit of high Reynolds numbers by weak Euler solutions 
exhibiting dissipative anomalies. As a matter of pure mathematics, several conjectures of \cite{onsager1949statistical} 
are now established by rigorous proofs based on ideas with unexpected connections to other areas 
of mathematics and others are partially demonstrated or reduced to empirically testable propositions. 
This does not, of course, mean that Onsager's theory is a correct description of physical reality, even apart from 
remaining mathematical gaps. For one thing, the basic assumptions of the theory could be invalid.  \cite{onsager1949statistical} 
began his deliberations by assuming that the incompressible Navier-Stokes model \eqref{NS} is valid 
for the turbulent flows at $Re\gg 1$, but even a more realistic model such as the Landau-Lifschitz equations 
\eqref{LLNS} might break down locally during extreme turbulent events \citep{bandak2022dissipation}. 
Furthermore, the  ``ultimate regime'' posited by the theory for the limit $\log Re\gg 1$ might not be 
attained in many practical circumstances and the observations on anomalous dissipation and drag 
could possibly have more mundane explanations.

The only way to really test the physical validity of the \cite{onsager1949statistical} theory is by 
careful laboratory experiments at high Reynolds numbers and, to a lesser extent, by numerical  
simulations. Some very basic predictions of the theory still remain unverified, such as the deterministic,
local $4/5$th-law \citep{duchon2000inertial,eyink2003local}, which are probably within the range 
of the current highest-resolution Navier-Stokes simulations. However, numerical simulations,
while very useful for their ability to access quantities that are difficult to measure by experiment, 
obviously incorporate too many theoretical assumptions and biases to provide definitive scientific tests. I hope
that one of the consequences of this essay may be a renewed interest in the experimental study of fluid 
turbulence at extremely high Reynolds numbers. It goes without saying that any experimental
study must involve solid walls and obstacles (such as the wire mesh or grid in a wind tunnel). In fact,
in my opinion, experimental investigations of turbulence have been too strongly influenced by the
bias of many leading theoreticians for ``simple'' or ``building block'' flows. Homogeneous and isotropic 
turbulence is one such flow that has obviously been in the past accorded a ``building block'' status, but 
to a lesser extent also the ``canonical wall-bounded flows''---plane-parallel channel flow, smooth pipe flow, 
and flat-plate boundary layer---have been accorded such a status. It is far from clear based on the considerations 
presented in this essay, however, that any of the previous ``simple'' flows are a good starting point for 
understanding more realistic turbulence. For example, in all of the ``canonical wall-bounded flows'', form drag 
from pressure forces vanishes identically and instead drag arises entirely from skin friction, whereas in most 
turbulent flows with walls the drag at high Reynolds numbers gets a non-negligible contribution from form drag, 
even in flows past streamlined bodies designed to minimize form drag. There are thus legitimate reasons to 
believe that the ``canonical flows''are not the best stepping-stones to understand wall-bounded turbulence in general. 


If I could be granted as a wish one new experiment, I would probably choose it to be an update of the   
experiment of \cite{achenbach1972experiments}, at much higher Reynolds numbers and taking advantage
of modern capabilities for measuring surface pressure, skin friction and near-surface flow characteristics.
Even better would be an experimental study of the transient regime of acceleration of the sphere from rest, 
at a sequence of ever higher Reynolds numbers. It could be somewhat simpler to investigate the flow past
a normal disk since, as illustrated in Fig.~\ref{fig_empirical_b}, there is no drag crisis and the drag coefficient 
appears to be nearly constant over four orders of Reynolds number \citep{roos1971some}.  A proper such study 
along these lines should vary also the external turbulence level, to determine its potential effect on all observations.  
Such a study could possibly detect the extreme events in the skin friction and the near-surface normal velocity 
that are predicted by breakdown of weak-strong uniqueness, under the assumption of a non-vanishing drag 
at asymptotically infinite Reynolds numbers (section \ref{sec:wkstrng}).  It is unclear what would be the exact 
nature of such events but they are plausibly associated to strong eruptions of fluid away from the surface. 
Otherwise, the experiments could verify the uniform vanishing condition \eqref{DN-cond} or more directly the 
local vanishing of skin friction. In that case, one might hope to verify the consequence \eqref{p-cont} on continuity 
of pressure at the wall. Or perhaps we would observe that drag eventually tends to zero after all! 
Either way, we would have more clarity about the mechanism by which drag appears and how it persists. 

However, any good experimentalist who reads this essay can doubtless think of their own 
tests that can be made of the predictions at very high Reynolds numbers. Such active and thriving experimental 
work is vital not only for Onsager's theory but for any theory of wall-bounded turbulence in the very high Reynolds regime. 

\subsection{Why Does It Matter?}\lb{sec:matter} 

If the ``ideal turbulence'' theory deals only with the presumed ultimate regime of near-infinite Reynolds 
numbers, then one might wonder how much it matters. Transition to turbulence and intermediate 
ranges of Reynolds numbers certainly have their own importance and not everything can be 
understood from a high-Reynolds-number theory. My own response to this important question has several
parts.

Ultimately, my basic interest is understanding the physics of how turbulence works and making sense 
of the diverse array of experimental observations. If there is indeed such an asymptotic high-$Re$ regime 
governed by weak Euler solutions, then describing this state has a fundamental importance for the 
understanding of turbulence. Not only would this infinite-Reynolds-number ``fixed point'' account for the 
high-Reynolds asymptotics but also it can be a suitable starting point for finite-Reynolds corrections, 
as discussed in  section \ref{sec:open}({\it c}), bridging the range of intermediate Reynolds numbers. 
As we saw there also, many of the arguments that have been developed are quite robust and do not depend 
upon the existence of a ``strong anomaly'', with merely a ``weak anomaly''sufficing. 

Another reason for the importance of the ``ideal turbulence'' theory is that there is already a large amount  
of discussion of the infinite-Reynolds-number limit in the fluid mechanics literature. Much of this work 
is based on less rigorous mathematics (e.g. \cite{cantwell2019universal,vankuik2022non}), or employs 
formal asymptotics whose time-range of validity is severely restricted by the underlying ansatz of smooth 
(and stable) boundary layers (see as examples \cite{barlev1975initial} or the \cite{prandtl1905ueber} theory, 
as recently reviewed by \cite{nguyenvanyen2018energy}). The Onsager theory can help to put this other 
body of work on better mathematical footing and, in some cases, to emend and correct it. 

In particular, the LES modelling community discusses frequently the infinite Reynolds number limit,
as mentioned already in section \ref{sec:RG-LES}. One segment of that community espouses 
the method of {\it implicit LES} which proposes to calculate turbulent flows at very high-Reynolds 
numbers via numerical discretizations of dissipative weak solutions of the Euler equations
\citep{hoffman2010computational,grinstein2011implicit,kronbichler2021efficient} and in fact the ideas of \cite{onsager1949statistical} 
are often cited explicitly in this context (e.g. see \cite{hoffman2015towards,fehn2022numerical}). The work on 
implicit LES is very interesting as a kind of ``computational experiment'' but the method still lacks a complete
mathematical and physical foundation. For example, in the presence of solid walls it is not yet clear from 
mathematical theory what are the appropriate boundary conditions for either velocity or for pressure. 
The open question regarding the pressure boundary was already mentioned in section \ref{sec:moment}. 
As to velocity, natural boundary conditions for smooth Euler solutions are vanishing wall-normal velocity, 
the so-called ``no-penetration condition''. However, we have seen that a sufficiently strong form of that condition 
for weak Euler solutions, \eqref{DN-cond}, implies weak-strong uniqueness and thus no limiting drag at infinite 
Reynolds number.  
Another issue in implicit LES which calls for serious attention by mathematicians is convergence.
So far as I know, none of the various numerical schemes that have been proposed to compute 
dissipative Euler solutions have been proved to converge under grid refinement. An alternative 
point of view is that Onsager's theory connects best to LES through the notion of ``coarse-grained 
solutions'' as in \eqref{wEuler} or \eqref{FEuler}, which may be interpreted as ``effective field theories''
with running parameters depending upon the scale $\ell.$ In that case, convergence is not a 
relevant issue, as discussed already in section \ref{sec:open}, and it is better to focus on getting 
answers independent of the arbitrary resolution scale $\ell.$ Methods from theoretical physics 
may be useful here, e.g. see \cite{goldenfeld1998block}. LES is an area that intersects strongly 
with Onsager's theory and provides fertile ground for interaction between modellers, mathematicians,
and physicists.

More broadly, the concept of ``ideal turbulence'' has importance beyond fluid turbulence proper, as it gives 
an example of an ``effective field theory'' in physics \citep{schwenk2012renormalization,liu2018lectures}, but of a 
novel type which corresponds to an ideal Hamiltonian system for which na\"{\i}vely conserved quantities are vitiated 
by anomalies. There is evidence to support such an ``ideal turbulence'' description also for systems beyond just 
high-Reynolds incompressible fluids, e.g. nearly collisionless plasma kinetics is plausibly described by dissipative 
weak solutions of the ideal Vlasov equation \citep{eyink2015turbulent,eyink2018cascades,bardos2020onsager}. 
Thus, Onsager's theory helps to deepen the connections between fluid mechanics and other areas of physics. 

\subsection{Last Words}\lb{sec:last}  

I leave this essay hopefully having explained what Onsager's ``ideal turbulence'' theory is all about,
which mathematical results are now established and which conjectures are still open,  and how observations
so far confirm the theory. As in any living area of science, this review may soon be outdated, either by new 
theoretical breakthroughs or by novel observations that call for revision of the theory. After working on the 
subject for more than thirty years, I am still fascinated by the high Reynolds-number limit of turbulence 
and I am looking forward to seeing further progress in the next couple of decades, possibly by readers of this essay! 

\vskip 0.1in


\vskip 0.1in

\noindent{\bf Acknowledgements.} I thank C. Bardos, D. Barkley, C. DeLellis, L. DeRosa, T. D. Drivas, B. Dubrulle, N. Fehn, 
A. Leonard,  A.~A. Mailybaev, J.~L. McCauley, Jr., C. Meneveau, G. Menon, K.~R. Sreenivasan, L. Sz\'ekelyhidi, Jr. and V. Vicol 
for very helpful comments.

\vskip 0.1in

\noindent{\bf Funding:}
We thank also the Simons Foundation for support of this work through the Targeted Grant No. MPS-663054, 
``Revisiting the Turbulence Problem Using Statistical Mechanics'', 

\vskip 0.1in

\noindent{\bf Declaration of interests.}  The authors report no conflict of interest.

\vskip 0.1in

\noindent{\bf Author ORCIDs.} G. Eyink \href{https://orcid.org/0000-0002-8656-7512}{https://orcid.org/0000-0002-8656-7512}.

\bibliographystyle{jfm}
\bibliography{jfm}

\end{document}